\documentstyle[12pt,epsf,epsfig,psfig]{article}
\newcommand{\gapproxeq}{\lower.7ex\hbox{$\;\stackrel{\textstyle>}{\sim}\;$}}
\newcommand{\lapproxeq}{\lower.7ex\hbox{$\;\stackrel{\textstyle<}{\sim}\;$}} 
\begin{document}

\def\3P0{$^3$P$_0$}
\def\A{{\rm A}}
\def\B{{\rm B}}
\def\C{{\rm C}}
\def\J{{\rm J}}
\def\L{{\rm L}}
\def\M{{\rm M}}
\def\P{{\rm P}}

\date{Aug. 2002}
\title{
\small \rm \begin{flushright} 
\end{flushright} 
\vspace{1.2cm}
\Large \bf Strong Decays of Strange Quarkonia \\
\vspace{0.8cm} 
}

\author{
T. Barnes$^{1,2}$\thanks{email: barnes@bethe.phy.ornl.gov, tbarnes@utk.edu},
N. Black$^2$\thanks{email: nblack@utk.edu}$\;$ and
P.R. Page$^3$\thanks{email: prp@lanl.gov}\\
{\footnotesize\it $^1$Physics Division, 
Oak Ridge National Laboratory,} \\
{\footnotesize\it Oak Ridge, TN 37831-6373}  \\  
{\footnotesize\it $^{2}$Department of Physics and Astronomy,
University of Tennessee, }\\
{\footnotesize\it Knoxville, TN 37996-1501} \\
{\footnotesize\it $^3$Theoretical Division, 
Los Alamos National Laboratory,} \\
{\footnotesize\it Los Alamos, NM 87545}  \\  
}

\vspace{1.5cm}
\maketitle

\begin{abstract}
In this paper we evaluate strong decay amplitudes and
partial widths of strange mesons (strangeonia and kaonia) 
in the \3P0 decay model.
We give numerical results for all 
energetically allowed open-flavor 
two-body decay modes of all $n\bar s$ and $s\bar s$ strange 
mesons in the
1S, 2S, 3S, 1P, 2P, 1D and 1F multiplets, comprising 
strong decays of a total of 43 resonances 
into 525 two-body modes, with 891 numerically evaluated amplitudes. 
This set of resonances includes
all strange $q\bar q$ states with allowed strong decays
expected in the quark model up to
{\it ca.} 2.2 GeV.
We use standard nonrelativistic SHO quark model wavefunctions to evaluate
these amplitudes, and quote numerical results for all amplitudes 
present in each decay mode. 
We also discuss the status of the
associated experimental candidates, and note which states and decay 
modes would be especially interesting for future experimental
study at hadronic, $e^+e^-$ and photoproduction facilities.
These results should also be useful in distinguishing 
conventional quark model mesons from exotica such as glueballs
and hybrids through their strong decays.
\end{abstract}
\newpage

\section{Introduction}

Strange quarkonia
are light ($u,d,s)$ mesons with at least 
one strange quark or antiquark
in their dominant $q\bar q$ valence component. These 
are known as kaonia if the dominant valence basis state is 
$n\bar s$ (where $n\equiv u,d$), antikaonia if
$s\bar n$, and strangeonia if $s\bar s$.

A principal goal of light meson spectroscopy 
is the identification of exotica, which
are resonances that are {\it not} dominantly $q\bar q$ states. These include
glueballs, hybrids, and multiquark systems. 
In the case of explicitly exotic quantum
numbers, such as the J$^{\rm PC}=1^{-+}$ exotic $\pi_1(1600)$ resonance, 
exotica can be
identified without a comparative study of the $q\bar q$ spectrum. Models 
of glueballs and hybrids predict however 
that the majority of light exotica will have 
nonexotic quantum numbers, and therefore 
must be identified against a background of 
conventional $q\bar q$ quark model mesons. 
In some cases, such as the scalar glueball,
there is evidence of strong mixing between the gluonic basis state
and $q\bar q$ quarkonium states. 
In these sectors it may be difficult to distinguish quarkonia from exotica,
although the overpopulation of experimental resonances 
relative to the naive $q\bar q$ quark model will
indicate the presence of the additional basis states. 

Searches for the expected rich spectrum of exotica 
with nonexotic quantum numbers will require
a well-established experimental meson spectrum over the relevant mass range of 
{\it ca.} 1.3-2.5~GeV, both to eliminate conventional 
quarkonium states and to study the 
possibility of a complicated pattern 
of mixing between exotica and conventional mesons.

The spectrum of meson resonances to 2~GeV is only moderately 
well determined at present, and
little is known regarding states above 2.2~GeV in any light quark sector.
The $n\bar n$ multiplets expected to $\approx 2.0$~GeV are 
1S, 2S, 3S, 1P, 2P, 1D and 1F, and of these
44 resonances, {\it ca.} 30 are now known.
Of the 22 corresponding kaonia expected to 2.1~GeV, {\it ca.} 13 are known. 
Strangeonia in contrast are a {\it terra incognita}. 
Counting the maximally mixed
$\eta$-$\eta'$ as one $s\bar s$ state, only 7 
probable
$s\bar s$ resonances 
of the 22 expected to 2.2~GeV are widely accepted, these being the 
$\eta$-$\eta'$, 
$\phi(1019)$, $h_1(1386)$, $f_1(1426)$,
$f'_2(1525)$, $\phi(1680)$ and $\phi_3(1854)$. As we shall see, 
there are controversies regarding
the nature of two of these states as well.

In this paper we give detailed theoretical 
predictions for the strong decay amplitudes 
of two-body decay modes of 
all the strange mesons 
expected in the quark model to {\it ca.} 2.2~GeV. 
These decay amplitudes and 
partial and total widths are derived in the $^3$P$_0$ model,
which (in several variants) is the standard model 
for strong decays. 
Since most experiments will rely on strong
decay modes and amplitudes to identify and classify 
meson resonances, we have derived decay
amplitudes 
to all open-flavor two body modes that are nominally
accessible. 
These results should be of use in establishing 
strange $q\bar q$ mesons, as well as in the  
identification of non-$q\bar q$ exotica.

Our results are presented in detailed tables 
of decay amplitudes, with entries for each resonance, decay mode
and amplitude. We also include a short discussion 
of each quark model state and associated
experimental candidates in the text, and
where possible we compare our theoretical decay amplitudes to the
data. We also
note especially interesting theoretical and experimental 
results. 

In most cases we assume pure $q\bar q$ mesons 
with definite J, L and S
as both initial and final states. In some cases,
such as kaonia with J = L, 
spectroscopic mixing is allowed and is known experimentally to be important, 
so we give results as functions of mixing angles. 
Finally, in certain channels
such as $0^{-+}$ and $0^{++}$ 
($\eta$-$\eta'$ and $f_0$ states)
mixing between basis states of different 
flavor appears to be a large effect, and our results in these channels
should be interpreted as decay amplitudes for initial ideal basis states, 
intended as a reference for 
contrast with
experimental decays of the more complicated mixed states.

The organization of the paper is as follows: After this introduction we 
summarize the \3P0
decay model used here; 
some additional technical details of the calculations are discussed in an
appendix.
After the decay model we discuss decays of strangeonia, 
and consider the status of states and give decay predictions
for all states in the 1S,
2S, 3S, 1P, 2P, 1D and 1F multiplets,
in that order. The following section carries out this exercise for kaonia. 
Our numerical results for these decay amplitudes and widths are presented 
in extensive 
decay tables following the text. 
Finally we give our summary
and conclusions, 
and suggest topics of interest for future studies of strong decays.

\section{The Decay Model\label{model}}

We employ the 
\3P0 decay model with SHO $q\bar q$ wavefunctions 
to evaluate two-body open-flavor strong decay amplitudes and
widths. 
This model of strong decays was introduced
over thirty years ago by Micu~\cite{Mic69}, 
and was applied extensively to meson decays in the 1970s by
LeYaouanc {\it et al.} 
\cite{LeY}. 
This decay model
assumes that strong decays take
place through the production of a $q\bar q$ pair with
vacuum quantum numbers ($0^{++}$, which corresponds to the 
\3P0 state of a
$q\bar q$ pair). 
After pair creation the $q^2\bar q^2$ system separates into two mesons 
in all possible ways, which corresponds to the two decay diagrams
shown in Fig.A1 of Appendix A. 
Hairpin diagrams are assumed absent, and in any case 
would not
be allowed by momentum conservation in this version of the 
\3P0 model. 

Since this model predates QCD and has no clear relation
to it, one might expect that a description of decays in terms of allowed
QCD processes such as one gluon exchange (OGE) might be more realistic. 
There is strong experimental evidence that the
$q\bar q$ pair created during the decay does have spin one ($S_{q\bar q}=1$), 
as is assumed in both the \3P0 and OGE decay models.
The strong experimental upper limit 
on the decay 
$\pi_2(1670)\to b_1\pi$ 
(from the VES Collaboration 
\cite{Ame99,PDG02}) of
\begin{equation}
B_{\pi_2 (1670)\to b_1\pi}\ 
< \; 1.9 \cdot 10^{-3}, \ \ 97.7\% \ c.l. 
\end{equation}
provides striking evidence in favor of
$S_{q\bar q}=1$.
In the $q\bar q$ quark model this is a
$1^1$D$_2 \to 1^1$P$_1 + 1^1$S$_0$ transition, and any 
$(q_i\bar q_i) \to (q_i\bar q_f) (q_f\bar q_i)$ transition
from a  
spin-singlet to spin-singlets has a vanishing
matrix element if the $q_f\bar q_f$ pair is created with spin one.

A recent detailed theoretical study of light meson 
decays from OGE 
pair production~\cite{Ack96} 
found that OGE decay amplitudes were typically rather smaller than required 
by experiment (the single exception found 
was $1^3$P$_0 \to 1^1$S$_0 + 1^1$S$_0$)
and hence are presumably masked by a dominant, nonperturbative
decay mechanism.
In addition, in certain 
decay amplitude ratios such as the D/S ratios in
$b_1\to \omega \pi$ 
(recently remeasured by the E852 Collaboration \cite{Noz02})
and
$a_1\to \rho \pi$ there is
a clear preference for $q\bar q$ production from
a \3P0 rather than an OGE source~\cite{Gei94}. 

It is widely assumed that the \3P0 model is successful because it 
gives a reasonably accurate description of
a nonperturbative $q\bar q$ pair production mechanism, such as
breaking of the gluonic flux tube between quark and antiquark sources
through production of a new $q\bar q$ pair along the path of the flux 
tube. 
Presumably, future studies of lattice QCD will lead to a more fundamental
description of this strong decay process. Here we simply
assume the $^3$P$_0$ model, 
because of its success as an approximate description of
much of the experimental data on strong decays.

Although the \3P0 model
is difficult to justify theoretically, it apparently does give a 
good description of many of the observed decay amplitudes and partial widths
of open-flavor meson strong decays. There have been many references published 
on the decays of light, strange mesons 
using variants of the \3P0 model 
(see Table \ref{calcs})
with different choices for the 
meson wavefunctions, the treatment of phase space, 
and the details of the \3P0 $q\bar q$ source.
The flux-tube decay model~\cite{Isg85a,Kok87} 
is one well-known generalization of the \3P0 model, 
in which
the source strength is assumed to
be largest along a path connecting the initial quark and antiquark.

\begin{table}[t]
\begin{center}
\begin{tabular}{||l|l|c|c|l|l|}
\hline 
Reference & Initial mesons considered & Amps. & Waves & p.s.& Wfns. \\
\hline 
\hline 
this work & 1S, 2S, 3S, 1P, 2P, 1D, 1F ($s\bar{s}\;\& \; K$) &
Yes & Yes & R &  SHO \\
Bur00a~\cite{Bur00a} & $f'_2(1525)$, $f_2(2010)$, $f_\J(2220)$, $f_2(2150)$, 
& No & No & R, M & SHO \\
& $f_2(2300), f_2(2340)$  & & & & \\
Bur00b~\cite{Bur00b} & $h_1(1386)$, $2^1$P$_1$\ $s\bar{s}$ 
& No & Yes & R, M & SHO \\
Bur98~\cite{Bur98a} & $1^3$P$_0$\ $s\bar{s}$ & Yes & Yes & R, M & SHO \\
Str98~\cite{Str98} & $f_0(1370)$, $f_0(1500)$, $f_0(1710)$ 
& No & No & R & SHO \\
Rob98~\cite{Rob98} 
& 
$\phi$, 
$f_1(1510)$, 
$f'_2(1525)$, 
$1^3$P$_0$\ 
$s\bar{s}$, 
$\phi(1680)$,
& No & Yes & R, M & O \\
& (1S, 2S \& 1P) $K$, $K_2(1580), K^*(1717),$ & & & & \\
& $ K_2(1773), K_3^*(1776)$ & & & & \\
Ams96~\cite{Ams96} & $K_0^*(1412)$,
$1^3$P$_0$\ $s\bar{s}$ 
& No & Yes & R & SHO \\
Blu96a~\cite{Blu96a} & 
$\phi, f'_2(1525), K^*, K_0^*(1412), K_2^*(1429),$
& No & Yes & R, M & SHO, O\\
& $K_3^*(1776)$, $1^3$F$_2$ and $1^3$F$_4$ \ $s\bar{s}$, $K_4^*(2045)$ 
& & & & \\
Blu96b~\cite{Blu96b} 
& $K_1(1273)$ and $K_1(1402)$ 
& Yes & Yes & M & SHO, O \\
Kok87~\cite{Kok87} 
& 1S, 2S, 1P, 2P, 1D, $1^3$F$_4$ ($s\bar{s}\;\& \; K$) &
Yes & Yes & M & SHO, O \\
God85~\cite{God85} & 1S, 2S, 1P, 1D, 1F ($s\bar{s}\;\& \; K$) &
Yes & Yes & M &  SHO \\
God84~\cite{God84} & $1^3$F$_2$ and $1^3$F$_4$\ $s\bar{s}$ &
No & No & M & SHO \\
Bus83~\cite{Bus83} & $\phi(1680)$  &
No & No & R & SHO \\
\hline 
\end{tabular}
\caption{\label{calcs} Some previous theoretical studies of
strange meson decays in the literature.
We indicate whether {\it amplitudes} are quoted, 
whether decay widths are displayed for the individual 
{\it waves}, and the 
{\it phase space} (relativistic (R) or 
mock meson (M)) and {\it wavefunctions} used (simple harmonic oscillator (SHO)
or other (O)).} 
\end{center}
\end{table}

We assume a fixed 
\3P0 source strength
(equivalent to the nonrelativistic limit of
an ${\cal L}_I = g\, \bar \psi_q\psi_q$ pair 
production interaction
Lagrangian \cite{Ack96}), simple harmonic oscillator (SHO)
quark model meson wavefunctions, and physical 
(relativistic) phase space.
The procedures we use to evaluate decay amplitudes and partial widths 
in this model are discussed in detail in~\cite{Ack96} and in 
\cite{Bar97a}; this paper is basically an application of the methods of 
the latter reference
to the strange sector. 
The decay model 
parameters assumed here (in the notation of Ref.\cite{Bar97a}) are
$q\bar q$ pair production amplitude
$\gamma=0.4$ and SHO wavefunction scale parameter $\beta=0.4$~GeV. We
assume physical, charge-averaged 
PDG values for the meson masses when there are clear
and relatively uncontroversial candidates for states, and otherwise 
use an estimated mass, based where possible on known states 
in the same multiplet or in the 
nonstrange flavor sectors. 
Further details of the decay calculations are presented in Appendix~A.

We use the \3P0 decay model to 
evaluate all decay amplitudes and partial and total widths
numerically for all the energetically allowed 
open-flavor two body modes of
all expected 1S, 2S, 3S, 1P, 2P, 1D and 1F $s\bar s$
and $n\bar s$
states. This is the most complete survey of strange meson
decays presented in the literature to date. For reference, 
in Table \ref{calcs} 
we summarize previous 
strange meson strong decay calculations.

Since we use a narrow resonance approximation, one should 
interpret our predictions carefully for modes that are close to 
nominal thresholds. 
Some near-threshold modes that are energetically 
forbidden 
may actually have significant branching fractions  
when width effects are included, as 
is noted in our discussions in several important cases.
One should also note that
amplitudes with large orbital angular momenta between the final state
mesons
are often
very sensitive to phase space, and hence to the assumed meson masses.

\newpage

\section{Strangeonia}

\subsection{General aspects} 

The study of strangeonia should enter a new era with
the advent of the new Hall D photoproduction facility 
GlueX at Jefferson Lab 
and the future 
upgraded $e^+e^-$ facilities VEPP (Novosibirsk) and DAPHNE
(Frascati).
In interactions with hadrons a photon beam can be regarded as 
a superposition of vector mesons with an important 
$s\bar s$ component, 
so studies of strange final states at GlueX
should lead to considerable improvement in our knowledge of 
the $s\bar s$ spectrum.
The study of diffractive photoproduction 
reactions, $\gamma p \to X p$,
should lead to the observation of
many 
C$=(-)$ $s\bar s$ states. At $e^+e^-$ facilities one of course 
makes 
only $1^{--}$ states significantly, which will provide an 
extremely interesting
case study of
a pure J$^{\rm PC}$ sector with 
broad overlapping resonances, presumably including vector hybrids as well 
as quarkonia. Central production has been shown 
at CERN and Fermilab to be very effective
in the production of candidate $s\bar s$ states such as axial vectors,
and it may be possible to use the STAR detector at RHIC similarly 
to study $s\bar s$ spectroscopy using pomeron and photon
processes. 

In previous experimental studies, 
strangeness-exchange reactions such as $K^-p\to X\Lambda$
were used as strangeonium production mechanisms. 
Unfortunately many of the more 
well-studied hadronic reactions, such as $\pi^- p$, 
have relatively weak $s\bar s$ production cross sections. 

Surprisingly little is known about the strangeonium sector 
experimentally, 
due largely to the weakness of experimentally accessible 
$s\bar s$ production cross sections.
Only three well-established  
resonances have been shown to be
dominantly $s\bar s$, these being the
$\phi, f'_2(1525)$ and $\phi_3(1854)$. 
(Negative searches or confirmations of
weak branching fractions 
to nonstrange final states are required to confirm
$s\bar s$ dominance.)
In this paper we hope to assist future
searches for strangeonia by giving detailed predictions 
for the strong
decay amplitudes of all $s\bar s$ mesons expected 
below {\it ca.} 2.2~GeV. 
For these calculations we employ the standard $^3$P$_0$
decay model, combined with SHO wavefunctions. This model has been tested
extensively in decays of light nonstrange mesons, and is known to 
reproduce the qualitative features of most strong decays reasonably
well, including relative amplitudes in several well-known test cases.
Although the model is not derived from QCD and is therefore of unknown
accuracy in its predictions in novel decay channels, it is the most accurate
description of strong decays available, and its predictions should 
at least serve as 
a useful guide in the search for higher-mass states. Once the model
proves to be accurate in a given channel, one can presumably trust the 
predictions in other flavor partners of that channel. Alternatively, 
a clear failure of the model may lead to important insights into the 
still poorly understood mechanism of strong decays.  

Although we consider all allowed open-flavor 
two-body decay modes, some of these
are especially characteristic of $s\bar s$ states.
Due to the OZI rule, the observation of a state with a large 
branching fraction to
$\eta\phi$, $\eta'\phi$ or $\phi\phi$ and small branches to
nonstrange final states can serve as a 
``smoking gun" for an initial $s\bar s$ state.
(This rule may need modification if gluonia are nearby in mass, 
as in the
scalar sector.)
The mode $\eta\phi$ is particularly attractive for identifying
C$=(-)$ $s\bar s$ candidates, 
and we strongly advocate 
the study of this final state in future experiments.
We emphasize that decays to open-strangeness final states
such as $KK$, $KK^*$
and $K^*K^*$ in isolation do {\it not} uniquely identify strangeonia,
since light-quark 
isosinglet mesons 
($(u\bar{u}+d\bar{d})/\sqrt{2}$) 
also decay to these open-strangeness final states. 

One might naively expect the higher-mass 
$s\bar s$ spectrum to simply
replicate the $n\bar n$ spectrum, {\it ca.} 200-250~MeV higher in mass. 
There is
already considerable evidence 
that this is not the case. First,
the near complete mixing of $n\bar n$ and $s\bar s$ states in the 
$\eta$ and $\eta'$ is very well established. Second, there is
circumstantial evidence that states in the scalar sector 
experience important
$n\bar n \leftrightarrow G \leftrightarrow s\bar s$
mixing,
specifically
in the unusual decay branching fractions of the three
states $f_0(1370)$,
$f_0(1500)$ and 
$f_0(1710)$. Similarly, the two known isosinglet 
$2^{-+}$ states $\eta_2(1645)$ and $\eta_2(1870)$ are both
observed in central production by WA102 
\cite{Bar00a}, 
with comparable cross sections into the nonstrange final state $\pi a_2$.
This suggests strong $n\bar n \leftrightarrow  s\bar s$
mixing in the $2^{-+}$ sector as well.
Thus we may find that the spectrum of states 
with hidden strangeness is rather more
complicated than a simple unmixed $s\bar s$ picture would suggest,
due to channel-dependent annihilation couplings of 
$n\bar n$ and $s\bar s$ basis states.

\subsection{1S States} 

There is a 
well-known problem with the decays of the 
lightest mesons that have allowed 1S$\to$~1S+1S strong transitions, such as 
$\rho\to\pi\pi$, $K^*\to \pi K$ (and here, $\phi\to KK$);
if we use parameter values fitted to a representative set
of higher-mass decays~\cite{Ack96,Bar97a}, these 
$ 1^3$S$_1 \to 1^1$S$_0 + 1^1$S$_0$ partial widths
are clearly underestimated. For  $\phi\to KK$ 
the predicted and observed
widths (Table S1) are
$\Gamma_{thy} = 2.47$~MeV 
and
$\Gamma_{expt} = 4.26\pm 0.05$~MeV \cite{PDG02}, so  
$\Gamma_{thy} / \Gamma_{expt} = 0.58$. 
Similarly, for the SU(3) partner decays
 $K^*\to \pi K$ and $\rho\to\pi\pi$ we find
$\Gamma_{thy} / \Gamma_{expt} = 0.42$ and $0.32$
respectively.
(These results follow from our
standard parameter set $\gamma=0.4$ and $\beta = 0.4$~GeV.)
The reason for this discrepancy 
relative to the decays of higher-mass states is not known; one
possibility 
involving reverse time-ordered ``Z-graph" diagrams 
has been discussed by Page, Swanson and Szczepaniak \cite{Pag99}. 

\subsection{2S States} 

\subsubsection{$\phi(1680)$} 

The $\phi(1680)$ is a natural candidate for the
$s\bar s$ radial excitation of the $\phi(1019)$, 
given its mass of $ca. \; 250$ MeV above the
2$^3$S$_1$ $n\bar n$ candidates 
$\rho(1465)$ and $\omega(1419)$ and the 
absence of an 
$\omega\pi\pi$ mode \cite{PDG02}. 
The
observation of the $\phi(1680)$ in $KK$ and $KK^*$ is sometimes
cited as evidence that this state is $s\bar s$. Of course this 
evidence is ambiguous, since $n\bar n$ states also populate these
modes; indeed, there is a danger of confusion of a
$\phi(1680)$ with an $n\bar n$ state such as the $\omega(1649)$
if one considers only open-strangeness decay modes.
True evidence for $s\bar s$ would be the observation of
large branching fractions to
hidden strangeness modes such as $\eta\phi$, or weak branching
fractions to all accessible 
nonstrange modes. 

Historically there has been considerable confusion about the
$\phi(1680)$, due in part to this ambiguity regarding the
$n\bar n$ versus $s \bar s$ origin of neutral $KK$ final states
\cite{Cle94}.
The first report of this state was by the DM1 collaboration
at DCI, 
in $e^+e^-\to K_LK_S$
\cite{Man81}, in which a rapid fall of the cross section
was interpreted as due to a new vector, the $\phi(1650)$. 
Similar behavior in 
$e^+e^-\to K^+K^-$ was also speculatively attributed to a
possible new vector meson by DM1
\cite{Del81}, and an excess of events in $e^+e^-\to K^+K^-$ above 1.15~GeV
invariant mass was noted by the VEPP-2M collaboration in Novosibirsk
\cite{Iva81}. This $e^+e^-\to K^+K^-$ reaction was subsequently studied
with slightly better statistics by DM2~\cite{Bis88}, who assumed
a $\phi(1680)$ to fit the cross section. 

Observation of a much larger signal in 
$e^+e^-\to K_SK^\pm\pi^\mp$~\cite{Man82} motivated fits with
interference between a $\phi(1680)$ and a $\rho'$ (which was needed
to explain the dominance of neutral over charged $KK^*$ states in this
channel); the fitted $\phi(1680)$ parameters were 
$\M=1677\pm 12$~MeV and $\Gamma=102\pm 36$~MeV.
A subsequent global fit by DM1 to these $KK$ and $KK\pi$ channels 
together with 
data on
$e^+e^-\to \omega \pi\pi$, 
$\rho \pi$, 
$\rho \eta$ and 
$\rho \pi\pi$ 
in a $\omega'$-$\rho'$-$\phi'$ model 
with interference gave $\phi(1680)$ resonance parameters 
$\M=1680\pm 10$~MeV and $\Gamma=185\pm 22$~MeV~\cite{Buo82}. 
DM2 next studied the reactions 
$e^+e^-\to KK\pi$~\cite{Bis91}
and
$e^+e^-\to \omega \pi^+\pi^-$~\cite{Ant92}
with improved statistics, and generally confirmed the DM1 results. Their
fitted $\phi(1680)$ parameters were
$\M=1657\pm 27$~MeV and $\Gamma=146\pm 55$~MeV,
and interference between this state and a $\rho'$ nearby in mass was
again used to explain the dominance of neutral over charged
$KK^* $ modes. A corresponding $\omega'$ near 1650~MeV was clearly evident 
in the 
$e^+e^-\to \omega \pi^+\pi^-$ cross section (see Fig.6 of Ref.\cite{Ant92}).
The only reported relative
strong branching fraction for the $\phi(1680)$ $e^+e^-$ state,
from DM1~\cite{Buo82}, 
is
\vskip 0.3cm
\begin{equation}
B_{\phi(1680) \to KK/KK^* } = 0.036\pm 0.004 
/ 0.49\pm 0.05.
\end{equation}
\vskip 0.3cm
\noindent
The PDG quote this as
$B_{\phi(1680) \to KK/KK^* } = 0.07\pm 0.01$. 

Photoproduction experiments have reached rather different conclusions 
regarding the ``$\phi(1680)$".
The CERN
Omega Spectrometer \cite{Ast81} found a $K^+ K^-$ enhancement
in $\gamma p \to K^+ K^- p$
centered at $\M_{K^+K^-} \approx 1.75$~GeV (see their Fig.4). A single
Breit-Wigner fit gave the parameters
$\M=1748\pm 11$~MeV and $\Gamma=80\pm 33$~MeV. 
They noted however that interference
effects can modify fitted resonance parameters, 
and in a model including interference
with light vector meson tails a lower mass was found,
$\M=1690\pm 10$~MeV and $\Gamma=100\pm 40$~MeV. 
A second 
Omega Spectrometer study of this process
by WA57~\cite{Atk85} advocated a single Breit-Wigner fit without interference,
which gave a mass and width of
$\M=1760\pm 20$~MeV and $\Gamma=80\pm 40$~MeV, 
consistent with the earlier 
photoproduction result.
Fermilab photoproduction experiment E401 
\cite{Bus89} studied 
photoproduction of $K^+K^-$ pairs at somewhat higher photon energies,
and confirmed the $\approx 1750$~MeV
enhancement; a Breit-Wigner fit gave the parameters
$\M=1726\pm 22$~MeV and $\Gamma=121\pm 47$~MeV. 
Finally, the FOCUS 
collaboration at Fermilab very recently reported a high-statistics
study of diffractive photoproduction of $K^+K^-$ \cite{Lin02}, 
and see a clear enhancement
with a fitted mass and width of
$\M=1753.5\pm 1.5 \pm 2.3 $~MeV 
and 
$\Gamma=122.2\pm 6.2\pm 8.0$~MeV,
again consistent with previous photoproduction experiments but with much 
smaller errors.
The $KK^*$ channel is also studied, and there is no evidence for the 
1750~MeV enhancement; in the neutral $KK^*$ channel a limit of
\begin{equation}
{ 
\Gamma_{X(1750)\to K^o K^{*o} \to \pi^+ K^- K_S +h.c.} 
\over
\Gamma_{X(1750)\to K^+K^-} 
} \ \
< 0.065 \ \ \ 90\% \ {\it c.l.}
\end{equation} 
is reported. Note that this is in striking disagreement with the
$KK^*$ dominance found for the $\phi(1680)$ state seen in $e^+e^-$.

In summary, $e^+e^-$ and
photoproduction experiments typically
find ``$\phi(1680)$" enhancements at masses that differ by 
$\approx 50$-$100$~MeV, with $e^+e^-$ reporting $KK^*$ dominance and
photoproduction reporting $KK$ dominance.
This may constitute evidence for two distinct states, although
interference with $n\bar n$ vectors may complicate a comparison of these
two processes.  
This issue can be addressed
by studying channels in which interference 
with $n\bar n$ vectors is expected to
be unimportant, notably $\eta\phi$, and by comparing the 
relative branching fractions to charged versus neutral modes in
decays 
to $KK$ and
$KK^*$.   

In our decay calculations (Table S2) 
we find that $KK^*$ is predicted to be the dominant
$2^3$S$_1$ $s\bar s$ decay
mode, 
as is observed for the $e^+e^-$ state 
$\phi(1680)$. 
We actually predict a $KK/KK^*$ 
branching fraction ratio of
$B_{\phi(1680) \to KK/KK^*}  \approx 0.35$, 
rather larger than the experimental ratio
$0.07\pm 0.01$. 
Our $2^3$S$_1$ $s\bar s$ decay predictions are 
in clear disagreement with the $KK$ dominance reported
by FOCUS 
for the $X(1750)$; evidently this state is not
consistent with the \3P0 model predictions for 
a $2^3$S$_1$ $s\bar s$ radial excitation.  

The $\eta\phi$ mode should be useful in establishing 
the true mass and width of the $2^3$S$_1$ $s\bar s$ state,
since interference with nonstrange vectors should be unimportant 
in this channel.
Our prediction of a 
branching fraction ratio 
of $B_{\phi(1680) \to \eta \phi /KK^* } \approx 0.18$ should
be reliable, since these decays are controlled by the same amplitude,
have similar phase space, and 
differ mainly through a flavor factor. We strongly encourage 
the study of the $\eta\phi$ channel in searches for evidence
of a $2^3$S$_1$ $s\bar s$ state in the 
``$\phi(1680)$" region.  

\subsubsection{2$^1$S$_0$ $s\bar s$ and the $\eta(1440)$ region} 

The 2$^1$S$_0$ $s\bar s$ state should theoretically have quite simple 
strong decay properties, 
assuming that $\eta$-$\eta'$ type flavor mixing
is unimportant in the radially-excited states. 
The only open-flavor mode is $KK^*$, 
which is a P-wave decay (Table S2). Since the 2$^1$S$_0$ $s\bar s$ 
state is presumably rather close
to $KK^*$ threshold, we find a total width that varies strongly
with mass; between $\M=1415$~MeV and 1500~MeV the 
predicted width increases from
11 to 100~MeV. Since the experimental $\eta(1440)$ is reported to
have an important $KK^*$ mode and has a total width (PDG estimate)
of 50-80~MeV, it appears plausible as a 2$^1$S$_0$ $s\bar s$
candidate.

Unfortunately the $\eta(1440)$ suffers from many complications
in the determination of its resonance parameters. One problem
is that the S-wave $f_1(1426)$ signal 
is typically present in the same reactions, and the 
$0^{-+}$ and $1^{++}$ contributions are difficult to separate. Another
problem is the strong $KK$ final state interaction, which distorts
the $\eta(1440) \to KK^*\to KK\pi$ invariant mass distribution
and leads to a low-mass $KK$ peak, 
which may be misidentified as a decay to 
$\pi a_0(980)$~\cite{Fra85}. 
If there actually is a strong $\pi a_0(980)$ mode, 
what this tells us about the $\eta(1440)$ is unclear because
the $a_0(980)$ itself is not well understood.
Finally, there are suggestions
of several $0^{-+}$ isosinglet states near this mass, 
because fits to the $\eta\pi\pi$ and $KK\pi$
final states give somewhat different masses for the parent resonances
\cite{PDG02}.
Of course this might also be due to final state interactions
or interferences that vary between channels.
The recent evidence from E852 (BNL) \cite{Ada01} for two resonances 
$\eta(1415)$ and $\eta(1485)$  
in the {\it same}
decay channel, $KK^*$, may be more significant.
If this is correct, the existence of the three states
$\eta(1295)$, $\eta(1415)$ and $\eta(1485)$ suggests the presence of
additional
degrees of freedom beyond the two I=0 $2^1$S$_0$ 
$q\bar q$ quark model 
states expected in this mass range.

The $\eta(1440)$ confusion may be dispelled through the study of different
production mechanisms and decay modes. 
Possibilities include 
$\gamma\gamma$ production (these rates can be calculated in the
quark model,
and checked against well-established $q\bar q$ states in this mass region) and 
flavor-tagging radiative decays such as 
$\eta(1440)\to \gamma \rho^o$,
$\gamma \omega$ and
$\gamma \phi$. 

There is a recent report from L3
\cite{Acc01a}
of a signal consistent with the $\eta(1440)$ in 
$\gamma\gamma \to K_S K^{\pm} \pi^{\mp}$, 
with a two-photon width of
\begin{equation}
\Gamma_{\gamma\gamma}(\eta(1440)) \cdot B_{KK\pi} = 212 \pm 50 \pm 23
\ \ {\rm eV} \ \cite{Acc01a},
\end{equation}
which is comparable to the larger of the theoretical expectations for the
two-photon width of a 
2$^1$S$_0$ $s\bar s$ state. 
(Scaling the Ackleh-Barnes result 
$\Gamma_{\gamma\gamma}(\pi(1300)) = 0.43$-$0.49$~KeV \cite{Ack92}
by $2/9$ for flavor and $(1.44/1.3)^3$ for phase space gives
$\Gamma_{\gamma\gamma}( 2^1{\rm S}_0 \ s\bar s) \approx  140$~eV.
Similarly scaling the M\"unz $\pi(1300)$ results, which use three
different models \cite{Mue96}, gives 
$\Gamma_{\gamma\gamma}( 2^1{\rm S}_0 \ s\bar s) \approx  30$-$100$~eV.)

Although little is known experimentally about the radiative transitions
of any higher-mass states, 
there is an early Mark III report of a 
large $\eta(1440)\to \gamma \rho^o$
partial width~\cite{Cof90} 
that, if confirmed, would invalidate the assumption that
this is a relatively pure $s\bar s$ state. Measurements of 
the radiative partial widths of the $\eta(1440)$ and other states through
high-statistics studies of $J/\psi \to \gamma \gamma V$
$(V =  \rho^o, \omega$ and
$\phi) $
would be very important experimental contributions, which
should be feasible at CLEO-c.

Until such data become available, we can summarize
the status of the $\eta(1440)$ (assuming that this is indeed a single state)
by noting that the reported total width and two-photon partial width
appear
consistent with expectations for a 
2$^1$S$_0$ $s\bar s$ state decaying dominantly to 
$KK^*$, but 
final state interactions 
may invalidate this agreement.

\subsection{3S States} 

\subsubsection{The unobserved $\phi(2050)$}

The 3$^3$S$_1$ $s\bar s$ vector state, to which we assign an estimated mass
of 2.05 GeV, is not known at present. This state should
be important in 
future spectroscopic studies, because with $1^{--}$ quantum numbers it can be 
made both in 
diffractive photoproduction and in $e^+e^-$ annihilation. 
A hybrid with the same quantum numbers and a similar mass is predicted
by the flux-tube model \cite{Isg85b,Bar95}, 
so overpopulation of this sector
may be anticipated. 

The \3P0 model
predicts that this will be a rather broad state, 
$\Gamma_{tot}\approx 380$~MeV (Table S3). 
In flux-tube decay models the corresponding $s\bar s$-hybrid is predicted
to be much narrower, $\Gamma_{tot}\approx 100$-$150$~MeV 
\cite{Pag99,Clo95}.
The dominant decay modes of the 3$^3$S$_1$ state
are predicted to be $K^*K^*$, $KK^*(1414)$ and $KK_1(1273)$, 
in order of decreasing branching fraction. 
All these lead to
important $KK\pi\pi$ final states. 
The large branching fraction for the 3S $\to $ 1S+2S 
transition to $KK^*(1414)$ may appear surprising, since this
decay amplitude has three nodes. These however are at 
$x=|\vec p_f|/\beta \approx 2.4,
4.5$ and $7.5$, 
rather far from the physical $x\approx 0.4$, so there is no
dramatic nodal suppression.
Assuming that the decay model is accurate, it will be very interesting to see
whether the problematical $K^*(1414)$ is indeed produced copiously in
$\phi(2050)$ decay, 
as expected if the $K^*(1414)$ is the 2$^3$S$_1$
state. 
Finally, the $KK$ mode is
near a node in the \3P0 decay
amplitude, and so is predicted to be very weak. 

A study of the $s\bar s$-signature
modes $\eta\phi$ and 
$\eta'\phi$ may be an effective experimental strategy for
identifying this state.
A 3$^3$S$_1$ $s\bar s$ 
$\phi(2050)$ is predicted to have 
significant branching fractions to both of these final states, whereas the
decay couplings of 
any $n\bar n$ state to {\it anything} $+ \phi$ should be weak.
Close and Page \cite{Clo95} anticipate that the $s\bar s$-hybrid vector
should also have a large $\eta\phi$ branching fraction, although
the $\eta'\phi$ mode of the hybrid should be weak.

We note in passing that since the $K^*$ and $\bar K^*$ are antiparticles,
neutral $(K^*\bar K^*)^o$ final states of definite
isospin have diagonal C-parity,
\begin{equation}
C\, |K^*\bar K^*\rangle_{L,S,I} 
= (-)^{L+S+I} \; |K^*\bar K^*\rangle_{L,S,I} \ .
\end{equation}
C-parity conservation forbids many transitions to VV states
that one might expect to appear in the
decay amplitude tables based on angular momentum alone.
The two C-forbidden amplitudes here are
$\phi(2050) \to K^*K^*$ ($^5$P$_1$)
and
$\phi(2050) \to K^*K^*$ ($^5$F$_1$).

\subsubsection{The unobserved $\eta_s(1950)$}

The \3P0 decay model predicts a relatively narrow 3$^1$S$_0$ $s\bar s$ state, 
with $\Gamma_{tot}\approx 175$~MeV, 
decaying dominantly to
$KK^*$ and $K^*K^*$ (Table S3). 
Experimental confirmation of this state
may be difficult 
despite the moderate width, 
due to small production cross sections
and the absence of 
characteristic $s\bar s$-signature decay modes such as 
$\eta\phi$.
Nondiffractive photoproduction of this C=$(+)$ state is expected to be weak,
since $\gamma \to V$ followed by nonstrange t-channel 
C=$(-)$ meson exchange 
does not lead to $s\bar s$ states (assuming the OZI rule).
As an $s\bar s$ state, the $\eta_s(1950)$ will
also have a small $\gamma\gamma$ coupling.

Radiative transitions from the $J/\psi$ may be a more
appropriate 
technique for identifying the $\eta_s(1950)$,
since 
$J/\psi \to \gamma\eta$ and $\gamma\eta' $ are both 
known to have relatively large branching fractions, and
no important $s\bar s$ suppression is expected
in this process.
Hadronic 
production of this state may also be 
effective in reactions with 
significant $s\bar s$ production cross sections.

\newpage

\subsection{1P States} 

\subsubsection{$f'_2(1525)$} 

This state is almost universally accepted as the $s\bar{s}$ member of the
$1^3$P$_2$ $q\bar q$ flavor nonet, together with the
$a_2(1318)$, $f_2(1275)$ and $K^*_2(1429)$. 
Although $n\bar n \leftrightarrow s\bar s$ mixing is allowed
in principle,
in practice the $f'_2(1525)$ appears to be close to pure $s\bar s$;
the mixing angle is strongly
constrained by the experimental $f'_2(1525)$ 
$\gamma\gamma$ coupling, which limits the
$n\bar n$ content to a few percent.

Our decay model predictions 
are in good agreement with the reported total width 
of $76\pm 10$~MeV (we predict 80~MeV) and the known 
partial widths, 
shown in Tables \ref{modes} and S4. There is a difficulty with this
comparison, however, which is that the PDG gives partial widths assuming 
that only the modes $KK$, $\eta\eta$ and $\pi\pi$ contribute
significantly. We find that the neglected mode $KK^*$ should actually 
be about as large as $\eta\eta$. 
There is only a weak experimental constraint on this mode at present,
$B_{KK^*}  < 0.35$ at 95\% $c.l.$~\cite{PDG02}. 

\begin{table}
\centering
\begin{tabular}{|l|c|c|c|c|}
\hline
mode: $\Gamma_i$ (MeV)
& $K K $
& $K K^* $
& $\eta \eta $
& $\pi\pi $
\\
\hline
\hline
$f'_2(1525)$ (expt)
&   $65 {+5 \atop -4} $ 
&   $- $ 
&   $7.6\pm 2.5 $ 
&   $0.60 \pm 0.12 $ 
\\
\hline
$f'_2(1525)$ (thy)
&   $61 $ 
&   $8.6 $ 
&   $10.4 $ 
&   $0$ 
\\
\hline
\end{tabular}
\caption{\label{modes}Experimental and theoretical 
partial widths of the $f'_2(1525)$. Note the unreported $KK^*$ mode.}
\label{table_f2}
\end{table}

\subsubsection{$f_1(1426), f_1(1510)$} 

The status of axial-vector states in this mass region has long been confused,
largely due to
the overlap of important $0^{-+}$,
$1^{++}$ and $1^{+-}$ amplitudes in $KK\pi$ hadroproduction 
near $KK^*$ threshold. 
Although some studies of phase motion of these amplitudes
have been reported \cite{Ada01,Bir88}, the statistics to date have not been 
sufficient to extract convincing individual 
resonance phase shifts in the pseudoscalar or axial-vector channels.

Three light, C $=(+)$ 
axial-vector isosinglets have been claimed experimentally, 
the $f_1(1285)$, $f_1(1426)$ and $f_1(1510)$. There is also evidence for the
$f_1(1285)$ and $f_1(1426)$ in $J/\psi$ radiative decays and 
$\gamma\gamma^*$, and some 
rather more controversial evidence 
in $J/\psi$ hadronic decays. The various reports 
of axial-vector signals were summarized recently by Close and Kirk 
\cite{Clo97}, who 
expressed skepticism regarding the existence of an $f_1(1510)$, and 
speculated that there might be 
significant $n\bar n \leftrightarrow s\bar s$ 
flavor mixing between the $f_1(1285)$ and $f_1(1426)$.

The historically confused experimental status of
light axial vectors has improved considerably 
with high-statistics 
central production experiments 
on $\eta\pi\pi$, $KK\pi$ and $4\pi$ states by WA102 (CERN)  
\cite{Bar97b,Bar98} 
and $KK\pi$ by E690 (Fermilab) \cite{Sos99}.
Central production 
of $KK\pi$ and $\eta\pi\pi$ 
in this mass region has been found to 
favor axial-vector quantum numbers strongly, and very clear 
$f_1(1285)$ and $f_1(1426)$ states are 
observed. There is no evidence of an $f_1(1510)$ in central production.

In view of their masses, the obvious assumption 
is that the $f_1(1285)$ is the light, dominantly $n\bar n$
$1^3$P$_1$
state, and the 
$f_1(1426)$ is its dominantly $s\bar s$ $1^3$P$_1$ partner. 
Since there is a controversy over the identification 
of the $f_1(1426)$ or the $f_1(1510)$ as 
the 
$1^3$P$_1$ $s\bar s$, in Fig.1 we show the \3P0-model
total width prediction for a range of 
$1^3$P$_1$ $s\bar s$ masses.
(The only open-flavor two-body mode below 1.77~GeV is $KK^*$.) The 
nominal threshold is 1390~MeV, however as 
the $f_1\to KK^*$ 
decay is dominantly S-wave we find that the width increases 
rapidly with increasing mass. At $\M = 1420$~MeV 
the predicted width is 
254~MeV, and the resonance envelope would obviously 
be strongly distorted by the nearby threshold,
which is at a $\Delta E << \Gamma_{tot}$. 
Other 
calculations of the $f_1(1426) \to KK^*$ width, also 
using the \3P0 model but 
taking threshold modification of the 
Breit-Wigner resonance shape into account, quote effective widths of
$\sim 70$ MeV~\cite{Rob98} and $\sim 120$ MeV~\cite{Kok87}. Thus the \3P0
model is roughly consistent with the reported 
width of the $f_1(1426)$, given the uncertainties in modeling the 
effect of the nearby $KK^*$ threshold.

\begin{figure}
$$\epsfxsize=4truein\epsffile{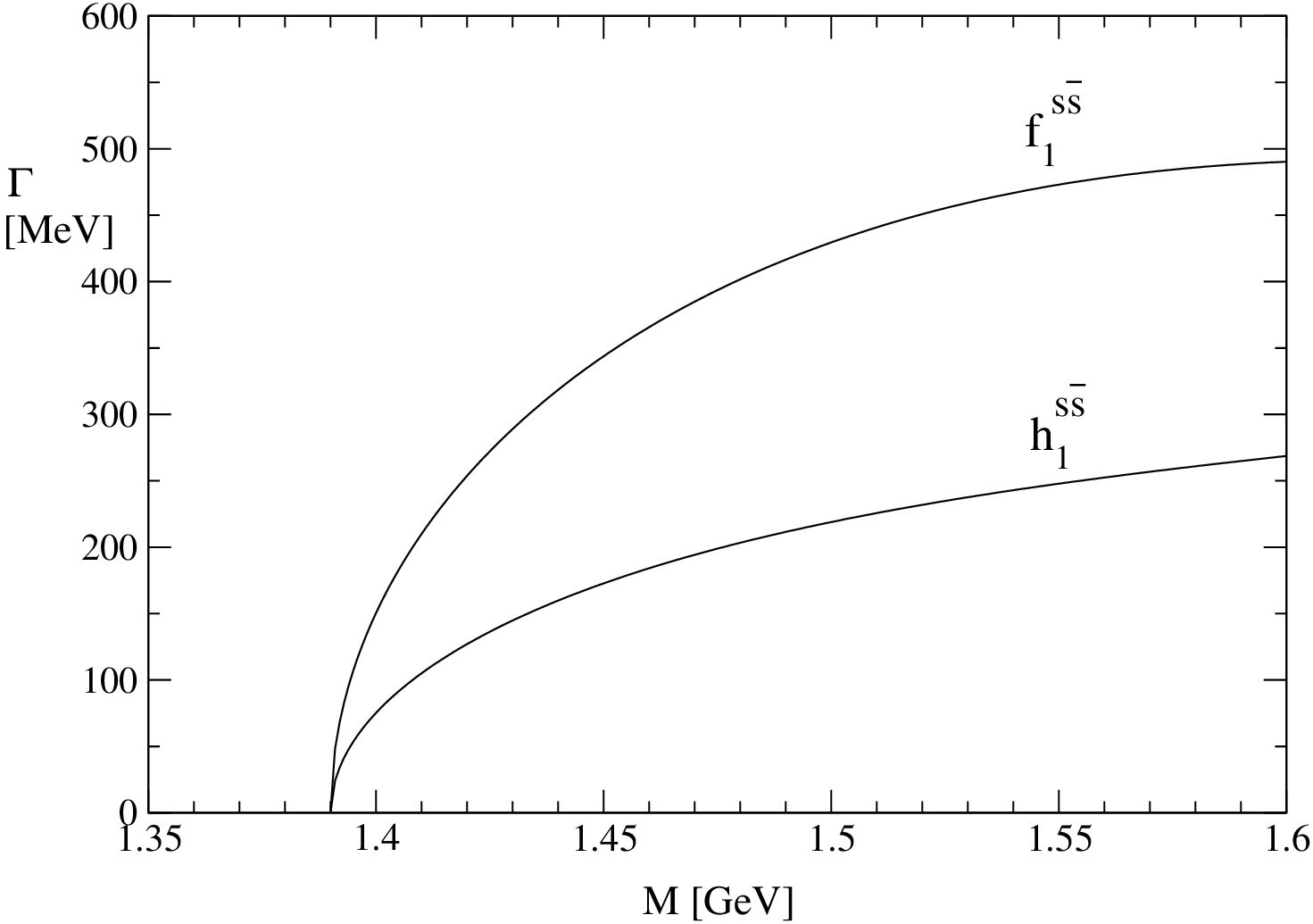}$$
\center
{
Figure~1.
Theoretical $KK^*$ widths of $1^3$P$_1$ $f_1$ and 
$1^1$P$_1$ $h_1$ $s\bar s$ states versus assumed mass.
}
\end{figure}

At the mass of 1530~MeV reported by LASS
\cite{Ast88a},
the theoretical width of a $1^3$P$_1$ $s\bar s$ 
$f_1(1530)$ is a very large 459~MeV. Assuming
the decay model is realistic for this channel, the 
relatively small reported width of $\Gamma = 100\pm 40$~MeV 
makes the $f_1(1530)$ appear implausible as a $1^3$P$_1$ $s\bar s$ state.

As the decay model predicts a quite strong coupling 
between the bare quark model $1^3$P$_1$ $s\bar s$ 
state  
and the $KK^*$ decay channel, and these are 
close to degenerate, it may be necessary
to 
treat this as a coupled $s\bar s$, $n\bar n$ and $KK^*$ system. This 
concern applies to the $1^1$P$_1$ $s\bar s$
sector as well. 

In future experimental work it will be important to test
the expected resonant phase motion of the
$f_1(1426)$ in channels in which this state 
and others are present with comparable amplitudes.
This is especially important here 
because of the possibility of 
misinterpreting 
a nonresonant threshold enhancement as a resonant state.

Future accurate measurements of radiative transition
rates of the 
$f_1(1426)$ and the other 
axial vectors will be of great importance 
in testing candidate $q\bar q$ assignments \cite{Don01,Clo02,Bon02}. 
Transitions 
such as 
$f_1(1426)\to \gamma \rho$ 
and
$\gamma \phi$
are flavor tagging, and will allow 
determinations of 
the amount of
flavor
mixing in the parent axial vectors.
(This is especially interesting because Close and Kirk 
\cite{Clo97} cite
evidence of important
$n\bar n \leftrightarrow s\bar s$ mixing in the 
axial vector system.) 
The absolute radiative transition rates are among the simplest and presumably 
most reliable quark model predictions for $q\bar q$ mesons, so a 
set of accurate measurements
of radiative partial widths to 
$\gamma \omega$, $\gamma \rho$ and $\gamma \phi$
could be definitive in establishing 
the nature of the axial vectors and other states in this
mass region.
A first 
measurement of the radiative transition
$f_1(1426)\to \gamma \phi $ has been reported by WA102 \cite{Bar97b}, 
who quote a relative branching fraction of
$B_{f_1(1426)\to \gamma \phi /  K K \pi} = 0.003\pm 0.001 \pm 0.001$, 
corresponding to
$\Gamma(f_1(1426)\to \gamma \phi) 
\sim 150$~keV (but clearly not yet well determined). Given the large errors,
this may be consistent with the theoretical 
expectation of $\Gamma_{thy}(f_1(1426)\to \gamma \phi) \approx 50$~keV
for pure $s\bar s$ initial and final mesons \cite{Godpri}. Evidently 
experimental accuracies of {\it ca.} 
10~keV  
will be required for definitive radiative transition tests 
of 
$s\bar s$ 
quark model assignments.

\subsubsection{$f_0(1500)$ and $f_0(1710)$} 

The scalar sector is of great interest, since
LGT predicts that the lightest glueball 
is a scalar 
with a mass 
near 1.7~GeV \cite{Mor99} 
(neglecting decays and mixing with quarkonia).
We also expect $1^3$P$_0$ 
$n\bar n$
and $s\bar s$ quark model scalars 
at masses of $\sim 1.4$~GeV and $\sim 1.6$~GeV respectively, so
the I=0 $0^{++}$ sector may be expected to 
show evidence of overpopulation relative to the $q\bar q$ quark model 
in this mass region. 

Ideally we might hope to distinguish a glueball from quarkonia
through anomalous decay or production amplitudes. 
Assuming unmixed $f_0$ $q\bar q$ states, we would expect the
$\pi\pi$ decay mode
to identify the $n\bar n$ state, whereas $KK$ and $\eta\eta$
final states 
should be populated by both 
$n\bar n$ and $s\bar s$.
To illustrate this,
in Table S4
we show the \3P0-model predictions for the decays of a pure
$s\bar s$ $f_0(1500)$; a total width of $\Gamma_{tot} = 279$~MeV
is predicted, with branching fractions of 
$B_{KK} = 76\% $ 
and
$B_{\eta\eta}  = 24\% $. 

In contrast to $s\bar s$, the
flavor-singlet decay amplitudes naively expected for an unmixed
glueball should populate {\it both} $\pi\pi$ and $KK$ modes.
The
relative flavor-singlet branching fractions (with phase space
removed) are
\begin{displaymath}
B({\bf \underbar 1}) /p.s.\  
(\pi\pi : KK : \eta\eta : \eta \eta' : \eta'\eta') =
\end{displaymath}
\begin{equation}
\hskip 1.7cm   3 \ :\ \; 4 \ \ :\  1\  :\ 0\ :\ \ 1 \ .
\end{equation}

Three I=0 scalar states are known in
this mass region, the $f_0(1370), f_0(1500)$ and $f_0(1710)$; 
the experimental status of these states was summarized recently
by Amsler \cite{Ams02a}. 
The branching fraction ratios 
reported by the Crystal Barrel and WA102 Collaborations 
for the $f_0(1500)$ and
$f_0(1710)$ 
are 
given in Table~\ref{f0}.
Neither of the higher-mass states shows the flavor-singlet decay pattern
expected for a scalar glueball; instead the $f_0(1500)$ strongly
favors $\pi\pi$ over $KK$, whereas the $f_0(1710)$ favors
$KK$ over $\pi\pi$.
Since the observed branching fractions of these states do not
match the expectations for decays of unmixed states,
several studies of $3\times 3$ mixing models have been 
carried out
in which the scalars are allowed 
$|n\bar n\rangle$,
$|s\bar s\rangle$ and 
$|G\rangle$
components; see for example Refs.\cite{Ams96,Lee00}. 
In these studies Amsler and Close 
\cite{Ams96,Ams02a} concluded that the 
$f_0(1370)$, $f_0(1500)$ and $f_0(1710)$ are dominantly
$n\bar n, G$ and $s\bar s$ respectively. 
In contrast, Weingarten {\it et al.}
\cite{Lee00} prefer the assignments
$f_0(1500)\approx s\bar s$ 
and
$f_0(1710)\approx G$.

\begin{table}
\centering
\begin{tabular}{|l|l||c|c|c|c||c|}
\hline
State & Experiment & $\pi\pi$ & $KK$ & $\eta\eta$ & $\eta\eta'$ & $4\pi$ \\
\hline\hline
$f_0(1500)$ & WA102 & $1$ 
& $0.33\pm 0.07 $ 
& $0.18\pm 0.003 $ 
& $0.096\pm 0.026$ 
& $1.36\pm 0.15$\\
& CBar  & $1$ 
& $0.184\pm 0.025 $
& $0.08\pm 0.04 $
& $0.065\pm 0.008 $  
& $1.62\pm 0.18 $\\
\hline
$f_0(1710)$ & WA102  & $1$ 
& $5.0\pm 0.7 $ 
& $2.4\pm 0.6$ 
& $< 0.18$ 
& $<5.4$ \\
\hline
\end{tabular}
\caption{Experimental branching ratios for $f_0(1500)$ and 
$f_0(1710)$ from the WA102~\protect\cite{Bar00b} and Crystal Barrel 
\protect\cite{Abe01a}
experiments, normalized to the $\pi\pi$ branching ratio.
} 
\label{f0}
\end{table}

There is evidence of an additional complication, which is that
the 
intrinsic strong decay amplitudes
of the basis states themselves are strongly model- and 
parameter-dependent.
Determination of the state mixing matrix from decay branching fractions,
{\it assuming slowly varying decay amplitudes} 
as is done in the mixing models may therefore lead to inaccurate results.
One concern is that the \3P0-model decay amplitude for 
$f_0^{q\bar q} \to $PsPs has a node at $|\vec p_f | = (3/\sqrt{2}) \beta
\approx 0.8$~GeV. This is close enough to the physical momenta of 
final pseudoscalars
to invalidate the use 
of simple relative flavor factors,
especially in $f_0(1710)$ decays. 
In addition, Ackleh {\it et al.} \cite{Ack96} 
found that the usually neglected
OGE decay amplitude is
anomalously large in $f_0^{q\bar q}\to $PsPs, so the \3P0 
decay amplitude may not be dominant in scalar decays. 
Finally,
there is evidence from LGT 
of violation of the naive
flavor-singlet G-PsPs 
coupling amplitude often assumed for a pure
glue state;
see Sexton {\it et al.} \cite{Sex96}.

Since these states may well have important flavor mixing, and the
strong decay amplitudes for scalars may have strong momentum dependence,
other aspects of these states should be studied for information regarding
their
Hilbert space decomposition.
In particular, 
radiative transitions 
may be a more appropriate approach for the 
identification of the $n\bar n$ and $s\bar s$ components of these states,
since one-photon transitions from $n\bar n$ basis states will populate
$\gamma \omega$ and $\gamma \rho^o$, whereas $s\bar s$ will populate
$\gamma \phi$ \cite{Ams95}. 
A simple study of the invariant mass distributions of
$\gamma \omega$ and $\gamma \phi$ should tell us a great deal about 
flavor mixing in the scalar sector.  

The two-photon couplings of these states may similarly be effective
in identifying their $q\bar q$ components, since the $1^3$P$_0$ 
$n\bar n$ scalar
is predicted to have a larger $\gamma\gamma$ width than any other
$q\bar q$ state. 
An $s\bar s$ state should naively have a two-photon width
about $2/25$ as large as its I=0 $n\bar n$ partner,
whereas a glueball should have a weak $\gamma\gamma$
coupling. 
(Vector dominance may modify this simple picture, for example 
if a glueball
has a large $\rho\rho$ coupling.) 
The recent strong L3 limit on the 
$\gamma\gamma$ partial width of the $f_0(1500)$ \cite{Acc01a}
may constitute evidence that
the $n\bar n$ component of this state is rather small.
In contrast, the $f_0(1710)$ may have been seen in 
$\gamma\gamma\to K_SK_S$
by L3 \cite{Acc95, Acc01b} and Belle \cite{Hua01}. 
(There is some disagreement between these experiments;
L3 favors dominance of $K_SK_S$ by J=2,
whereas Belle favors J=0.) 
Future experimental
studies of two-photon widths
should prove very interesting as tests of the nature of the
scalar states.

\subsubsection{$h_1(1386)$} 

The $h_1(1386)$ has been reported by only two experiments, 
LASS~\cite{Ast88a}
and
Crystal Ball~\cite{Abe97}. It is nonetheless a convincing candidate for 
the $s\bar s$ partner
of the $1^1$P$_1$ states $h_1(1170)$ and $b_1(1230)$, in view of its
mass and dominant decay to $KK^*$. ($KK^*$ is the only 
open-flavor decay channel available
to a $1^{+-}$ $s\bar s$ state at this mass.) The total width of
$91\pm 30$~MeV reported by the PDG is problematic because the state 
lies at
$KK^*$ threshold, so the  $KK^*$ mass distribution and effective
width will not be well described by a Breit-Wigner form.
We may compare the reported total width 
with expectations for a $1^1$P$_1$ $s\bar s$ state 
in a qualitative manner by 
varying the 
assumed $h_1$ mass. 
As we increase the mass from 1390 to 1440 MeV
(by roughly $\Gamma_{expt}/2$),
the predicted width varies from 0 to 160 MeV (Fig.1).
Since this range is qualitatively similar to the experimental
$91\pm 30$~MeV, the assignment of this state to $1^1$P$_1$ $s\bar s$
appears plausible.

Theoretical 
modeling of the $s\bar s$ state and $KK^*$ continuum 
as a coupled-channel 
problem, including the effect of the nonzero $K^*$ width,
should allow predictions of the expected $KK^*$ distributions for both
resonant ($s\bar s \leftrightarrow KK^*$) 
and non-resonant ($KK^*$ threshold enhancement) descriptions of the 
$h_1(1386)$.

\subsection{2P States} 

\subsubsection{$f_2(2000)$}

The 2$^3$P$_2$ $s\bar s$ tensor $f_2(2000)$ is predicted to be a
broad state, with a total width near 400 MeV, decaying dominantly 
to $KK^*$ and $K^*K^*$. (See Table S7 for 
decays of $2^3$P$_{\J}$ states.)
The $K^*K^*$ mode has 
three nonzero amplitudes, and the \3P0 model anticipates 
nontrivial 
relative strengths; the dominant S- and D-wave
spin-quintet amplitudes are predicted to
be comparable, $^5$D$_2$ / $^5$S$_2 = -0.59$, and the 
quintet and singlet 
D-wave amplitudes are in the ratio $^5$D$_2$ / $^1$D$_2 = -\sqrt{7}$.
(The spin-triplet amplitude $^3$D$_2$ is 
identically zero due to C-parity.) 

Unfortunately there are no $s\bar s$-signature modes open to
this state, given our assumed mass of 2000~MeV. However 
for this broad state we 
would expect to observe some coupling to the $s\bar s$-signature mode 
$\phi\phi$ above threshold,
and the intrinsic strength of this mode 
is quite large; at a mass of 2100~MeV,
the theoretical partial width is $\Gamma_{\phi\phi} = 143$~MeV.
The $^5$S$_2$ amplitude 
is dominant in $f_2(2100)\to \phi\phi$, 
however the D-waves should be observable
($^5$D$_2$ / $^5$S$_2 = -0.12$ given this mass) and 
have the same characteristic pattern as in $K^*K^*$, 
$^1$D$_2$ :  $^5$D$_2$ = 1 :  $-\sqrt{7}$.
($^3$D$_2$ is forbidden to $\phi\phi$ states by Bose symmetry.)

Since the experimental spectrum at this high mass is poorly 
established, it is not possible to identify clear
experimental candidates for this state. There is a 
LASS report \cite{Ast91} of a resonance in $K^* K^*$ 
with a mass and width of
$\M = 1950\pm 15$ MeV and $\Gamma = 250\pm 50$ MeV,
which might be this 2$^3$P$_2$ $s\bar s$ state. 
However, little is known about this
state at present;
possible J$^{\rm PC}$ quantum numbers include $1^{+-}$ and
$2^{-+}$ in addition to 
$2^{++}$, and the isospin has not yet been determined.

In view of the predicted strong coupling 
of the 2$^3$P$_2$ $s\bar s$ state to $\phi\phi$, the 
signals reported in
this channel in previous glueball searches 
should be assessed as possibly due to 
this state. 
(The other $2^{++}$ $s\bar s$
state expected near this mass, the $1^3$F$_2$ $f_2(2200)$, is
predicted to have a very weak $\phi\phi$ coupling.)
The reaction $\pi^- p \to \phi\phi n$  was studied at BNL \cite{Etk88}, 
and 
evidence for three $2^{++}$ states at masses of
$2011{+62\atop -76}$,
$2297\pm 28$
and
$2339\pm 55$ MeV was reported. 
The first of these BNL states is an obvious
candidate for the theoretical $2^3$P$_2$ $s\bar s$ $f_2(2000)$. 
The experimental $f_2(2011)$ was found to have a $\phi\phi$
partial width decomposition of  
$B(^5{\rm S}_2) = 98{+1 \atop -3}\% $,
$B(^5$D$_2$) = $0{+1 \atop {}}\% $ and
$B(^1$D$_2$) = $2{+2 \atop -1}\% $ 
\cite{Etk88}.
The S-wave is clearly dominant as predicted for
2$^3$P$_2$ $s\bar s$, however this is unsurprising given 
the lack of phase space. One might test a 2$^3$P$_2$ $s\bar s$
assignment for the $f_2(2011)$ by searching for this state in
$K^*K^*$ and $KK^*$ final states.

\subsubsection{The unobserved $f_1(1950)$}

The 
2$^3$P$_1$ 
$s\bar s$ state is predicted to be moderately
broad, with $\Gamma_{tot} \approx 300 $~MeV.
It may be most easily identified in the $KK^*$ mode,
in which it has a very characteristic dominance of D-wave
$KK^*$ final states over S-wave. 
Evidence for this unusual amplitude ratio has been reported
for the $a_1(1700)$ \cite{Ame95,Chu02}, 
which is a candidate 2$^3$P$_1$ I=1 partner 
of the $f_1(1950)$.

A nonexotic $n\bar n$-hybrid with J$^{\rm PC} = 1^{++}$ is
predicted at a similar mass in the flux-tube model \cite{Isg85b,Bar95}.
The Isgur-Paton 
flux-tube decay model predicts that this will
be a very broad state
\cite{Clo95}, however a $^3$S$_1$ variant of the flux-tube decay model 
studied by
Page, Swanson and
Szczepaniak \cite{Pag99} suggests that this hybrid might be rather narrow. 
In the latter case overpopulation 
of the $1^{++}$ sector of the quark model
near this mass might easily be confirmed.
The hybrid, unlike the 2$^3$P$_1$ $s\bar s$ state,
is predicted by Page {\it et al.} \cite{Pag99} to have a dominant S-wave
amplitude in its $KK^*$ decay mode.

\subsubsection{The unobserved $f_0(2000)$}

The 2$^3$P$_0$ $s\bar s$ $f_0(2000)$ is predicted to be
very broad, with a total width of
$\Gamma_{tot} \approx 800$~MeV. This is the largest total
width predicted for any of the states considered in this paper.
The dominant mode is expected to be
$KK_1(1273)$; this mode is also predicted to dominate the decays 
of another broad state,
the $1^3$D$_1$ $s\bar s$ $\phi(1850)$. The 
$f_0(2000)$ theoretically has sufficiently
strong couplings to $KK$ and $K^*K^*$ to be identified in those channels,
especially if the coupling to $KK_1(1273)$ and resulting very
large total
width are overestimated by the \3P0 decay model.
Unfortunately there are no characteristic $s\bar s$-signature modes
open to this state, with the possible exception of the very 
problematical channel $\eta \eta_s(1415)$.

\subsubsection{The unobserved $h_1(1850)$}

Unlike the other 2P $s\bar s$ states, the 2$^1$P$_1$ $h_1(1850)$
is predicted to be moderately narrow, with $\Gamma_{tot} = 193$ MeV.
(See Table S8.)
Only four open-flavor modes are accessible to the
$h_1(1850)$, and of these 
one ($KK_1(1273)$) is predicted to be 
numerically unimportant. The modes $KK^*$ and
$K^*K^*$ are largest, but the relatively large 
branching fraction predicted to the 
$s\bar s$-signature mode $\eta\phi$ ($B_{\eta\phi}\approx 15\%$) 
and the smaller backgrounds expected in this channel suggest that
$\eta\phi$ should be ideal for identifying the $h_1(1850)$.

The large photoproduction cross section reported for the
$1^1$P$_1$ 
$h_1(1170)$ \cite{Atk84}
makes the 2$^1$P$_1$ $h_1(1850)$ 
an attractive target for 
diffractive photoproduction. Since the flux-tube model
predicts nonexotic hybrids with these quantum numbers nearby in mass 
\cite{Isg85b,Bar95}, 
it will be important to identify this state as a ``background"
quarkonium resonance.

\subsection{1D States}

\subsubsection{$\phi_3(1854)$}

The $\phi_3(1854)$ was first reported 
in $K^-p \to \phi_3 \Lambda$
in
a 1981 CERN bubble-chamber experiment~\cite{Alh81}. It was reported in 
$KK$ and $KK^*$, with 
a total width of 50-120~MeV and a
relative branching fraction of
$B_{KK^*/KK} = 0.8\pm 0.4$. 
Subsequently in 1982 the Omega Spectrometer Collaboration 
\cite{Arm82} 
at CERN observed the $\phi_3$ in $K^+K^-$, and reported a mass and width of
$\M=$1850-1900 MeV and $\Gamma = $110-250 MeV. 
More recently the LASS Collaboration~\cite{Ast88b} observed the 
$\phi_3$ in $K^+K^-$ and $K_SK^\pm\pi^\mp$, 
and in several fits found masses and widths of
$\M \approx 1855$ MeV and $\Gamma \approx 60\pm 30$ MeV.
The PDG gives averaged masses and widths of $\M=1854\pm 7$
MeV and $\Gamma=87{+28\atop -23}$ MeV~\cite{PDG02}.    
A branching ratio of
$B_{KK^*/KK} = 0.55{+0.85\atop -0.45}$ was quoted by LASS.

In the \3P0 model with our parameters we predict  
a total $\phi_3(1854)$ width of 104~MeV and a $B_{KK^*/KK}$ 
branching fraction of 0.52 (Table S9), consistent with 
experimental estimates.
We also predict a large 
$K^*K^*$ mode,
with a relative $B_{K^*K^*}/B_{KK}$ branching fraction of 0.70. 
The $K^*K^*$ mode is interesting in that four independent amplitudes are
allowed; the \3P0 model predicts the $^5$P$_3$ $K^*K^*$ amplitude to be 
dominant and $^5$H$_3$ to be zero.
(Decay to the $^3$F$_3$ $K^*K^*$ state is forbidden by C-parity.)

\subsubsection{The unobserved $\phi_2(1850)$} 

The identification of this state would be very interesting,
as no $2^{--}$ states are known at present.
The $\phi_2(1850)$ is attractive experimentally because
the mass of the 1D $s\bar s$ multiplet 
is well established by the $\phi_3(1854)$,
and the total width is predicted to be relatively small,
$\Gamma_{tot}$ = 214 MeV. Only two decay modes are predicted to have large
branching fractions, $KK^*$
and $\eta \phi$. The latter is a very attractive $s\bar s$-signature
mode, which we expect to coupling strongly only to states
with large
$s\bar s$ components.

The $\phi_2(1850)$ can be diffractively photoproduced, 
although the strength of the $2^{--}$ photoproduction amplitude is not known.
The dominant $KK^*$ and $\eta \phi$
final states will allow tests of the \3P0 model, since these modes
are predicted to have significant
$^3$P$_2$ {\it and} $^3$F$_2$ amplitudes. 
We predict $^3$F$_2$~/~$^3$P$_2$ amplitude ratios of 
$-0.34$ for $KK^*$ and  $-0.19$ for $\eta\phi$ (Table S9).
A measurement of
this ratio in either decay would provide an important test
of the \3P0 model in a new angular channel; the 
existing accurate amplitude ratio
tests have only considered decays of L=1 mesons. 

\subsubsection{The unobserved $\phi(1850)$}

The $1^3$D$_1$ $s\bar s$ 
$\phi(1850)$ is predicted to be a very broad resonance, 
$\Gamma_{tot} \approx 650$ MeV, 
due to a very large S-wave coupling to the $KK_1(1273)$ decay channel.
Although this appears discouraging experimentally, one should note that
there has been no experimental confirmation of the theoretically 
very large
$1^3$D$_1\to\; 1^1$S$_0 + 1^3$P$_1$ 
and
$1^3$D$_1\to\; 1^1$S$_0 + 1^1$P$_1$ 
decay amplitudes in any flavor channel;
if the \3P0 model has significantly
overestimated these amplitudes, the $\phi(1850)$
might be considerably narrower.
Rather smaller couplings to
$KK$ and $KK^*$ are predicted, with branching fractions of
$\approx 10\%$. The branching fraction to the
$s\bar s$-signature
mode $\eta\phi$ is expected to be
$\approx 5\%$.

The 
very strong coupling predicted to $KK_1(1273)$ may be tested independently,
assuming that the $\omega(1649)$ and $\rho(1700)$ are the I=0 and I=1 
$1^3$D$_1$ 
$n\bar n$ 
partners of the hypothetical $\phi(1850)$. These $n\bar n$ states
are predicted to have analogously large decay amplitudes in
$\omega(1649)\to \pi b_1$ 
and
$\rho(1649)\to \pi a_1, \pi h_1$, which will presumably 
be studied in $e^+e^-$ at VEPP and DAPHNE.

Determination of the excited vector spectrum is of interest in
part because the flux-tube model anticipates vector hybrids
\cite{Isg85b,Bar95}, which
the existence of the $\pi_1(1600)$ suggests 
may be in this mass region. The vector $s\bar s$-hybrid is predicted
to have a rather smaller total width
than this $s\bar s$ quark model state  
\cite{Pag99,Clo95}.

\subsubsection{$\eta_2(1850)$}

Assuming a mass of 1850~MeV for the $1^1$D$_2$ $s\bar s$ state, 
only three open-flavor modes are accessible given our nominal masses,
$KK^*$, $K^*K^*$ and $KK_1(1273)$ (see Table S10). 
$KK^*$ is predicted to be 
dominant, with a branching fraction of $\approx 90\% $
and a rather large F-wave component, 
$^3$F$_2$~/~$^3$P$_2 = +0.52$.
The
remaining decays are expected to 
populate $K^*K^*$ almost exclusively. The predicted
total width is rather small, $\Gamma_{tot} = 129$~MeV, due to 
few open modes, limited phase space,  and the centrifical barriers
present in all cases.

Experimentally there are two known resonances with these quantum numbers,
the $\eta_2(1617)$ and $\eta_2(1842)$. In view of the mass of the I=1
$\pi_2(1670)$, these two $\eta_2$ states would appear to be 
$n\bar n$ and $s\bar s$ $1^1$D$_2$ candidates, although 
the total width of the
$\eta_2(1842)$, $\Gamma_{expt} = 225\pm 14$~MeV, is somewhat larger than
our estimate for the $1^1$D$_2$ $s\bar s$ state.  
Although LASS did not claim an isoscalar
$2^{-+}$ resonance, 
their data suggest an enhancement
at 1.8-1.9 GeV in the $2^{-+}$ $K^0_S K^{\pm}\pi^{\mp}$ 
partial wave 
in $K^-p\to K^0_S K^{\pm}\pi^{\mp}\Lambda$
(see Fig.2e of
Ref.~\protect\cite{Ast88a}). Since this production process enhances
$s\bar s $ relative to $n\bar n$, 
LASS may have evidence that the higher-mass $\eta_2$
is dominantly an $s\bar{s}$ state.
The $K^0_S K^{\pm}\pi^{\mp}$ final state
can arise from 
$KK^*$, which we predict to be the principal decay mode of the 
$1^1$D$_2$ $s\bar s$ state.

There are problems with unmixed $n\bar n$ and $s\bar s$ 
$2^{-+}$ assignments.
The $\eta_2(1842)$ has only been reported in 
$4\pi$ and $\eta\pi\pi$ modes, which are inaccessible to pure $s\bar s$
states in the \3P0 decay model. Both $\eta_2$ states were reported 
in
double diffraction 
to $\pi a_2$ 
by WA102, 
with comparable strengths 
(see Fig.3e of Ref.\cite{Bar00a}), which suggests important
$n\bar n \leftrightarrow s\bar s$ mixing if both states are 
indeed $q\bar q$.
The lighter $\eta_2(1617)$ has been reported by WA102 in both $\pi a_2$ and
$KK\pi$ \cite{Bar97b}, and the experimental ratio
$B_{KK\pi /\pi a_2} = 0.07\pm 0.03$ is not far from our prediction of
0.14 
for a pure $n\bar n$ $1^1$D$_2$ state \cite{Bar97a}. 
This suggests 
that flavor
mixing in the $\eta_2$ system is not very large, 
contrary to what is implied by the relative $\pi a_2$ strengths.

\begin{figure}
$$\epsfxsize=4truein\epsffile{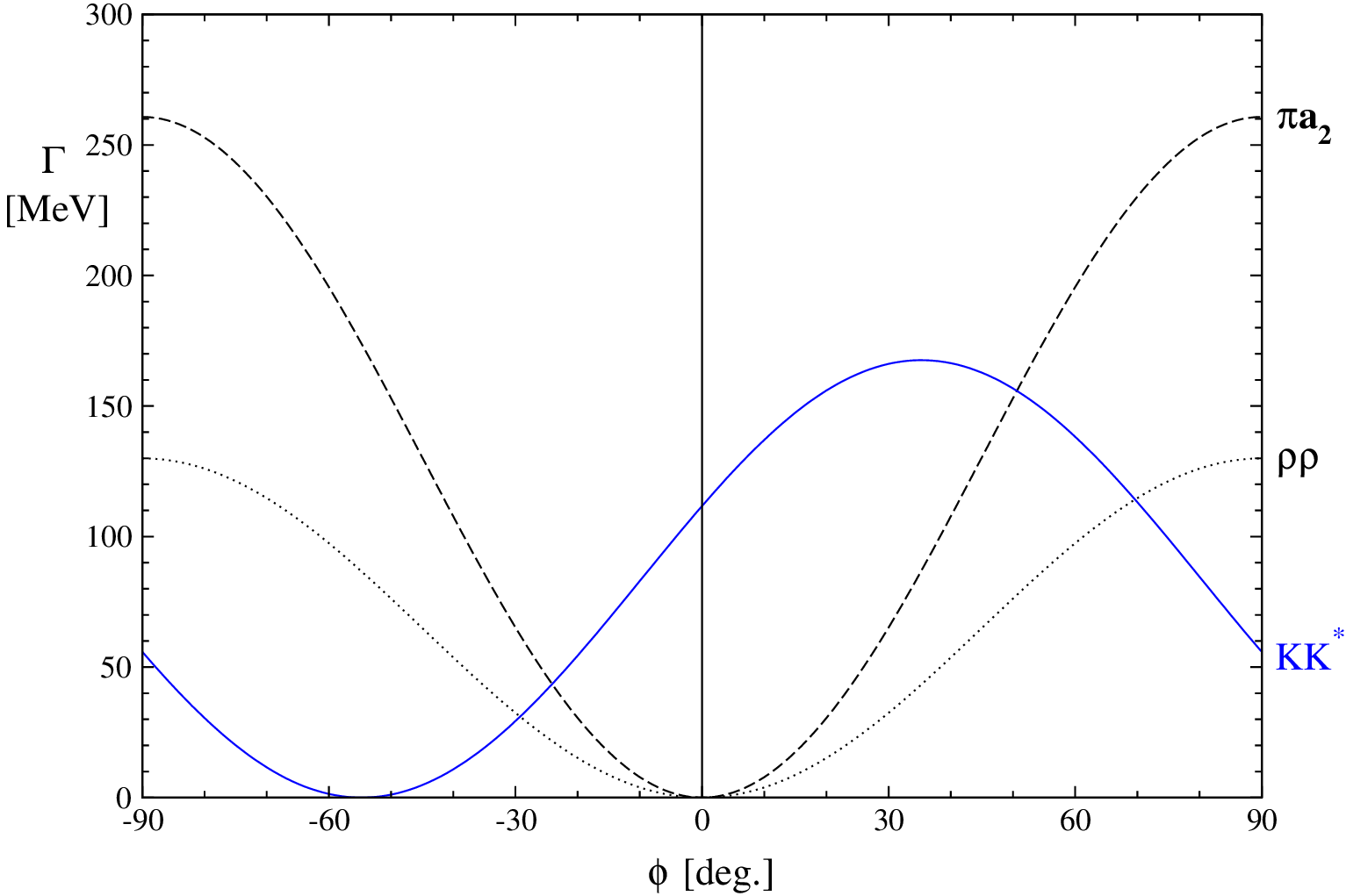}$$
\center
{
Figure~2.
Theoretical widths of the three leading modes of a 
flavor-mixed $1^1$D$_2$ $\eta_2(1842)$.  
}
\end{figure}

\begin{figure}
$$\epsfxsize=4truein\epsffile{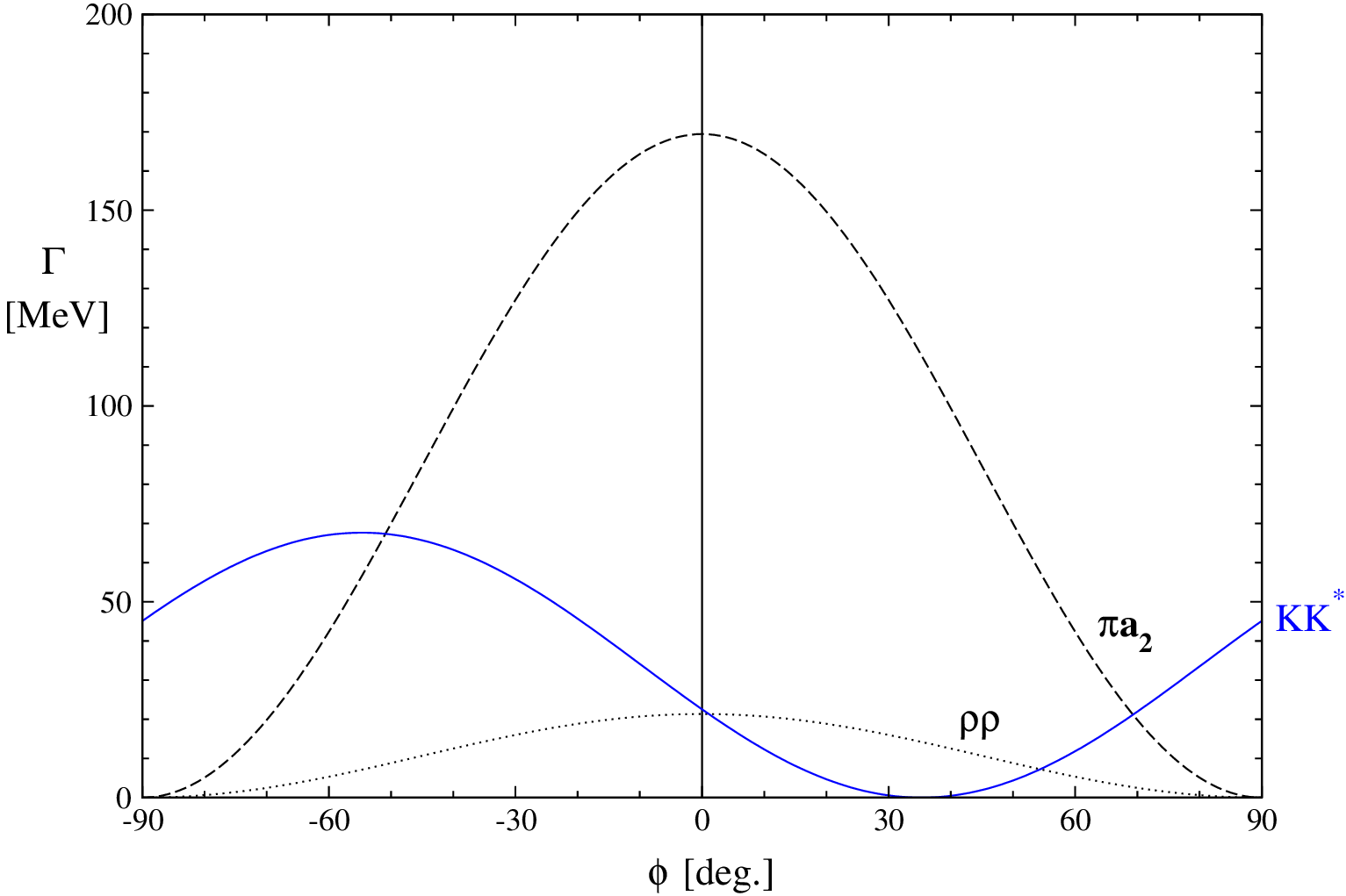}$$
\center
{
Figure~3.
Theoretical widths of the orthogonal partner 
$1^1$D$_2$ $\eta_2(1617)$.  
}
\end{figure}

We can test the possibility of significant $n\bar n \leftrightarrow s\bar s$
flavor mixing in the $\eta_2$ system by generalizing our \3P0 decay
calculations to mixed initial states
\begin{equation}
|\eta_2(1617)\rangle = 
\cos(\phi) \, 
|n\bar n \rangle_{_0}
-
\sin(\phi)\,  |s\bar{s}\rangle
\end{equation}
and
\begin{equation}
\hskip 0.28cm
|\eta_2(1842)\rangle =
\sin(\phi)\,  
|n\bar n \rangle_{_0}
+
\cos(\phi)\,  |s\bar{s}\rangle \ ,
\end{equation}
\vskip 0.5cm
\noindent
where we have assigned these the PDG experimental
masses. The resulting decay amplitudes
and partial widths are given in Tables S11 and S12.
The partial widths of the $\eta_2(1842)$ to the three
important modes $\pi a_2$, $KK^*$ and $\rho\rho$ 
are shown in Fig.2 
as functions of the flavor mixing angle $\phi$.
Evidently, large couplings to $\pi a_2$ 
and $\rho\rho$ follow from
moderate mixing, which could explain the WA102 observation of the
$\eta_2(1842)$
in $\eta\pi\pi$ and $4\pi$. 
Since the ratio $B_{KK^*}/B_{\pi a_2}$
is strongly dependent on the flavor mixing angle $\phi$, this ratio may 
be useful in determining $\phi$
if the $\eta_2(1842)$ is indeed a quark model state.

At present however the assignment of the $\eta_2(1617)$ and $\eta_2(1842)$
to a mixed-flavor quark model pair appears implausible, due to the
$KK^*$ final state.
The dominant decay modes of
an orthogonal partner state $\eta_2(1617)$ are shown in
Fig.3 and given in Table~S12. The facts that the
$\eta_2(1617) \to KK^*$ branching fraction observed by WA102 
is rather small (see Fig.2e of Ref.\cite{Bar97b}),
$B_{\eta_2(1617)\to KK\pi /\pi a_2} = 0.07\pm 0.03$, and that
there is no indication of the $\eta_2(1842)$ in this data,
argues against assigning both reported states to an 
$n\bar n \leftrightarrow s\bar s$ mixed pair;
our Fig.2 and Fig.3 show that there should be a fairly large $KK^*$
mode evident in the combined $\eta_2(1617)$ and $\eta_2(1842)$ signals,
whatever the mixing angle $\phi$. Only the quite weak $\eta_2(1617) \to KK^*$
transition is evident. 

An alternative possibility is that the higher-mass WA102 state 
$\eta_2(1842)$ is
an $n\bar n$-hybrid rather than a mixed $n\bar n \leftrightarrow s\bar s$
quark model state, and the PsV coupling of the hybrid is
rather small; for some reason the PsT mode $\pi a_2$ is preferred. 
Assuming the hybrid assignment,
we would expect to find evidence of
an I=1 $2^{-+}$ 
partner hybrid at a similar mass. 
There have been several reports of possible $\pi_2$ states in
this mass region, notably a D-wave $\pi f_2$ signal reported
by ACCMOR in 1981 \cite{Dau81}
that peaks near 1850~MeV. Several other possible 
higher-mass $\pi_2$ signals
are discussed in Ref.\cite{Bar97a}.
Quite recently a state with these quantum
numbers and resonant phase motion was reported
by the E852 Collaboration in $\rho^-\omega$ \cite{Pop01}, with a mass
and width of $\M = 1890 \pm 10 \pm 26$~MeV and 
$\Gamma = 350 \pm 22 \pm 55$~MeV. This exciting result may imply that
a flavor nonet of nonexotic 
$2^{-+}$ hybrids exists at a mass of 1.8-1.9~GeV (for $n\bar n$ flavor),
just as anticipated by the flux tube model \cite{Isg85b,Bar95}.

\subsection{1F States} 

\subsubsection{1F $s\bar s$ and the ``$\xi(2230)$" region} 

The 1F $s\bar s$ multiplet has long 
been of interest because of the
Mark III~\cite{Bal86} and BES~\cite{Bai96} 
reports of a possible very narrow $\xi(2230)$
in $J/\psi$ radiative decays. This evidence is
controversial because DM2~\cite{Aug88} did not see
this state, although they had slightly 
better statistics than Mark III.
The JETSET Collaboration studied 
$K_SK_S$ \cite{Eva97} and $\phi\phi$ \cite{Eva98}
final states in $p\bar p$ annihilation at LEAR,
and found no evidence for a narrow resonance with the
reported $\xi(2230)$ mass and width.
The Crystal Barrel Collaboration~\cite{Abe01b} also saw 
no evidence for this 
narrow state in $p\bar p \to \eta \eta$, although the
BES results on $p\bar p$ and $\eta\eta$ 
imply that they should have seen a large signal. 
The most recent experimental 
developments are extremely strong limits on a narrow $\xi(2230)$
in $\gamma\gamma\to K_SK_S$ from L3 \cite{Acc01b} and Belle \cite{Hua01},

\begin{equation}
\Gamma_{\gamma\gamma}(\xi(2230)) \cdot B_{\xi(2230)\to K_S K_S }\ < \ \
\cases{
1.4\!\!\!\!\!  &eV,\ \  95\% {\it c.l.} (L3)\cr
1.17\!\!\!\!  &eV,\ \  95\% {\it c.l.} (Belle).
}
\end{equation}

Motivated by the original Mark III results,
Godfrey {\it et al.}
\cite{God84} 
calculated a subset of $1^3$F$_2$ and $1^3$F$_4$ $s\bar s$ decay modes
(thought to be all the important ones), and found relatively small
total widths for these states. 
(These particular $s\bar s$ states were considered 
because the reported signal had a mass consistent with 
expectations for the 1F $s\bar s$ multiplet,
and had to have {\it even}$^{++}$ quantum numbers since it
was reported in $K_SK_S$.) These results suggested that the 
surprisingly narrow $\xi(2230)$, if real, might 
simply be a conventional
$s\bar s$ meson rather than a more unusual state such as a 
glueball
or hybrid. 
Subsequent work by Blundell and Godfrey~\cite{Blu96a} greatly modified
these conclusions. In the $1^3$F$_2$ $s\bar s$ case an
orbitally excited mode 
that had previously been neglected ($KK_1(1273)$) was found to be
dominant, making this $2^{++}$ state rather broad;
this eliminated the 
tensor $s\bar s$ option for the $\xi(2230)$, provided that
the \3P0 decay model
is reasonably accurate. The $1^3$F$_4$ $s\bar s$ state
was confirmed to couple primarily to $K^*K^*$, $KK^*$ and $KK$ in 
the Blundell-Godfrey work, although a total width of over 100~MeV
was found. This was an order of magnitude larger than the 
$\xi(2230)$ widths reported by Mark III and BES, so the explanation
of the $\xi(2230)$ as a 1F $s\bar s$ state now 
appears implausible.

There is experimental evidence of a somewhat wider state in this mass region.
A state 
with a mass and width of $\M = 2231\pm 10$ MeV and $\Gamma = 133\pm 50$ MeV
(with J undetermined)
was reported in $\phi\phi$
by WA67 (CERN SPS) \cite{Boo86},
and the LASS Collaboration reported a $4^{++}$ resonance 
with a mass and width of
$\M=2209^{+17}_{-15}\pm 10$ MeV and
$\Gamma = 60^{+107}_{-57}$ MeV in $K^- p\to K^-K^+\Lambda$
and  $K^- p\to K_SK_S\Lambda$ \cite{Ast88c,Ast88f}. 
Very recently, E173 (Serpukhov) also reported
an enhancement 
in $K_SK_S$, 
with $\M = 2257$~MeV and  
$\Gamma = 56$~MeV  \cite{Noz01}.

\subsubsection{$f_4(2200)$} 

The $f_4(2200)$ is predicted to be the narrowest of the 1F 
$s\bar s$ states, 
with an expected total width of about 150 MeV (Table~S13).. 
Our results for this state are quite similar to those found by 
Blundell and Godfrey~\cite{Blu96a} in their variant of the
\3P0 model using Kokoski-Isgur
phase space. We also find that the three important modes are
$KK$, 
$KK^*$ and 
$K^*K^*$. Our partial widths for these modes are
comparable, although the precise values are rather sensitive 
to kinematics 
because $KK^*$ and $KK$ are G-wave final states, with a 
resulting threshold
behavior of $|\vec p_f|^9$. 
The observation of the $f_4(2200)$ in both $KK$ and 
$K^*K^*$ would be
interesting in part because of the rather inaccurate prediction
of the SU(3) partner decay $f_4(2040)\to\pi\pi$~\cite{Bar97a}
(which may be due to this strong $|\vec p_f|^9$ momentum dependence)
and the lack of information regarding $f_4(2040)\to\rho\rho$,
which is predicted to have a large branching fraction.
There are also (relatively weak) analogous $\phi\phi$ and $\eta\eta$
modes, which measure the same decay amplitudes at different momenta
and thus would 
provide useful information. The multiamplitude VV mode $K^*K^*$ 
is predicted unsurprisingly to be dominated by the lowest-L
amplitude, $^5$D$_4$. Nonetheless an experimental study of the
three higher-L amplitudes predicted
to be weak or zero
would provide an interesting test of the \3P0 model.

Identification of the $1^3$F$_4$ $s\bar s$ 
and a
determination of its decay
parameters would be an important contribution to our understanding 
of this historically controversial region of the spectrum.
The broader experimental states in the 
2200~MeV region, which are discussed at the
end of the previous
section, are possible candidates for the $f_4(2200)$ 
$1^3$F$_4$ $s\bar s$ state.

\subsubsection{The unobserved $f_3(2200)$}

The $f_3(2200)$ $1^3$F$_3$ $s\bar s$ state is 
predicted to have a total width of about 300~MeV, and to decay 
dominantly to
$KK^*_2(1429)$, 
with a branching 
fraction of $\approx 40\% $.
(The VV and PsV modes have an L=2 barrier, whereas 
the $KK^*_2(1429)$ mode is dominantly P-wave.)
Next in importance is 
$KK^*$, with a branching fraction of $\approx 25\% $
and a $^5$G$_3$/$^5$D$_3$ amplitude ratio of $-0.62$. 
The 
branching fractions to $K^*K^*$, 
$KK_1(1273)$, 
$K^*K_1(1273)$
and the unusual mode $\eta f'_2(1525)$ are each $\approx 5$-$10 \% $.
The $K^*K^*$ channel has two allowed amplitudes,
and the 
\3P0 model predicts the
$^5$G$_3$/$^5$D$_3$ amplitude ratio to be $+0.51$.
Interesting measurements here include the G/D ratio in $KK^*$, a test of the
predicted dominance of $KK^*_2(1429)$, 
and the presence of this state in $\eta f'_2(1525)$,
which in the decay model is due to an 
$(s\bar s) \to (s\bar s) +(s\bar s)$ transition.

\subsubsection{The unobserved $f_2(2200)$}
  
The $f_2(2200)$ is predicted to have a very large decay coupling to
$KK_1(1273)$, which would make this a rather broad state;
the expected total width is 425 MeV, with $B_{KK_1(1273)} \approx 60 \%$.
The other decay modes of this state have theoretical branching fractions of 
$< 10\% $ and are not especially characteristic of $s\bar s$ states. 
It may be possible
to identify the $f_2(2200)$ in $KK$ or $KK^*$, or perhaps in
the $\eta\eta$ or $\eta\eta'$ modes. 

One might hope to identify $s\bar s$ states in $\phi\phi$,
which has previously been studied experimentally in searches for
glueball resonances, notably in
$\pi^- p \to \phi\phi n$ at BNL \cite{Etk88}. 
Three tensor states were
reported in $\phi\phi$ at BNL, and the
two near our assumed 1F $s\bar s$ mass of
2200~MeV were at 
$2297\pm 28$
and
$2339\pm 55$~MeV. 
These
{\it a priori} appear to be natural candidates
for the $1^3$F$_2$ $s\bar s$ quark model state, 
and the reported $f_2(2297)\to \phi\phi$ 
strengths in different $\phi\phi$ waves 
are similar to the pattern
predicted for $1^3$F$_2(s\bar s) \to \phi\phi$; 
Etkin {\it et al.} \cite{Etk88} reported 
$B(^5{\rm S}_2) = 6{+15 \atop -5}\% $,
$B(^1$D$_2$) = $69{+16 \atop -27}\% $ 
and
$B(^5$D$_2$) = $25{+18 \atop -14}\% $,
whereas for a $1^3$F$_2$ $s\bar s$ 
$f_2(2300)$ (note the increased mass) we predict 
$B(^5$S$_2$) = $0\% $,
$B(^1$D$_2$) = $49\% $, 
$B(^5$D$_2$) = $28\% $
and
$B(^5$G$_2$) = $23\% $. The theoretical ratio of D-wave 
partial widths 
in $1^3$F$_2(s\bar s) \to \phi\phi$ is $B(^1$D$_2) / B(^5$D$_2) = 7/4.$

In the \3P0 model however 
the $1^3$F$_2$ $(s\bar s) \to \phi\phi$ branching fraction 
is predicted to be very small ($0.5 \% $ for $\M = 2200$~MeV),
and unless this small branching fraction is confirmed,
identification of any of the resonances seen in 
$\phi\phi$
with the $1^3$F$_2$ $s\bar s$ is questionable. Of course 
one should not eliminate the possibility that this tiny decay
coupling may
simply be an inaccurate prediction of the \3P0 decay model; this 
predicted small coupling should
be checked against the VV decays of other members of the $1^3$F$_2$ 
flavor nonet, once these are identified.

\subsubsection{The unobserved $h_3(2200)$} 

The spin-singlet $h_3(2200)$ (Table~S4) is predicted to have a moderate
total width of $\Gamma_{tot}\approx 250$~MeV. This 
C$=(-)$ state can be 
diffractively
photoproduced, 
and the production amplitudes of a higher-L state may
provide interesting information about the nature of diffraction.
The decay modes 
$KK^*_2(1429)$ and $KK^*$
are predicted to be dominant. 
The $s\bar s$-signature
mode $\eta\phi$ is more
attractive experimentally; 
the $h_3(2200)$ is predicted to have a rather large,
{\it ca.} $10\% $ branching fraction to $\eta\phi$, with comparable
strengths in D- and G-waves.
The ratios we find for these 
amplitudes are
$^3$G$_3$~/~$^3$D$_3$~$ = +0.59$ for $\eta\phi$ and 
$+0.83$ for the partner open-strangeness mode $KK^*$.

\newpage

\section{Kaonia}

\subsection{General aspects} 

The kaon sector is interesting for several reasons.
One notable feature is that
the usual
kaon and antikaon states do not have diagonal C-parity, 
so
there are no J$^{\rm PC}$-exotics in the kaon spectrum.
The kaon-flavor analogues of $n\bar n$ and
$s\bar s$ J$^{\rm PC}$-exotic hybrids should
instead appear as a rich overpopulation of states
in the conventional excited kaon spectrum.

A
detailed comparison between the kaon and I=1 $n\bar n$ 
spectra 
may therefore be 
useful for the identification of hybrids
using overpopulation.
For this comparison one should specialize to
J$^{\rm P} = 0^-, 0^+, 1^-, 2^+, 3^- \dots $,
for which one C-partner is exotic.
For example, the J$^{\rm P} = 1^-$
kaon spectrum will have an overpopulation of states relative
to 
J$^{\rm PC} = 1^{--} $
I=1 $n\bar n$, 
due to the
presence 
of both
J$^{\rm PC} = 1^{--} $ 
quarkonium and 
J$^{\rm PC} = 1^{-+} $ hybrid basis states
in the J$^{\rm P} = 1^-$ kaon mixing problem.

Not only will there be ``too many states" in a given
kaon J$^{\rm P}$ sector relative to I=1 $n\bar n$,
we also anticipate 
irregularities between the kaon
and I=1 $n \bar n$ spectra, due to mass shifts from
kaon mixing with 
J$^{\rm PC}$-exotic hybrid basis states that cannot 
mix in the I=1 $n \bar n$ problem.
The anomalously low mass of the
$K^*(1414)$ relative to the $\rho(1465)$ may be an example
of this effect.

The absence of C-parity 
also implies that the physical J$^{\rm P}$ kaon states
are admixtures of spin-singlet and spin-triplet 
$q\bar q$ basis states with different C
for J$^{\rm P} = 1^+, 2^-, 3^+ \dots$, 
unlike their neutral I=1 $n\bar n$ partners.
The $K_1$ system is a familiar example of this mixing;
the physical $K_1(1273)$ and $K_1(1402)$ are
strongly mixed linear combinations of 
$| 1^1{\rm P}_1\rangle $
and
$| 1^3{\rm P}_1\rangle $
basis states.
The precise mechanism of this mixing of different 
$S_{q\bar q}$ states is an interesting open question in the
kaon system. Mixing has been attributed to
coupling through decay channels 
(originally by Lipkin \cite{Lip77})
as well as to 
$q\bar q$ spin nonconservation in the OGE spin-orbit
interaction (because $m_s\neq m_{u,d}$), although this effect does
not appear large enough to explain the observed 1P mixing angle
\cite{God91}. The dominant 
mechanism of singlet-triplet mixing has evidently
not yet been definitively established, and can presumably be clarified
through additional theoretical studies and 
measurements of the corresponding mixing angles in the
2P, 1D and 1F systems.
Experience with the $K_1$ system suggests that 
strong decays of the higher-mass states will allow determination of
these mixing angles; to assist in this exercise 
we give the mixing-angle dependence of strong partial widths and decay
amplitudes for general mixed states in the decay tables.

The fact that the kaon sector has no valence annihilation
may also make it useful for the identification
of large mixing effects between 
I=0 $n\bar n$, $s\bar s$ and glueball basis states.
A comparison of the spectra in these different flavor
sectors may show irregularities 
where valence annihilation is important, 
as may be the case in the scalar sector.

In summary, a comparison between 
the kaon and 
I=1 $n\bar n$ spectra should provide evidence for hybrids
through kaon overpopulation, and a comparison with the
I=0 $n\bar n$ and $s\bar s$ spectra may provide evidence
of $q\bar q$-glueball mixing.
Establishing the spectrum of excited 
kaon states may thus be important for 
searches for both types of gluonic hadrons expected in the meson 
spectrum. 
A determination of singlet-triplet mixing angles in higher-mass kaon states
through measurements of branching fractions and decay amplitudes
is also an interesting experimental exercise, since these 
angles have not yet
been determined except in the light $K_1$ sector, and the 
mechanism that drives this mixing is not yet understood.

Experimentally there are few plans to study 
the excited kaon spectrum with improved statistics. 
This is unfortunate in view of the importance of
the kaon spectrum for studies of overpopulation, mixing and 
valence annihilation effects. Hadronic reactions such as
$K^\pm p \to (K n\pi)^\pm p$ could be explored 
with a medium-energy RF-separated kaon beam, as is
now under construction at 
Serpukhov. This would also be possible at JHF, although 
this is not part of their current physics program.
(At lower beam energies 
a $K^+$ beam may be preferable to the usual $K^-$, to 
avoid a large background of
s-channel strange baryon resonances.)
Other possibilities
include photoproduction and $e^+e^-$ facilities, 
which could study higher-mass kaon spectra through the sequential 
decays of initial $s\bar s$ states, and $p\bar p$ annihilation in flight
at GSI \cite{GSI01}. 
In $p\bar p$ annihilation one may 
extract higher-mass kaon resonances for example
from partial wave analyses of $p\bar p \to K + (Kn\pi )$.

Finally, the high-statistics studies of heavy-quark physics and CP 
violation at D and B factories can contribute to the study of excited
kaon spectroscopy, through the identification of 
resonances in final states with a kaon. The excited kaons already
reported in heavy-quark nonleptonic weak decays
are the
$K_1(1273)$, $K_1(1402)$, $K_0(1412)$ and $K^*(1717)$ (in
D decays, typically at the $1\%$ level) 
\cite{PDG02}. 
Unfortunately, D decays are limited by phase space
to kaon resonances with 
$M < 1.73$~GeV.
B~decays to $K n\pi$ final states have ample phase space but
are limited by small branching fractions, for example
$B_{B^+\to K^+\pi^+\pi^-} = ( 5.6 \pm 1.0) \cdot 10^{-5}$ \cite{PDG02}.
B decays to $J/\psi + Kn\pi $ may be more attractive, since they
have much larger branching fractions; 
$B_{B^+\to J/\psi K^+\pi^+\pi^-} = ( 1.4 \pm 0.6) \cdot 10^{-3}$ 
and
$B_{B^o\to J/\psi K^+\pi^-} = ( 1.2 \pm 0.6) \cdot 10^{-3}$ 
\cite{PDG02},
and the available phase space of 2.18~GeV is adequate for the study
of many of the excited kaons discussed here. 

There is also interest in the decay systematics of D and B mesons to
final states such as $\eta K$ and $\eta' K$. Lipkin \cite{Lip81,Lip97}
has 
noted that some heavy-quark weak decay processes involve the strong decay
of an intermediate excited kaon, which 
can lead to
unusual branching fractions for these modes. 
There is already some evidence favoring the resulting selection rules, 
for example the counterintuitive result
$
B_{B^+ \to \eta' K^+} = ( 7.5 \pm 0.7) \cdot 10^{-5}
>>
B_{B^+ \to \eta K^+} <  6.9  \cdot 10^{-6}, \ 90\%  \ c.l.$
(See App.B for a discussion and
generalization of Lipkin's results.)

\subsection{1S States} 

The predicted partial width 
for the transition $K^*\to K\pi$ is somewhat underestimated
by the \3P0 decay model, as usual for 
$ 1^3$S$_1 \to \ 1^1$S$_0 + 1^1$S$_0$ 
decays; this discrepancy was discussed in the section
on 1S $s\bar s$ decays.

\subsection{2S States} 

\subsubsection{$K^*(1414)${\it , a problematical state}} 

The $K^*(1414)$ is an especially interesting state for future
experimental study, since its properties are clearly in
disagreement with the expectations of the quark model for
a first radial excitation of the $K^*(894)$. 

This state was first
reported at CERN in 1976 as a $1^-$ enhancement 
in $\bar K^o\pi^+\pi^-$ near 1450~MeV~\cite{Ver76}. 
The phase difference between
the $1^-$ 
$\pi K^*$ and $\rho K$ waves in this mass region was 
observed to be approximately
constant, as required if both arose from a single resonance. 
However these phases 
did not show resonant phase motion 
relative to the 
clear $2^+$ $K_2^*(1429)$ $\pi K^*$ and $\rho K$ amplitudes, 
which argued against a resonance interpretation 
of the $1^-$ enhancement.  

A BNL $K^-p$ experiment next studied this $1^-$ enhancement 
in the final state $K_S\pi^+\pi^-$, again in the reaction
$K^- p \to \bar K^o\pi^+\pi^- n$ 
\cite{Etk80}. Depending on the fit assumptions, the mass and width 
of the enhancement were
found to be 
$\M \approx 1450$-$1500$~MeV and $\Gamma \approx 170$-$210$~MeV. 
Both
$\pi K^*$ and $\rho K$ modes were reported, $\pi K^*$ being dominant;
$B_{\pi K^* / \rho K} \sim 5$-$7$. Possible evidence for resonant phase
motion was reported (Fig.12f of Ref.\cite{Etk80}), 
but the statistics were clearly
insufficient for definitive conclusions.

This was followed by a CERN study of $K^-p\to \bar K^o\pi^+\pi^- n$
\cite{Bau82}, which confirmed a large $1^-$ signal in $\pi K^*$, 
and gave a fitted mass and width of
$\M = 1474\pm 25$~MeV and $\Gamma = 257\pm 65$~MeV. 
(See their Table~3; note that this width is
reported by the PDG as 275~MeV \cite{PDG02}.)
This reference concluded that this signal did not seem to be due to a resonance,
because the $\rho K$ amplitude did not show the expected 
resonant phase motion relative to the strong $2^+$ $\pi K^*$ 
and $\rho K$ waves. (See 
Figs.~12c and 13 of Ref.\cite{Bau82}). 

In 1984 the LASS collaboration also
reported a study of $K^-p\to \bar K^o\pi^+\pi^- n$ \cite{Ast84}; they found
a large $\pi K^*$ $1^-$ signal with a mass and width of
$\M = 1412\pm 9\pm 2$~MeV and $\Gamma = 196\pm 18\pm 12$~MeV,
and saw no evidence for this state in $\rho K$. The slowly-varying relative
$\pi K^*$ $1^-$ and $2^+$ phase was attributed to the presence of both
$1^-$ and $2^+$ resonances, with similar masses and widths. The novel result
of this experiment was the lack of a $\rho K$ $K^*(1414)$ signal, and
it was
also noted that the $\pi K$ coupling of the $K^*(1414)$ must be very weak.
In 1987 LASS reported another study of $K^-p\to \bar K^o\pi^+\pi^- n$ 
\cite{Ast87}; the conclusions regarding the $K^*(1414)$ enhancement
and the fitted resonance parameters were quite similar to their earlier
results in Ref.\cite{Ast84}. A 1988 LASS study of
$K^-p\to K^-\pi^+ n$ \cite{Ast88e} found that the $K^*(1414)$ was weakly
coupled to $\pi K$, with a branching fraction of only 
$(6.6\pm 1.0 \pm 0.8 )\% $. 
Despite the weak coupling, there was evidence
that the $1^-$ $\pi^- K^+$ phase motion in this mass region was better
described by assuming a $K^*(1414)$ resonance (Fig.17 of Ref.\cite{Ast88e}).
The weak but resonant 
$\pi K$ coupling of the $K^*(1414)$ was also reported by LASS
in an unpublished study of $K^-p\to \bar K^o\pi^- n$ \cite{Bir89}.

The $K^*(1414)$ seems an obvious candidate for the 2$^3$S$_1$
radial excitation of the $K^*(894)$, since it is the first
strange $1^-$ vector resonance 
observed above the $K^*$.
On closer inspection however 
there are problems with this identification. 
First, the $K^*(1414)$ mass appears
too light if we also accept the $\omega(1419)$ and $\rho(1465)$
as 2$^3$S$_1$ $n\bar n$ states; a mass for their strange partner 
of $ca.$ 1.55~GeV would appear more plausible. (For example, 
Godfrey and
Isgur~\cite{God85} found a mass of 1.58~GeV for their 2$^3$S$_1$ kaon.)

A second problem with identifying the 
$K^*(1414)$ with the 2$^3$S$_1$ kaon
is that the reported $\pi K$ branching fraction is 
rather smaller than
the \3P0-model prediction.
In Tables~K2 and K3 we give predictions for the branching 
fractions and decay amplitudes of a 2$^3$S$_1$ kaon, 
assuming masses of 
1414 and 1580 MeV. The $K^*(1414)$ option 
predicts large and comparable branching fractions to 
$\pi K$,
$\eta K$,
$\rho K$
and
$\pi K^*$. Although the total width is consistent with that
of the 
$K^*(1414)$, the LASS \cite{Ast88e} $\pi K$ branching fraction
of $(6.6\pm 1.0 \pm 0.8 )\% $ is well below our predicted
$28 \% $. 

A third problem with identifying the $K^*(1414)$ with a
2$^3$S$_1$ kaon, probably the most serious, 
is the reported strong experimental preference
for $\pi K^*$ over $\rho K$, $B_{K^*(1414)\to \rho K / \pi K^*}
< 0.17, \ \ 95\% \ \ {\it c.l.}$ \cite{Ast84}. These are both
2$^3$S$_1 \to ^1$S$_0 + ^3$S$_1$ transitions and are within
the same SU(6) multiplets; theoretically these amplitudes 
are the same function of momenta, and up to phase space corrections
these branching fractions should be identical. (Our predicted branching 
fraction ratio of $B_{K^*(1414)\to \rho K / \pi K^*} = 0.61$ 
in Table~K2 only departs from unity because of 
phase space differences.) Although there is a node in this radial transition
amplitude,
in the \3P0 model 
it is at $|\vec p_f | = \sqrt{15/2}\; \beta \approx 1.1$~GeV, far from the   
physical final momenta of $\approx 300$-$400$~MeV. It is difficult to
see how the reported branching fraction ratio can be accommodated given a
simple 2$^3$S$_1$ kaon assignment for the $K^*(1414)$.

In the tables we also give results for an alternative $K^*(1580)$ 
2$^3$S$_1$ state; this higher mass resonance should be rather 
broad (total width $\approx 350$ MeV), but will again be dominated
by decays to
$\pi K$,
$\eta K$,
$\rho K$
and
$\pi K^*$, with comparable branching fractions.

This disagreement in mass 
for the $K^*(1414)$ as a 2$^3$S$_1$ kaon 
is the clearest discrepancy between theory and experiment we find 
in any of the strange mesons we have considered. If this state is 
indeed a real $1^-$
resonance, the low mass may be due to the presence of additional
hybrid mixing states, as we noted in the introduction. 
(The mixing problem for kaons is different from the nonstrange 
sector because of C-parity.)
Given the different set of hybrid states available
for mixing in the kaon flavor sector, if the mixing is large
we would
not expect the mass or decay properties to
be consistent with the 
2$^3$S$_1$ $n\bar n$-flavor candidates $\omega(1419)$ and $\rho(1465)$.
A comparison of the $\pi K$, $\pi K^*$ and $\rho K$ 
branching fractions of the $K^*(1414)$
with the $\pi\pi$ and $\pi \omega$  branching fractions of
the $\rho(1465)$, for example, would be a very interesting test
of whether these states appear to belong to the same SU(3) flavor multiplet. 
In view of the anomalously low mass of this state,
establishing resonant phase motion and
accurately determining its decay branching
fractions should be 
a high priority in
future experimental studies of the spectrum of strange mesons.

\subsubsection{$K(1460)$} 

This state was first reported in 1976 at SLAC by Brandenberg {\it et al.}
\cite{Bra76} in a PWA of $K^\pm \pi^+\pi^-$ final states produced in 
$K^\pm p \to K^\pm \pi^+\pi^- p$ at 13~GeV. The fitted mass and width were
$\M = 1404\pm 12$~MeV 
and 
$\Gamma = 232\pm 16$~MeV, and the dominant coupling was found to be
``$\epsilon K$", with some evidence for $\pi K^*$ (with a poorly understood 
$0^-$ contribution near 1.23 GeV) and $\rho K$ (which was about $30\% $ as large in
intensity as $\epsilon K$, and peaked at a rather higher mass, about
1.5-1.6 GeV; see their Fig.2c). 

This discovery was followed by the ACCMOR 
analysis of about 200K $K^- p \to K^- \pi^+\pi^- p$ events at
63 GeV~\cite{Dau81}. The $K^- \pi^+\pi^-$ $0^-$ amplitude was fitted 
assuming 
the same three modes, $\epsilon K$, $\pi K^*$ and $\rho K$. The 
estimated mass 
and width of the $0^-$ resonance were $M\sim 1.46$ GeV and 
$\Gamma \sim 260$ MeV, with partial widths into each mode of
$\Gamma_{\epsilon K} = 117$ MeV,
$\Gamma_{\pi K^*} = 109$ MeV and
$\Gamma_{\rho K} = 34$ MeV. 
Again it was found that that the $\rho K$ signal peaked at
a higher mass than  $\pi K^*$ and $\epsilon K$ (see their Fig.18).
Daum {\it et al.}
noted that the ``$\epsilon K$" mode would also include any
contribution from ``$\pi\kappa $". 
The Particle Data Group~\cite{PDG02} 
has attributed {\it all} of this
ACCMOR ``$\epsilon K$" partial width to $\pi K^*_0(1412)$, although Daum
{\it et al.} do not make this claim. 

The mass of this state is consistent with expectations for
a 2$^1$S$_0$ radial excitation of the $K$, assuming that the 
$\pi(1300)$ is the corresponding pion radial excitation.
Similarly, the reported total width 
of $\sim 250$-$260$ MeV
is comparable to the \3P0 model 
expectation of $\Gamma_{tot} \approx 200$ MeV (Table~K2). 

We have not included the broad $\pi\pi K$ modes 
``$\epsilon K$" or ``$\pi\kappa $" in Table~K2
because they 
are closed given our assumed 
$f_0(1370)$ and
$K^*_0(1412)$ masses. 
(See also Ref.\cite{Kop01} regarding the $\kappa$.)
We can test whether these modes are important by
assigning a lower mass to the scalars as a width effect. 
If we assume an $f_0(700)$ and a $K^*_0(1100)$ to model 
a broad ``$\epsilon$" and ``$\kappa$", we find
very small widths; 
$\Gamma_{K(1460)\to \epsilon K} = 0.2$ MeV
and
$\Gamma_{K(1460)\to \pi \kappa } = 1.5$ MeV. 
There is a node close to the $f_0(700) K$ physical 
point, but the amplitude is nonetheless intrinsically quite small, 
as is evident from the small $\pi K^*_0(1100)$ width.
Thus the \3P0 model
is inconsistent with the reports of large $\epsilon K$ or $\pi \kappa$
modes. 

The possibility that the ``$\epsilon K$" mode might arise from a
nonresonant Deck effect was rejected by Brandenburg {\it et al.},
as clear resonant phase motion was evident in this channel.
A similar situation is found in the decays of the $\pi(1300)$;
the \3P0 decay model predicts a dominant $\pi\rho$ mode
\cite{Bar97a}, with
$\pi (\pi\pi)_S$ only making a small contribution. There 
actually is a large $0^-$ $\pi (\pi\pi)_S$ signal in the 
1.2 GeV region, which
if resonant disagrees with the decay model. 
The VES collaboration has argued however 
that the $\pi (\pi\pi)_S$ signal may arise 
from a
Deck effect rather than from the $\pi(1300)$ resonance
\cite{Ame95}. 

Daum {\it et al.}~\cite{Dau81} reported branching fractions of 
$B_{\pi K^*}\sim 42 \% $,
$B_{\rho K}\sim 13 \% $ and 
$B_{\pi K^*}\sim 45 \% $; this preference for $\pi K^*$ 
over
$\rho K$ is predicted by the \3P0 model, but is expected to be less 
pronounced. 
The $\omega K$ mode
has not been studied; it would be interesting to study this mode
because
$\omega K$ is ``cleaner" than the modes that have
been reported, and the relative strengths of the $\omega K$ and 
$\rho K$ 
should be close to the SU(3) flavor factor of
1/3 of these final states arise dominantly from the $K(1460)$. 

\subsection{3S States} 

\subsubsection{The unobserved $K^*(1950)$} 

The masses of the 3S states are not yet well established in
any of the light flavor sectors. Here we assume the mass of
the experimental $K(1830)$ for our 
3$^1$S$_0$ 
state, and with a spin-spin splitting suggested by the 
$\rho(1465)$ and $\pi(1300)$ 2S candidates we assume a
rounded mass of 1950~MeV for the $3 ^3$S$_1$ kaon.

This state has many open two-body decay modes, as shown in 
Table~K4. The dominant mode is predicted to be
$\rho K^*$, with a strong preference for the $^5$P$_1$ final
state. The $\omega K^*$ mode is also important, suppressed by
a flavor factor of 1/3 relative to $\rho K^*$. One surprise is
that the second mode after $\rho K^*$ is 
predicted to be 
$\pi K^*(1414)$, assuming
that this problematical state is indeed the 2S kaon.
The very weak $\pi K$ mode is due to a node in the \3P0 
decay amplitude that is accidentally quite close to the physical point;
this also suppresses $\eta K$ and $\eta' K$.

The dominance of the 3$^3$S$_1$ $s\bar s$ coupling to
$\eta' K^*$ over $\eta K^*$ is the consequence of
an interesting 
interference between the $n\bar n$ and $s\bar s$ components of the 
$\eta$ and $\eta'$, coupled to the spin-one $K^*$.
This system has a selection rule opposite to that of
the more familiar $\eta K$ and $\eta' K$ final states, as explained in
App.B.
With our parameters we predict a branching ratio of
$B_{\eta' K^*} / B_{\eta K^*} = 24$. 

\subsubsection{$K(1830)$} 

The \3P0 model predicts that this state has a total width
of only about 200~MeV, and the dominant decay modes are
$\rho K^*$ and (again rather surprisingly) $\pi K^*(1414)$, both
with branching fractions of $\approx 20\% $. A 
large $\eta K^*$ branching fraction is also predicted.

There is an experimental candidate for this state from the CERN Omega
Spectrometer \cite{Arm83}, reported in their partial wave analysis of the 
$\phi K$ final state in 
$K^- p \to K^+ K^- K^- p$. A pseudoscalar amplitude with
resonant phase motion was observed at a mass and width
of $\M\sim 1830$~MeV and $\Gamma\sim 250$~MeV,
consistent with our theoretical total width. 
The 
predicted branching fraction 
to this mode is  
$B_{\phi K} = 9\% $. 

\subsection{1P States} 

\subsubsection{$K^*_2(1429)$}

Given the success of the \3P0 model in describing the strong
decays of the $f_2(1275)$~\cite{Bar97a}
and $f'_2(1525)$,
one would expect that the decays of 
their kaonic
partner $K^*_2(1429)$ would also be well described.
This is qualitatively the case;
the predicted ordering of partial widths
$\pi K> \pi K^*> \rho K> \omega K$ is in agreement with experiment,
and the predicted and observed values are roughly consistent. 
(See Table~\ref{table_K2}.)

The detailed agreement 
with experimental $K^*_2(1429)$ partial widths however does not 
appear as impressive
as 
for its $s\bar s$ partner $f'_2(1525)$ 
(Table~\ref{table_f2}). 
This is due to
a mismatch between
the scales of widths to PsPs and PsV final states
that has not been tested in $f'_2(1525)$ decays, since the partial width
for $f'_2(1525)\to KK^*$ has not been measured. 

Note that the partial width to
$\eta K$ is very small. 
This
mode
is suppressed by destructive interference
between the $n\bar n$ and $s\bar s$ components of the 
$\eta$, due to the comparable size and opposite sign 
of the
$K^*_2\to (n\bar n) K$ 
and 
$K^*_2\to (s\bar s) K$
transition amplitudes. 
The coupling to $\eta' K$ in contrast has constructive interference
and should be large; unfortunately this mode has no phase space 
in $K_2^*(1429)$ decays.
Observation of both these modes is 
possible
in decays of higher-mass excited kaons, and the dominant
mode depends on the angular quantum numbers of the initial kaon.
(See App.B).

\begin{table}
\centering
\begin{tabular}{|l|c|c|c|c|c|c|}
\hline
mode: $\Gamma_i$ (MeV)
& $\pi K $
& $\eta K $
& $\rho K $
& $\omega K $
& $\pi K^*$
& $\pi\pi K^*$
\\
\hline
\hline
$K^*_2(1429)$ (expt)
&   $49.1\pm 1.8  $ 
&   $0.15$${+0.33 \atop -0.10} $ 
&   $8.5\pm 0.8 $ 
&   $2.9\pm 0.8 $ 
&   $24.3 \pm 1.6  $ 
&   $13.2 \pm 2.2  $ 
\\
\hline
$K^*_2(1429)$ (thy)
&   $56 $ 
&   $0.57 $ 
&   $4.4 $ 
&   $1.2 $ 
&   $12.8  $ 
&    -
\\
\hline
\end{tabular}
\caption{\label{tensor}Experimental and theoretical 
partial widths of the 1$^3$P$_2$ tensor kaon
$K^*_2(1429)$.}
\label{table_K2}
\end{table}
\subsubsection{$K^*_0(1412)$}

This state is especially interesting due to the controversial status of
light scalar mesons in other flavor channels.
The $K^*_0(1412)$ has only been observed in 
the $\pi K$ mode. The LASS Collaboration
\cite{Ast88e} found a mass and width of
$\M = 1412 \pm 6$~MeV and
$\Gamma = 294\pm 23$~MeV, 
and 
determined a branching fraction of $B_{\pi K} = 0.93\pm 0.04 \pm 0.09$
by assuming that the reaction $K^- p \to \pi^+ K^- n$ was dominated
by one pion exchange.
This branching fraction is consistent with the 
\3P0 model, which
predicts that the other open channel, $\eta K$, has a branching fraction
of $\approx 5\% $. 

There is also evidence for the $K^*_0(1412)$ in $p\bar p$
annihilation at rest to $KK\pi$ 
in several channels, 
$K_LK_L\pi^o$ \cite{Abe96},
$K_LK^{\pm}\pi^{\mp}$ \cite{Abe98}, 
$K^+K^-\pi^o$ \cite{Abe99}
and
$K_SK^{\pm}\pi^{\mp}$ \cite{Ber98}, 
as summarized in Table~2 of Ref.\cite{Abe99}.
Fits to the $K^*_0$ resonance parameters \cite{Abe96,Abe98} 
(specifically T-matrix poles) gave a mass and
width of 
$M\approx 1.42$-1.43~GeV and 
$\Gamma\approx 0.28$~GeV,
very close to the LASS results.
It is interesting that the fitted $KK^*_0(1412)$ 
contribution to the $p\bar p\to KK\pi$ 
Dalitz plots is comparable to the $KK^*$ contribution; this suggests that
$p\bar p$ annihilation could be an effective approach for the production
of higher-mass excited kaon states, perhaps in future 
annihilation in flight experiments at GSI \cite{GSI01}.  

Our predicted total $K^*_0(1412)$ width is 
rather smaller than is observed, 
$\Gamma_{thy} \approx 120$~MeV 
versus the LASS result
$\Gamma_{expt} = 294\pm 23$~MeV \cite{Ast88e}. The amplitude for
$1^3$P$_0 \to 1^1$S$_0 + 1^1$S$_0$ however varies rapidly with 
wavefunction parameter $\beta$ and has a node near
$\beta = 0.3$~GeV, so this disagreement is rather sensitive to parameters.
We also note that the OGE decay amplitude was
found to be especially
large in this channel~\cite{Ack96}. 
Since the {\3P0}-model decay amplitude may not be
dominant in the decays of light scalar mesons, 
a comparison 
to experiment may not be justified in this case.

It is also notable that the $K^*_0(1412)$ was observed in
charmed meson nonleptonic decays by
E691 \cite{Anj93}  
and
E687 \cite{Fra94}, 
with a relatively large branching fraction.
The PDG \cite{PDG02} reports
$B_{D^+\to {K^*_0(1412)} {\pi^+}} 
= (3.7\pm 0.4) \% $ , compared for example to a total 
$K^-\pi^+\pi^+ $ branching fraction of
$B_{D^+\to {K^-} {\pi^+} {\pi^+}} = (9.1\pm 0.6) \% $,
most of which is nonresonant. 
In $D^o\to K^-\pi^+\pi^o$ a recent CLEO study found that the
largest kaon isobar contribution above the $K^*$ was
due to the $K^*_0(1412)$ \cite{Kop01} (see their Table~VIII).
Of all the excited kaon states above the $K^*$ only the 
low-spin states 
$K^*_0(1412)$,
$K_1(1273)$,
$K_1(1402)$ and the
$K^*(1717)$ have been reported in D and B decays; 
presumably the restriction of
the final state to total J = 0 suppresses higher-L$_{q\bar q}$ 
strange states.

\subsubsection{$K_1(1273)/K_1(1402)$} 

The axial kaons $K_1(1273)$ and $K_1(1402)$ are among the most interesting
states in the kaon spectrum. Unlike their $n\bar n$ and $s\bar s$ flavor
partners, the kaons do not have diagonal C-parity, so the 
spin-singlet n$^1$P$_1$ 
and
spin-triplet n$^3$P$_1$ basis states mix. 
This leads to a nontrivial mixing angle
$\theta$ for each n,L kaon multiplet, which for 1P we define by 
\begin{equation}
|K_1(1273)\rangle = 
+\cos(\theta) |1^1{\rm P}_1\rangle
+
\sin(\theta) |1^3{\rm P}_1\rangle \phantom{\ .}
\end{equation}
and
\begin{equation}
|K_1(1402)\rangle = 
-\sin(\theta) |1^1{\rm P}_1\rangle
+
\cos(\theta) |1^3{\rm P}_1\rangle \ .
\end{equation}
Although an apparently equivalent mixing angle formula is quoted 
by Blundell and Godfrey (Eq.(10) of Ref.\cite{Blu96b}), 
our angles are actually opposite
in sign because their definition assumes a heavy quark 
(hence $s\bar n = \bar K_1$, 
antikaons),
whereas we assume this mixing matrix for kaons ($n\bar s = K_1$). 
This implies opposite signs for $\theta$ 
in the two conventions because the charge conjugation 
operator ${\cal C}$ gives opposite
phases when applied to 
$|^1{\rm P}_1\rangle$ and
$|^3{\rm P}_1\rangle$ basis states. 
 
\begin{figure}
$$\epsfxsize=4truein\epsffile{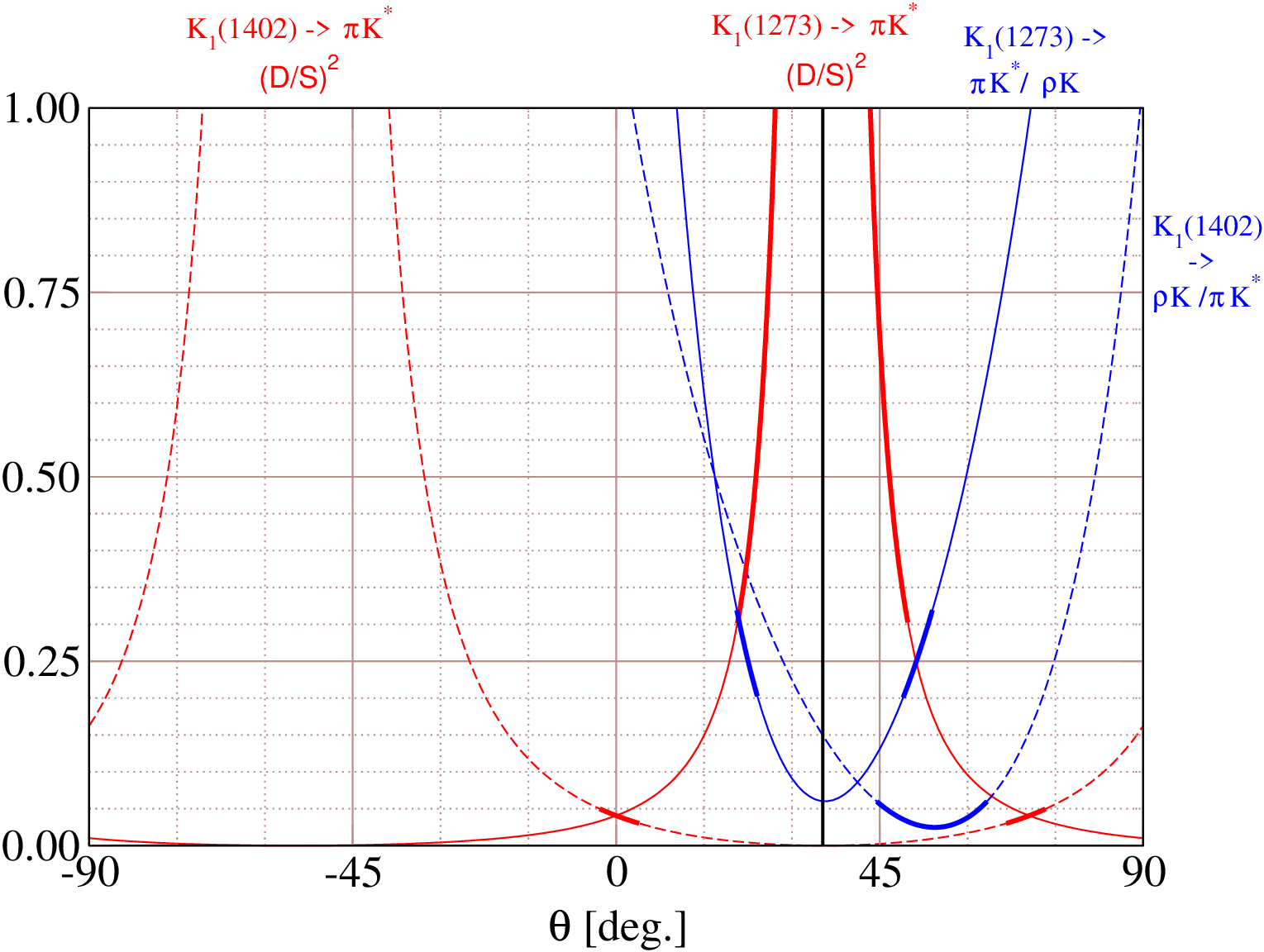}$$
\center
{Figure~4.
Mixing angle dependence of some $K_1$ decay ratios. 
Regions within $\pm 1\sigma$ of experiment are indicated by
thick lines, and the HQET points 
$\theta\approx +35.3^o$ 
and 
$\theta\approx -54.7^o$ 
are shown as dark verticals.
} 
\end{figure}

In Tables~K6 and K7
we give results for decay amplitudes and widths 
of the two $K_1$ states as functions
of $c = \cos(\theta)$ and $s = \sin(\theta)$. 
Clearly the decay amplitudes and branching fractions depend strongly 
on this mixing angle. Since the $^3$P$_0$ model is known to give reasonably
accurate results for the decay amplitudes of nominally pure $^3{\rm P}_1$
and $^1{\rm P}_1$ states (specifically $a_1\to\rho\pi$ and $b_1\to\omega\pi$,
both of which have nontrivial D/S amplitude ratios), we can apply the
$^3$P$_0$ model to the determination of this mixing angle with some 
confidence.

There is no theoretical consensus regarding the origin of the mixing 
angle $\theta$.
One speculation, originally due to Lipkin \cite{Lip77}, 
is that it might be
determined by the coupling of the two
$|K_1\rangle$ states through their decay channels. 
With sufficiently
strong decay couplings the physical resonances can be driven into
near ``mode eigenstates", which would explain the separation into
``$\rho K$" and ``$\pi K^*$" resonances.
Under certain simplifying assumptions this picture 
suggests a singlet-triplet 
mixing angle
of $\theta \approx 45^o$, essentially the value required by experiment
(see Fig.4).
Presumably the decay
mixing model 
can be elaborated and applied to the 1D $K_2$ and 2P $K_1$ systems 
as well, and can be tested
when quantitative information becomes available on the strong decays of 
these states.

Alternatively, it has been noted that the spin-orbit interaction 
also drives
singlet-triplet mixing given unequal quark and antiquark masses, and
in the 
HQET limit $m_Q/m_q\to \infty$ one finds 
``magic mixing angles" of 
$\theta = \tan^{-1}(1/\sqrt{2})
\approx +35.3^o$ and 
$\theta = -\tan^{-1}(\sqrt{2})
\approx -54.7^o$. The first value is not far from the $\theta$ suggested
by $K_1$ data (Fig.4). 
This approximate agreement may be spurious, however, as a
mixing angle 
of $\theta = + 5^o$ is
found in the $K_1$ system with realistic quark masses \cite{God91}, which 
is far from both the HQET value and experiment.

Experimentally, the pattern of decay branching fractions of the 
$K_1(1273)$ and
$K_1(1402)$ is striking. (See Table~\ref{table_K1}.) 
Of the three nominally rather 
similar PsV modes 
$\rho K$,
$\omega K$ and 
$\pi K^*$,
the $K_1(1273)$ shows a strong preference for $\rho K$, 
$B_{K_1(1273) \to \pi K^* / \rho K} = 0.26\pm 0.06$, 
whereas the 
$K_1(1402)$ decays almost exclusively to $\pi K^*$,
$B_{K_1(1402)\to \rho K/\pi K^* } = 0.03\pm 0.03$~\cite{PDG02}. 
Comparison with the theoretical \3P0 model branching fraction ratios
in Fig.4 shows that this can be satisfied by a $K_1$ 
singlet-triplet mixing angle of
$\theta \approx +45^o$.
 
The D/S ratios for the $K_1$ states also depend strongly
on the singlet-triplet mixing angle. 
The theoretical {\3P0}-model $|{\rm D/S}|^2$ width ratios for
$K_1(1273)\to \pi K^*$ 
and 
$K_1(1402)\to\pi K^*$ 
are shown in Fig.4; 
these are singular at the two HQET points.
The 
experimental $|{\rm D/S}|^2$ ratios of 
$1.0\pm 0.7$ for the $K_1(1273)$
and 
$0.04\pm 0.01$ for the
$K_1(1402)$ 
\cite{PDG02}
are also indicated, and 
the data show a strong preference for 
$\theta = +35^o$ over $-55^o$. 
A more accurate measurement
of the D/S ratio in $K_1(1273)\to \pi K^*$ and a measurement of
the sign of D/S in $K_1(1402)\to\pi K^* $ would be very useful
for constraining the singlet-triplet mixing
angle.

\begin{table}
\centering
\begin{tabular}{|l|c|c|c|c|c|c|c|c|c|c|}
\hline
mode: $\Gamma_i$ (MeV)
& $\rho K $
& $\omega K $
& $f_0(1370) K$
&  $\pi K^*$
& $\pi K^*_0(1412)$
\\
\hline
\hline
$K_1(1273)$ (expt)
&   $38\pm 10   $ 
&   $10\pm 3  $ 
&   $3\pm 2 $ 
&   $14\pm 6  $ 
&   $25\pm 7   $ 
\\
\hline
$K_1(1273)$ (thy)
&   $58  $ 
&   $ - $ 
&   $ - $ 
&   $ 3  $ 
&   $ -  $ 
\\
\hline
\hline
$K_1(1402)$ (expt) 
&   $5\pm  5 $ 
&   $2\pm  2  $ 
&   $3\pm  3$ 
&   $164\pm 16  $ 
&   not seen 
\\
\hline
$K_1(1402)$ (thy)
&   $30  $ 
&   $ 10 $ 
&   $ - $ 
&   $ 203  $ 
&   $ -  $ 
\\
\hline
\end{tabular}
\caption{Experimental and theoretical partial widths of the axial kaons
$K_1(1273)$ and $K_1(1402)$. The theoretical numbers assume  
an HQET mixing angle 
$\theta = \tan^{-1}(1/\sqrt{2})\approx +35.3^o$. 
}
\label{table_K1}
\end{table}

\subsection{2P States} 

Recent experimental work, especially from the VES, E852
and Crystal Barrel Collaborations, has established
several likely members of the 2P $n\bar n$ multiplet, specifically
the 
$a_2(1726)$ \cite{Ams02b}
(also notable as the first radial excitation reported in $\gamma\gamma$,
by L3 at LEP \cite{Acc97}),
$a_1(1700)$ \cite{Ame95,Chu02,Bak99} and 
$h_1(1594)$ \cite{Eug01}. Comparison with
their 1P analogues suggests that the 2P-1P separation is 
$\approx 450$~MeV. Presumably the splittings in the kaon system
are similar, so we expect the 2P kaon multiplet at about
1850 MeV for unmixed 
$2 ^3$P$_\J$ states 
and about 1800 MeV for
the mean $2 ^3$P$_1$-$2 ^1$P$_1$ mass.

\subsubsection{The unobserved $K^*_2(1850)$} 

We predict that the 2P tensor state $K^*_2(1850)$ is rather broad,
$\Gamma_{tot}\approx 370$~MeV, 
with no strong preference
for any one decay mode. The four largest branching fractions are
predicted to be to 
$\rho K^*$,
$\pi K^*$,
$\rho K$ and
$\pi K$,
each in the $10$-$20 \% $ range. 
Interference between the $|n\bar n\rangle $ and $|s\bar s\rangle$
components of the $\eta$ and $\eta'$
leads to 
the prediction that
$B_{\eta' K} >> B_{\eta K}$ 
(see App.B).
Note that there is an inverted rule
for the coupling of the 
2$^3$P$_2$ $s\bar s$ to $\eta K^*$ relative to 
$\eta' K^*$, so we also predict
an important $\eta K^*$ mode. 
The theoretically suppressed 
mode $\eta' K^*$ is unfortunately 
not easily accessible in $K^*_2(1850)$ decays 
due to the lack of phase space.
The predictions 
that 
$B_{\eta' K}$ and $B_{\eta K^*}$ are each $\approx 5\% $, 
but that the branching fraction to the lower-mass mode $\eta K$ 
is much weaker, may serve as useful signatures
for this  
state. 

\subsubsection{The unobserved $K^*_0(1850)$}

The 2$^3$P$_0$ scalar $K^*_0(1850)$ is predicted to have a 
total width of
$\Gamma_{tot}\approx 450$~MeV, comparable to
the 2$^3$P$_2$ tensor. 
Although the decays of the scalar 
typically have smaller centrifical barriers, many of the tensor decay model
are forbidden to the scalar. These compensating effects lead
to comparable total widths.

The important decays are again distributed over several modes,
but in this case decays to radially and orbitally excited
states are expected to dominate. The largest mode
is predicted to be $\pi K_1(1273)$ (about $30\% $), with 
{\it ca.} $10\% $ branching fractions to $\rho K^*$, $b_1 K$ and
$\pi(1300) K$. The relative strength into $\pi K_1(1273)$ 
versus $\pi K_1(1402)$ is strongly dependent on the
1P mixing angle $\theta$, here assigned the HQET value.
The VV modes $\rho K^*$ and $\omega K^*$ are predicted to have
similar $^1$S$_0$ and $^5$D$_0$ amplitudes.
None of the resulting $K+n\pi$ final
states is especially attractive experimentally, although
$b_1 K \to \pi \omega K$ might be interesting as a flux-tube model decay
mode expected to show strange hybrids. The 
$K^*_0(1850)$ may be observable in 
its relatively weak $\eta' K$ decay, which is also
expected to show evidence of 
the $K^*_2(1850)$ partner. A much weaker 
$\eta K$ mode is expected, due to destructive interference
in the $\eta$ flavor state (App.B).

The $K^*_0(1945)$ reported by LASS \cite{Ast88e} in $K\pi$ 
at a mass of
$1945\pm 10\pm 20$ MeV (actually this is an average of
two LASS solutions, see their Table~3) is
a possible experimental candidate for the 2$^3$P$_0$ state.
A recent reanalysis of
the data found the K-matrix pole at $1885^{+50}_{-80}$
MeV, consistent with the LASS analysis, 
but the physically more relevant 
T-matrix pole was found at a mass of $1820\pm 40$ MeV, with a
width of $250 \pm 100$ MeV~\cite{Ani97c}. These parameters are
consistent with our expectations for a $2^3$P$_0$ state.
The strength of the $\pi K$ coupling reported by LASS 
is however much larger than our expectations for the 
2$^3$P$_0$
quark model state;
experimentally $B_{K^*_0(1945)\to \pi K} = (52 \pm 8 \pm 12 )\% $ 
(again an average of two solutions), whereas 
the \3P0 model predicts a much smaller $B_{\pi K} = 6 \% $
for the 2$^3$P$_0$ kaon.

\subsubsection{2$^3$P$_1$-2$^1$P$_1$ $K_1(1800)$ states}

Motivated by the well-known $1^1$P$_1$-$1^3$P$_1$ mixing in the lighter
1P $K_1$ states, we quote decay amplitudes and partial widths 
for 2P $K_1(1800)$ states as functions of a 
similar singlet-triplet mixing angle $\theta$.
Our definition of the 2P mixing angle is 
\begin{equation}
|K_1^a(1800)\rangle = 
+\cos(\theta) |2^1{\rm P}_1\rangle
+
\sin(\theta) |2^3{\rm P}_1\rangle \phantom{\ .}
\end{equation}
and
\begin{equation}
|K_1^b(1800)\rangle = 
-\sin(\theta) |2^1{\rm P}_1\rangle
+
\cos(\theta) |2^3{\rm P}_1\rangle \ , 
\end{equation}
as was assumed for 1P states.

It is evident from Table~K10 that searches for these 
resonances might most usefully concentrate 
on the modes $\rho K$ and $\pi K^*$.
These branching fractions are intrinsically large, and as their 
$\cos(\theta)\sin(\theta)$ cross terms have opposite signs, a
state that is accidentally suppressed in one mode should
be clearly evident in the other. The 
somewhat weaker $\omega K$ and $\phi K$ modes
are rather cleaner to reconstruct, and will be useful as independent 
checks 
of the observation of these states. The $\omega K$ 
partial width is related to
$\rho K$ by a trivial isospin factor of $1/3$ (with
minor phase space differences).
The $\phi K$ mode
is experimentally 
attractive because the $\cos(\theta)\sin(\theta)$ cross term 
in this branching fraction is 
relatively weak, so we expect both states to be evident in
$\phi K$,
independent of the mixing angle. 
This might 
explain the PDG
report of a ``$K_1(1650)$" state in $\phi K$ 
at inconsistent masses of $1650\pm 50$~MeV and $\sim 1840$~MeV.

The VV modes $\rho K^*$ and 
$\omega K^*$ are interesting because the three 
subamplitudes 
$^3$S$_1$,
$^3$D$_1$ and 
$^5$D$_1$
are comparable and are individually proportional to 
$\cos(\theta)$
{\it or}
$\sin(\theta)$, and thus may be useful in determining this mixing angle.
The mode $\pi K_1(1273)$ may also be useful for establishing the
2P angle, since it couples strongly to the spin-triplet component in the
initial state.

Experimental candidates for these 2P axial-vector states exist, but
are rather poorly established.
The PDG
``$K_1(1650)$" entry summarizes three experimental reports of states
at
masses of $1650\pm 50$~MeV, $\sim 1800$~MeV and $\sim 1840$~MeV, and
only the lowest is inconsistent with our assumed 2P mass. In view
of the known experimental splitting of 
$ca. 130$~MeV between the two 1P $K_1$ states, these 
``$K_1(1650)$" reports may well represent observations of the two 
$\J^{\P} = 1^+$ 2P states.

\subsection{1D States} 

\subsubsection{$K^*_3(1776)$} 

In view of the reasonably successful \3P0-model description of
$\phi_3(1854)$ decays, one expects a similarly good description
of the decays of its
kaonic partner $K^*_3(1776)$. The relatively 
small total
width of the $K^*_3(1776)$ is indeed reproduced by the model;
experimentally it is 
$\Gamma_{tot} = 159\pm 21$~MeV, compared to a 
theoretical 
$\Gamma_{tot} = 148$~MeV.

The PDG reports
experimental 
branching fractions
for the $K^*_3(1776)$, based largely on constrained fits to
LASS data. The resulting partial widths are shown in Table~\ref{k3},
together with our predictions. 
Although discrepancies between 
theory and experiment appear possible, they are not 
especially significant at present accuracy. 
It is notable that the mode with the largest theoretical branching
fraction,
$\rho K^*$, has not been incorporated in the PDG fit. 
Neglect of this mode 
will lead to overestimated partial widths for the remaining modes,
as the branching fractions are assumed to sum to unity.

In addition to the nine modes given in the summary table, 
there are several other numerically
unimportant ones that are listed
in Table~K11.

The $\eta K/ \eta'K$ selection rule (see App.B) 
is clearly
evident theoretically; constructive interference between
$n\bar n$ and $s\bar s$ components of the $\eta$ in this odd-L
$\eta K$ state makes $\eta K$ an important mode, whereas
$K_3^*\to \eta' K$ suffers destructive interference and hence is
strongly suppressed. (Compare this to the even-L 
decay mode $K_2^*(1429) \to \eta K$ 
in Table~\ref{tensor}.) 
The PDG quotes their fitted branching fraction of 
$B_{K^*_3(1776)\to \eta K} = (30\pm 13)\% $, which combined with their
total width gives the $\eta K$ partial width of $48\pm 21$~MeV in
our Table~6. 
This width is consistent with the $\eta K$ selection rule,
{\it albeit} with large errors. 
We note however that there is a better determined 
$K_3^*(1776)$ 
branching fraction ratio published elsewhere by 
LASS \cite{Ast88d}, which is
$B_{\eta K} / B_{\pi K} = 0.50 \pm 0.18$. This gives
a partial width of $\Gamma_{K^*_3(1776)\to \eta K} = 15 \pm 6$~MeV,
which is also quoted in Table~6. 
Finally, there is an unpublished LASS result of
$B_{\eta K} / B_{\pi K} = 0.41 \pm 0.053$ \cite{Bir89}, which agrees
quite well with our theoretical ratio of 0.48.

\begin{table}
\centering
\begin{tabular}{|c|c|c|c|c|c|c|c|c|c|}
\hline
mode: $\Gamma_i$ (MeV)
& $\pi K$
& $\eta K$
& $\eta' K$
& $\rho K $
& $\omega K $
&  $\pi K^*$
& $\rho K^* $
& $\omega K^* $
&  $\pi K^*_2$
\\
\hline
\hline
$K^*_3(1776)$ 
(expt)
&   $30\pm 4   $ 
&   $48\pm 21  $ 
&    -
&   $49\pm 16 $ 
&    -
&   $32\pm 9   $ 
&    -
&    -
&   $< 25 $ 
\\
\hline
''
&      
& \ \ \ \  $15\pm 6\; \cite{Ast88d}  $ 
&   
&     
&   
&       
&    
&    
&    
\\
\hline
$K^*_3(1776)$ 
(thy)
&   $ 40 $ 
&   $ 19 $ 
&   $ 0.05  $ 
&   $ 10 $ 
&   $ 3.2  $ 
&   $ 14  $ 
&   $ 42  $ 
&   $ 12  $ 
&   $ 1.1  $ 
\\
\hline
\end{tabular}
\caption{\label{k3}Experimental and theoretical partial widths of the 
$K_3^*(1776)$ $1^3$D$_3$ kaon candidate.} 
\label{table_K3}
\end{table}

\subsubsection{$K^*(1717)$} 

The PDG considers only 
$\pi K$,
$\rho K$ and 
$\pi K^*$ modes for this ``$K^*(1680)$" state, and 
previous experimental studies indicate comparable
branching fractions to each. Since the $\pi K$ branching fraction was
determined by LASS \cite{Ast88e} to be $0.388\pm 0.014 \pm 0.022$,
these
three modes 
would appear to account for most of the decays of this state.

Our decay calculations suggest
that this is not correct; we find large
couplings to $0^- 1^+$ modes, and the largest branching fraction is
predicted to be to $\pi K_1(1273)$, with
$B_{\pi K_1(1273)}\approx 40\%$. 
The 
$\pi K_1(1402)$ mode in contrast is predicted to be weak, but this result
is strongly dependent on the $K_1$ mixing angle $\theta$,
here assumed to be equal to the HQET value $\approx 35.3^o$. If 
accurately measured,
these branching fractions might strongly constrain $\theta$.
Unfortunately, the $\pi K_1$ modes of the
$K^*(1717)$ have not been studied experimentally. 

This prediction of the $^3$P$_0$
model is familiar in the context of the $1^3$D$_1$ candidate
$\rho(1700)$, which is predicted to have very large couplings to
$\pi a_1$ and $\pi h_1$ \cite{Bar97a}. Since this large 
$1^3$D$_1\to 0^- 1^+$ coupling has not been confirmed experimentally in
any flavor sector, the predicted dominance of $\pi K_1(1273)$ 
found here should be 
considered an interesting future test of the $^3$P$_0$ decay model.

A large $\eta K$ branching fraction and a suppressed $\eta' K$ one
are predicted, as expected for an odd-L final state (App.B).
The more important 
or interesting $K^*(1717)$ decay modes 
(larger than $2\% $ branching fraction, and the
suppressed $\eta' K$ mode) 
are shown in the summary table. 
For the three reported modes we predict the ordering 
$\pi K> \rho K\approx \pi  K^*$, consistent with experiment.

\begin{table}
\centering
\begin{tabular}{|l|c|c|c|c|c|c|c|c|c|}
\hline
mode: $\Gamma_i$ (MeV)
& $\pi K$
& $\eta K$
& $\eta' K$
& $\rho K $
& $\omega K $
& $\phi K $
&  $\pi K^*$
&  $h_1 K$
& $\pi K_1(1273) $
\\
\hline
\hline
$K^*(1717)$ 
(thy)
&   $ 45 $
&   $53  $ 
&   $ 1.0 $
&   $ 26 $ 
&   $ 8.5 $ 
&   $8.6  $ 
&   $25  $ 
&   $33  $ 
&   $145  $ 
\\
\hline
\end{tabular}
\caption{Important theoretical partial widths of a
$^3$D$_1$ $K^*(1717)$ kaon.} 
\label{table_K3D1}
\end{table}

We note in passing that the $K^*(1717)$ has been reported in D-meson
nonleptonic weak decays,
$B_{D^+ \to {K^{*o}(1717)}\pi^+} = (1.45\pm 0.31)\% $ 
in the $K^- \pi^+$ mode \cite{PDG02}, so this approach might allow observation
of the interesting modes $\eta K$ (comparable to $\pi K)$ and $\eta' K$ in 
future (see App.B).
 
\subsubsection{$K_2(1773)$ and $K_2(1816)$} 

The $K_2$ sector is especially interesting because it allows
tests of models of mixing
between spin-singlet and spin-triplet states, as is seen in the $K_1$
system. If this is a short-distance effect we might expect to
find much stronger mixing in the P-wave $K_1$ system than in the
D-wave $K_2$ states. The smaller mass 
splitting in the $K_2$ sector 
suggests that the mixing angle $\theta$ may well be smaller here.

Since the $K_2$ states are 400-500 MeV higher in
mass, one can measure their couplings to many decay modes 
that are inaccessible to the 1P
$K_1$ states. This will allow many checks of the \3P0 decay
model, since if it is accurate a single value of the mixing 
angle $\theta$ should correlate a large number of decays.
In Tables~K12 and K13 we give results for the
decay amplitudes and partial widths of the $K_2$ states with
general mixing angles,

\begin{equation}
|K_2(1773)\rangle = 
+\cos(\theta)  | 1^1{\rm D}_2\rangle
+ 
\sin(\theta) | 1^3{\rm D}_2\rangle\phantom{\ .} 
\end{equation}
and
\begin{equation}
|K_2(1816)\rangle = 
-\sin(\theta)  | 1^1{\rm D}_2\rangle
+ 
\cos(\theta) | 1^3{\rm D}_2\rangle \ .
\end{equation}

One can see in the decay tables that many $K_2$ partial 
widths are strongly 
dependent on the mixing angle $\theta$. The relatively
clean modes $\omega K$ and $\phi K$ are especially interesting 
because their $\sin(\theta)\cos(\theta)$ cross terms have opposite
signs, so the ratio 
$B_{K_2\to \phi K / \omega K}$ 
depends strongly on the $K_2$ mixing angle.
The odd-L $\eta K^*$ mode is strongly $\theta$-dependent as well,
and is predicted to couple dominantly to the spin-singlet $^1$D$_2$ 
component of
the initial $K_2$ state 
(App.B, Table~B2).
The VV modes $\rho K^*$ and $\omega K^*$ are interesting because there
is no $\sin(\theta)\cos(\theta)$ cross term in the partial widths;
the individual
subamplitudes are proportional to $\sin(\theta)$ or $\cos(\theta)$
only. A determination of the relative 
$ ^3$P$_2$
and 
$ ^5$P$_2$ 
VV amplitudes 
would be an excellent independent check of $\theta$,
although these modes may be too weak to allow this measurement.
The $\pi K_2^*(1429)$ mode is also interesting, because it could be weak 
or dominant, depending on the value of 
$\theta$.

Unfortunately the experimental data on the $K_2$ states is not yet
sufficiently quantitative to be compared usefully to our decay
predictions. The PDG claims that $\pi K_2^*(1429)$ is the dominant 
$K_2(1773)$ decay mode, but the individual experiments are not all in
agreement about this. The $\pi K^*$, $f_2(1275) K$, $\rho K$ and
$\omega K$ modes of the $K_2(1773)$ are all ``seen", which is 
at least encouraging
for our proposed future determination of $\theta$ from 
$B_{K_2(1773)\to \phi K / \omega K}$.  

The $K_2(1816)$ data is even less constraining, with only two experimental
references. The modes $\pi K_2^*(1429)$, $\pi K^*$, 
$f_2(1275) K$ and $\omega K$
are again ``seen" in the PDG summary. 
Daum {\it et al.} \cite{Dau81} actually report a strong preference for
$\pi K_2^*(1429)$, 
$B_{K_2(1816)\to \pi K_2^* / \pi\pi  K} \sim 0.77$.
In comparison they quote
$B_{K_2(1816) \to f_2(1275) K / \pi\pi  K} \sim 0.18$
and
$B_{K_2(1816)\to \pi K^* / \pi\pi  K} \sim 0.05$.
This large $\pi K_2^* /  \pi K^* $ 
ratio is not consistent with the \3P0-model prediction
that 
these two modes have comparable strengths.

The LASS observation \cite{Ast93}
of both P- and F-wave contributions to the transition
$K_2(1773) \to \omega K$ (Table~2 of Ref.\cite{Ast93}) 
is quite interesting, as we find that the F/P amplitude
ratios for both $K_2\to \omega K$ transitions vary 
rapidly with the singlet-triplet mixing angle $\theta$. 
Although the LASS results are not very statistically
significant (the F-waves are $\approx 1\sigma$ and $2\sigma$ from zero),
they do show that F/P is quite small in $K_2(1816)\to \omega K$.
This argues in favor of a sizeable and negative $K_2$ mixing angle;
a vanishing $K_2(1816)\to \omega K$ F-wave requires
$\theta_2 = - \tan^{-1}(\sqrt{2/3}\, ) \approx -39^o$.

\subsection{1F States}

\subsubsection{$K_4^*(2045)$}  

The $K_4^*(2045)$ is the single well-established 
member of the 1F kaon
multiplet. It is assumed to be the flavor partner of the 
$n\bar n$ states $f_4(2025)$ and $a_4(2011)$ and perhaps the 
LASS $s\bar s$ candidate  
$f_4(2209)$. The reported mass is much closer to the $n\bar n$ 
states than the $s\bar s$ candidate, which is surprising, and is
reminiscent of the $K^*(1410)$. 

The PDG total width of $198\pm 30$ MeV is somewhat larger
than our theoretical expectation of
$\Gamma_{tot}\approx 100$ MeV. The \3P0 model predicts that only a few 
low-lying two-body modes of the $K_4^*(2045)$ 
have branching fractions larger than
a few percent. 
This weakness of higher-mass modes is typical of a high-L, high-J 
state,
since the angular threshold barriers for decays combined with smaller 
phase space leads to smaller branching fractions.

The largest modes are predicted to be $\rho K^*$ and $\pi K$,
with branching fractions of
$\approx 30\% $ and $\approx 20 \% $ respectively. 
The predicted partial
width $\Gamma_{\pi K} = 21$~MeV is consistent with the reported LASS
$\pi K$ branching fraction of $(9.9\pm 1.2) \% $~\cite{Ast88e}, 
given their total width 
of about 200 MeV. We note however that this agreement is rather fortuitous,
since this is a G-wave final state and as such has 
very strong $|\vec p_f|^9$ threshold behavior. 
Only three other modes are 
predicted to be larger than $5\% $, these being $\omega K^*$, 
$\rho K$ and $\pi K^*$.
The $\phi K^*$ mode, with a reported branching fraction of $(1.4\pm 0.7) \% $, 
can be used to test the assumed flavor independence
of the $n\bar n$ and $s\bar s$ pair production amplitudes in the novel
VV channel. Our predicted
branching fraction of
$2.7\% $ is consistent with experiment at the current limited accuracy. 
A comparison of the $\phi K^*$ mode with 
$\rho K^*$ or $\omega K^*$ would constitute an interesting
direct test of 
the assumed flavor (quark mass) independence of the $q\bar q$ 
pair-production amplitude in the \3P0 model, since $\phi K^*$ requires
$s\bar s$ pair production whereas $\rho K^*$ and $\omega K^*$ require
$n\bar n$.

It is interesting that the reported PDG branching fractions 
only account for about half
of the $K_4^*(2045)$ decays. The $^3$P$_0$ model does not anticipate
any additional modes with sufficient strength to explain this 
discrepancy.

\subsubsection{$K_2^*(2050)$}

We assume a mass of 2050 MeV for the $1^3$F$_2$ kaon, which is
a rounded $K_4^*(2045)$ mass. 
Since this is a high-mass state with low $J$, we find that many
two-body final states are predicted to have significant couplings.
Axial-vector plus pseudoscalar modes are among the most important;
$\pi K_1(1273)$, $b_1K$, $a_1K$ and $\eta K_1(1273)$
(in decreasing order of branching fraction) are all predicted
to be in the $\approx 10-30 \% $ range. 
The S-wave mode 
$\pi K_2(1773)$ is also predicted to have a large 
($\approx 20 \% $) branching fraction, although this 
is strongly dependent on the 2P singlet-triplet mixing angle; we have
assumed HQET values for the two $K_2$ states,
analogous to the $K_1(1273)$ and $K_1(1402)$, and if this 
is inaccurate
there may be a large $\pi K_2(1816)$ mode.

The ``standard" light modes such as $\pi K$ are predicted to couple 
rather weakly to this state. $\pi K$, $\rho K$ and $\pi K^*$ 
have predicted branching fractions of only about $5\% $. 
One attractive approach to identifying this state would be to observe it
in 
$\eta' K$ (also a {\it ca.} $5\% $ branch), but not in
$\eta K$, which is a signature for decays to 
these even-L final states (App.B).

Should this state be identified, there is an interesting $^3$P$_0$
decay model prediction that the light VV modes 
$\rho K^*, \omega K^*$ and $\phi K^*$ 
will couple quite
weakly, since they are predicted to
have zero coupling 
in S-wave. We might {\it a priori} have expected the S-wave 
to be the largest
VV amplitude. 

There is a possible LASS candidate
for this state at $1973\pm 8 \pm 25$ MeV \cite{Ast87}, reported
in $\rho K$ and
$\pi K^*$,
with a total width of $\Gamma_{tot} = 373\pm 33\pm 60$~MeV 
and
a relative branching fraction of 
$B_{K_2^*(1973) \to \rho K /\pi K^*} = 1.49 \pm 0.24 \pm 0.09$.
These results are consistent with our expectations for 
a $1^3$F$_2$ kaon. 

\subsubsection{The unobserved $K_3(2050)$ states}

The two J$^{\rm P} = 3^{+}$ 1F states 
will provide an independent test of models of the mixing between
spin-singlet and spin-triplet kaon
states, such as is observed in the 1P $K_1(1273)$-$K_1(1402)$ system.
We can expect the predicted 1F 
mixing angle to depend rather strongly on the assumed 
mechanism. If it is a short distance effect it should be much smaller in the 
L=3 $K_3$ states than in 1P, whereas if it is simply a mixing angle chosen by
heavy-quark symmetry, the 1F $K_3$ and 1P $K_1$ values should
be similar. Since the $K_3$ states have much higher masses, there are
many more decay modes that can be used to determine this mixing angle.

We will assume a mass of 2050~MeV for both 1F $K_3$ states, so we need 
only quote results for one linear combination, 
which we take to be 
\begin{equation}
|K_3^a(2050)\rangle =
+
\cos(\theta) |^1{\rm F}_3\rangle
+
\sin(\theta) |^3{\rm F}_3\rangle
\ .
\end{equation} 
The decay amplitudes and partial widths 
are
given in Table~K15
as functions of $\theta$. 
Note that the total width is
not strongly dependent on the mixing angle; we expect these 
states to have total widths of {\it ca.} 200-250 MeV whatever the value of
$\theta$. 

The PsV modes
$\rho K$,
$\omega K$,
$\pi K^*$
and
$\phi K$ 
have significant partial widths,
and it is notable that the sign of the 
$\sin(\theta) \cos(\theta)$
cross term
is channel-dependent; thus 
there is especially strong $\theta$-dependence
in ratios such as $B_{\phi K} / B_{\omega K}$.

The VV modes have the interesting feature that their
partial widths have no $\sin(\theta) \cos(\theta)$ cross term, because
the individual L, S subamplitudes are proportional to either 
$\sin(\theta)$ or $\cos(\theta)$. (This was also noted for 
mixed 2P state decays.) Thus measurements of the $\rho K^*$ or
$\omega K^*$ subamplitudes directly access $\sin(\theta)$ and $\cos(\theta)$.

In the higher-mass final states we find 
large branching fractions to 
$a_2 K$, $f_2 K$, $\pi K^*_2(1429)$ and $\pi K_3^*(1776)$,
again with strong $\theta$-dependence. 
The final state $\pi K_3^*(1776)$ is interesting in that it is
the only open $K_3(2050)$ mode
with an S-wave amplitude.

\newpage

\section{Summary and Conclusions}

In this paper we have presented 
a detailed survey of the status and strong decays
of all strange mesons expected in the quark model up to {\it ca.}
2.2~GeV. This includes the 1S, 2S, 3S, 1P, 2P, 1D and 1F multiplets
of strangeonia and kaonia, making a total of 44 states. 42 of these
have strong decays (43 since we consider
$\eta_2$ flavor mixing),
and we have carried out calculations 
of all the energetically allowed open-flavor decays of all these
states
in the \3P0 model.
All independent decay amplitudes and partial and total widths
were evaluated numerically and presented in detailed decay tables.
In total we have given numerical results for 
525 two-body decay modes and 891 decay amplitudes.

This work is intended as a guide for future experimental studies of
meson spectroscopy, to indicate what modes and amplitudes are
expected to be important and are theoretically interesting, 
as well as to allow the identification of unusual states
such as glueballs and hybrids through their anomalous decay properties.

We have identified several very
interesting issues 
for future experimental studies 
involving the conventional quark model states.
As one example, in the $s\bar s$ sector
we predict two rather narrow states that have not been identified, 
the 
$1^3$D$_2$ 
$\phi_2(1850)$ with $\Gamma_{tot} \approx 210$~MeV 
(with large $KK^*$ and $\eta\phi$ modes) and
the 
$1^1$D$_2$ 
$\eta_2(1850)$ (assuming it is pure $s\bar s$; see below) 
with $\Gamma_{tot} \approx 130$~MeV,
decaying mainly to $KK^*$. 
The $\eta_2$ states at 1617 and 1842~MeV are
also very interesting because the higher-mass state is only seen in $\pi a_2$.
We consider the effect of a large $n\bar n\leftrightarrow s\bar s$
mixing angle, and note that this implies important $KK^*$ modes 
that 
are not evident in the data; the possibility that the  
higher-mass $\eta_2(1842)$ 
is a nonstrange hybrid rather than a quarkonium state
certainly merits consideration.
Future searches for C=$(-)$ $s\bar s$ states
might exploit the 
$\eta\phi$ 
and
$\eta'\phi$ ``$s\bar s$-signature modes",
which are not directly accessible to 
light $n\bar n$ mesons. 

There are many interesting issues 
in the kaon sector. 
One is the amount
of spin singlet-triplet mixing in the series of 
$\J^{\rm P} = 1^+, 2^-, 3^+ \dots $ kaons. The $K_1(1273)$-$K_1(1402)$
system is known to have a large singlet-triplet mixing angle, 
and the physical origin is not well established. Similar mixing is considered
in the 2P, 1D ``$K_2$" and 1F ``$K_3$" systems, and it is noted that the
decay amplitudes and partial widths of these states 
are often very sensitive to
these mixing angles. Quantitative studes of these strong decay amplitudes
and branching fractions
will allow the determination
of these mixing angles, and can also provide tests of the accuracy
of the \3P0 decay model.

Kaons are much better established experimentally than $s\bar s$ states;
of the 21 theoretical excited kaon levels we consider, 
just eight do not have plausible associated
experimental candidates. The eight unknown kaon states are predicted to have
total widths in the $\Gamma_{tot}\approx 300$-400~MeV range, and the modes
$\rho K$, $\rho K^*$ and $\pi K^*$ should be useful for the identification
of most of these states. An interesting exception is the
$1^3$F$_2$ $K_2^*(2050)$, which is predicted to have large branching 
fractions to the unusual modes $\pi K_1(1273)$,
$\pi K_2(1773)$ and $b_1 K$.   

Kaon decays to modes with an $\eta$ or $\eta'$ are especially interesting, 
in that an interference takes place between the 
$|n\bar n\rangle$  
and
$|s\bar s\rangle$ components of the final $\eta$ or $\eta'$. This 
interference is 
strongly constructive
or destructive depending on the channel and angular quantum numbers, and
there is strong experimental evidence of this effect
in $K^*_2(1429)$ decays. 
The associated  
selection rules have also been applied to D and B meson weak decays
to $\eta K$ and related final states, where unusual branching fractions
have been observed. We derived these selection rules 
from our strong decay amplitudes in Appendix B, and noted that there is
a nontrivial generalization to modes such as $\eta K_1$ in B decays.

Finally, this work should be useful in searches for glueballs and hybrids,
assuming configuration mixing is not large, as one should eliminate the
$q\bar q$ quark model ``background" in any search for new, unconventional
meson resonances. 
We also note that 
the spectrum of kaons will appear rather different from the spectrum
of $n\bar n$ or $s\bar s$ states 
if mixing between quarkonia and
hybrids {\it is} important,
because kaonia mix with more
hybrid basis states due to the absence of C-parity. This may
lead to irregularities in 
relative level positions in the $n\bar n$ and excited
kaon spectra, as perhaps is already evident in the low mass of the
strange $K^*(1414)$ relative to the $n\bar n$ state
$\rho(1465)$.      
Irregularities between the kaon and I=1 $n\bar n$ spectra
may thus signal the presence of hybrid basis states.

\section{Acknowledgments}

We are pleased to thank
W. Bugg, S.U. Chung,
F.E. Close, A. Donnachie, W. Dunwoodie, A. Dzierba, S. Godfrey, T. Handler,
H. J. Lipkin,
R. Mitchell, C.A. Meyer, M. Nozar, W. Roberts, M. Selen, E.S. Swanson,
S. Spanier, E. Vaandering and D. Weygand
for useful discussions and communications,
and C. Salgado for enthusiastic support.
This work was supported in part by
the United States Department 
of Energy under contract DE-AC05-96OR22464 managed by UT Battelle at 
Oak Ridge National Laboratory, by the University of Tennessee, 
and by the United States Department of Energy 
under contract W-7405-ENG-36 at Los Alamos National Laboratory.

\newpage

\renewcommand{\theequation}{A\arabic{equation}}
\setcounter{equation}{0}  
\section*{Appendix A: \3P0 decay model conventions}  

In this appendix we discuss some 
details of the \3P0 decay 
calculations that are presented in this paper.
Our diagrammatic, momentum-space formulation of
the \3P0 model is described in Ref.\cite{Ack96},
and our results are essentially an extension of the
decay model calculations of Ref.\cite{Bar97a},
which considered only $n\bar n$ mesons in detail.

The \3P0 model describes open-flavor meson strong decay as
a $q\bar q$ pair-production process, in which the new $q\bar q$
pair separate into final $q\bar q$ mesons B and C.
The pair is assumed to be produced in a 
J$^{\rm PC} = 0^{++}$ state (hence ``\3P0 model"), corresponding
to vacuum quantum numbers. This choice is supported
by experimental amplitude ratios, notably the D/S
ratios in $b_1\to \pi \omega$ \cite{Noz02} and $a_1\to \pi \rho$.
As noted in Ref.\cite{Ack96}, the usual \3P0 decay
amplitude is equivalent to the nonrelativistic 
limit of the interaction Lagrangian 
${\cal L}_I = g\, \bar \psi_q\psi_q$, with the identification
$\gamma = g/ 2 m_q$. (The dimensionless $\gamma$ is the 
pair-production amplitude, which is taken to be a free
parameter in the \3P0 model.) 
In this first detailed survey of strange meson decays
we have chosen to avoid the 
complications of moderate parameter variations and the
effect of the larger strange quark mass on the meson wavefunctions,
and present results
that follow from the previously assumed 
$n\bar n$ SHO wavefunctions.
Thus the analytical results for amplitudes given in App.A of Ref.\cite{Bar97a}
are valid for this paper as well.
We assume the same SHO wavefunction width parameter and pair-production
amplitude as Ref.\cite{Bar97a},
\begin{eqnarray}
&\gamma &=\ 0.4 \\
&\beta &=\ 0.4 \ {\rm GeV} \ .
\end{eqnarray}
Comparison of partial widths to experiment
for light decays of strange states shows that the predictions 
are indeed a useful
guide with these parameters.
One should of course expect a slight decrease in length scale 
and hence a slightly larger $\beta$ in the strange mesons, 
due to the heavier strange quark.

This interaction leads to 
the two Feynman diagrams of Fig.1 for the process
${\rm A} \to {\rm B} + {\rm C}$.
As shown in Ref.\cite{Ack96}, the
T-matrix element for each diagram in a given decay is
the product of separate factors for flavor, spin, and
a convolution integral involving the three mesons' 
spatial wavefunctions. There is an additional overall ``signature" phase
of $(-1)$ due to quark operator anticommutation.
The color degree of freedom, which would lead to a common
overall multiplicative color factor, is suppressed.

The meson 
flavor states 
follow the conventions of Ref.\cite{Bar97a}. 
The fundamental quark flavor-${\bf 3}$ and antiquark flavor-${\bf \bar 3}$ are
$q = (+d, +u; +s)$ and $\bar q = (+\bar u, -\bar d; -\bar s )$, so for 
example 
$|\pi^-\rangle = + |d\bar u\rangle$ and $|K^-\rangle = + |s\bar u\rangle$
but
$|\phi\rangle = -|s\bar s\rangle$, $|\rho^+\rangle = -|u\bar d\rangle$,
and $|K^+\rangle = - |u\bar s\rangle$. 
Unless otherwise stated we take the 
$\eta$ and $\eta'$ to be maximally
mixed flavor states,
\begin{equation}\label{seq1a}
|\eta\rangle \;  = 
{1 \over \sqrt{2}} \Big(|n\bar n \rangle_{_0}  - |s\bar{s}\rangle \Big) 
\end{equation}
\begin{equation}\label{seq1b}
\hskip 0.05cm
|\eta'\rangle   = 
{1 \over \sqrt{2}} \Big(|n\bar n \rangle_{_0}  + |s\bar{s}\rangle \Big)
\end{equation}
where the I=0 state $|n\bar n \rangle_{_0} $ is
\begin{equation}
|n\bar n \rangle_{_0} = 
{1 \over \sqrt{2}} \Big(|u\bar u \rangle + |d\bar{d}\rangle \Big)\ .
\end{equation}
For cases in which we consider the dependence on the 
$\eta$-$\eta'$ mixing angle we use an expansion in
$q\bar q$ flavor states, with a flavor mixing angle $\phi$;

\begin{equation}\label{seq2}
|\eta\rangle = 
\cos(\phi) \, 
|n\bar n \rangle_{_0}
-
\sin(\phi)\,  |s\bar{s}\rangle
\end{equation}
and
\begin{equation}\label{seq3}
\hskip 0.28cm
|\eta'\rangle =
\sin(\phi)\,  
|n\bar n \rangle_{_0}
+
\cos(\phi)\,  |s\bar{s}\rangle \ .
\end{equation}
\vskip 0.5cm
\noindent
The more common expansion 
in SU(3) octet and singlet flavor states is
\begin{equation}\label{seq2a}
|\eta\rangle =
\cos(\theta_P)\,  
|\eta_8\rangle
- 
\sin(\theta_P)\,  
|\eta_1\rangle
\end{equation}
and
\begin{equation}\label{seq2b}
\hskip 0.32cm
|\eta'\rangle =
\sin(\theta_P)\, 
|\eta_8\rangle
+
\cos(\theta_P)\, 
|\eta_1\rangle \ .
\end{equation}
\vskip 0.5cm
\noindent
Our conventions for these SU(3) basis states are
\begin{equation}\label{seq2c}
|\eta_8\rangle
=
\sqrt{1\over 6}\ 
\Big(
|u\bar u \rangle
+
|d\bar d \rangle
-
2\;
|s\bar s \rangle 
\Big)
\end{equation}
and
\begin{equation}\label{seq2d}
|\eta_1\rangle
=
\sqrt{1\over 3}\ 
\Big(
|u\bar u \rangle
+
|d\bar d \rangle
+
|s\bar s \rangle 
\Big) \ .
\end{equation}
These expansions imply a relation between $\phi$ and $\theta_P$,
\begin{equation} 
\theta_P = \phi - \tan^{-1}(\sqrt{2})\ \  \approx \ \phi - 54.7^o \ .
\end{equation}
Our maximally mixed states, with $\phi = 45^o$, correspond to
the familiar value $\theta \approx -10^o$.

The strange mesons $K_1(1273)$ and
$K_1(1402)$ also require careful phase definitions, since various conventions
have appeared in the literature. We define the 
singlet-triplet mixing angle $\theta$ 
for $n \bar s$ axial kaon states as 

\begin{eqnarray}
\hskip 0.5cm
&|K_1(1273)\rangle &= 
+ \cos(\theta) \, | 1^1{\rm P}_1 \rangle 
+ \sin(\theta) \, | 1^3{\rm P}_1 \rangle \\ 
&|K_1(1402)\rangle &= 
- \sin(\theta) \, | 1^1{\rm P}_1 \rangle 
+ \cos(\theta) \, | 1^3{\rm P}_1 \rangle \ .
\end{eqnarray}
As we noted in the section on $K_1$ mesons, our mixing angle $\theta$
is opposite in sign to that of Blundell and Godfrey \cite{Blu96b},
because they apply Eqs.(A13,A14) to $s\bar n$ antikaons, whereas we apply
it to $n\bar s$ kaons. 

Two physically independent 
values of $\theta$ follow from the HQET limit $m_s\to\infty$, which are
$\theta = +\tan^{-1}(1/\sqrt{2}) \approx +35.3^o$ 
and $\theta = -\tan^{-1}(\sqrt{2})\approx -54.7^o$.
Reference to Fig.4 shows that the data strongly prefer
$\theta = \tan^{-1}(1/\sqrt{2})$,  
which gives 
the HQET $K_1$ states 
\begin{eqnarray}
&|K_1(1273)\rangle &= 
+\sqrt{ 2/ 3 } \ | 1^1{\rm P}_1 \rangle 
+ 
\sqrt{ 1/ 3 } \ | 1^3{\rm P}_1 \rangle \\ 
&|K_1(1402)\rangle &= 
-\sqrt{ 1/ 3 } \ | 1^1{\rm P}_1 \rangle 
+ 
\sqrt{ 2/ 3 } \ | 1^3{\rm P}_1 \rangle \ . 
\end{eqnarray}
This choice assigns the lighter $K_1(1273)$ state
to the $j_q=3/2$ multiplet, which may appear surprising since
the $j_q=3/2$ axial  
is expected to be the higher-mass
state in the HQET limit. Of course the HQET limit is difficult
to justify for strange quarks; this limit also anticipates
a higher-mass $1^3$P$_2$ state relative to the $1^3$P$_0$, 
whereas these are approximately degenerate in the 
experimental excited kaon spectrum.
Our antikaon $K_1$ states are taken to be
\begin{eqnarray}
\hskip 0.5cm
&|\bar K_1(1273)\rangle &= 
- \cos(\theta) \, | 1^1{\rm P}_1 \rangle 
+ \sin(\theta) \, | 1^3{\rm P}_1 \rangle \\ 
&|\bar K_1(1402)\rangle &= 
+ \sin(\theta) \, | 1^1{\rm P}_1 \rangle 
+ \cos(\theta) \, | 1^3{\rm P}_1 \rangle 
\end{eqnarray}
with the corresponding HQET states
\begin{eqnarray}
&|\bar K_1(1273)\rangle &= 
-\sqrt{ {2/ 3} } \; | 1^1{\rm P}_1 \rangle 
+ 
\sqrt{ {1/ 3} } \; | 1^3{\rm P}_1 \rangle \\ 
&|\bar K_1(1402)\rangle &= 
+\sqrt{ {1/ 3} } \; | 1^1{\rm P}_1 \rangle 
+ 
\sqrt{ {2/ 3} } \; | 1^3{\rm P}_1 \rangle \ . 
\end{eqnarray}
Note the change of the relative sign of the singlet and triplet
basis states relative to Eqs.(A15,16). 
We use these HQET states in calculating decays to $K_1$ final states
unless otherwise specified. For other excited kaon states
with allowed singlet-triplet mixing we treat the mixing angle
as a free parameter.

The flavor factors that result from contracting these 
explicit flavor states using diagrams ${\rm d}_1$ and ${\rm d}_2$ of Fig.A1
are given in Table~A1 for the strange decays of interest here.

\begin{center}
\begin{tabular}{||l|l|c|c|c||}  \hline
\multicolumn{5}{||c||}{Table A1. 
Flavor weight factors for strange meson decays.} \\ \hline
$\ $ Generic Decay $\ $  & $\ $ Example $\ $ & $\ I_{flavor}(d_1)\ $
& $\ I_{flavor}(d_2)\ $  &
$\ {\cal F}\ $
\\ 
\hline
\hline
$(s\bar s)  \to (n\bar s)(s\bar n)$ 
& $\phi \to K^+ K^- $ & $+1$ & $0$    &  $2$
\\  \hline
$(s\bar s) \to (n\bar s)(s\bar n)'$ 
& $\phi(1680) \to K^+ K^{*-} $ & $+1$ & $0$    &  $4$
\\  \hline
$(s\bar s) \to \eta \eta $ & $f_2'(1525) \to \eta \eta $ & 
$-1/2$ & $-1/2$    &  $1/2$
\\  \hline
$(s\bar s) \to \eta \eta' $ & $f_2'(1525) \to \eta \eta' $ & 
$+1/2$ & $+1/2$    &  $1$
\\  \hline
$(s\bar s) \to \eta' \eta' $ & $f_4(2200) \to \eta' \eta' $ & 
$-1/2$ & $-1/2$    &  $1/2$
\\  \hline
$(s\bar s) \to \eta (s\bar s) $ & $\phi(1680) \to \eta \phi $ & 
$-1/\sqrt{2}$ & $-1/\sqrt{2}$    &  $1$
\\  \hline
$(s\bar s) \to \eta' (s\bar s) $ & $\phi(2050) \to \eta' \phi $ & 
$+1/\sqrt{2}$ & $+1/\sqrt{2}$    &  $1$
\\  \hline
$(s\bar s) \to (s\bar s)(s\bar s) $ & $f_4(2200) \to \phi \phi $ & 
$-1$ & $-1$    &  $1/2$
\\  \hline
$(s\bar s) \to (s\bar s)(s\bar s)' $ & $-$ & 
$-1$ & $-1$    &  $1$
\\  \hline
$(n\bar s) \to (n\bar n)_{I=1} (n\bar s) $ & $K^{*+}\to \pi^o K^+$
& $0$ 
& $+1/\sqrt{2}$  
&  $3$  
\\  \hline
$(n\bar s) \to (n\bar n)_{I=0} (n\bar s) $ 
& $K_3^{*+}\to \omega K^+$
& $0$  
& $+1/\sqrt{2}$ 
&  $1$  
\\  \hline
$(n\bar s) \to \eta (n\bar s) $ & $K_3^{*+}\to \eta K^+$
& $-1/\sqrt{2}$  
& $+1/2$ 
&  $1$  
\\  \hline
$(n\bar s) \to \eta' (n\bar s) $ & $K_3^{*+}\to \eta' K^+$
& $+1/\sqrt{2}$  
& $+1/2$ 
&  $1$  
\\  \hline
$(n\bar s) \to (s\bar s) (n\bar s) $ 
& $K_3^{*+}\to \phi K^+$
& $-1$  
& $0$ 
&  $1$  
\\  \hline
\end{tabular}
\end{center}

The amplitudes quoted in the detailed decay tables are just the 
$\{ {\cal M}_{LS} \}$ amplitudes of Ref.\cite{Ack96},
in units of [GeV$^{-1/2}$].
Amplitude ratios allow sensitive tests of the nature of a resonance,
so it is important to determine these with 
well-defined relative phases.
To quote specific amplitudes in the decay tables we 
have specialized
to particular charge states, which are illustrated by 
the examples in Table~A1.
Note that the BC ordering is important;
if we exchange mesons B and C in Table~A1 or in
the decay tables, we change the 
phases of the decay amplitudes. To obtain a unique set of phases
we define $\hat \Omega$ as the recoil direction of
meson B (with C along
$-\hat \Omega$), and the amplitudes are taken to be 
the coefficients of
the T-matrix expansion in orthonormal angular momentum
eigenfunctions $\{ f_{\rm J L S}(\hat \Omega) \} $. For 
spinless final mesons B and C, these amplitudes are the 
coefficients of a spherical harmonic expansion. 

The total decay rate is given by the T-matrix amplitude squared,
integrated over final angles, summed over all final 
spin and charge states, 
and multiplied by the physical, relativistic 
phase space; again this procedure is described in
detail in Ref.\cite{Ack96}. (We note in passing that the 
$\rho^+\to \pi^+\pi^o$ example in that reference 
has a typographical error in
Eq.(A17); the factor of $M_\rho$ should be  
$E_B E_C / M_A = M_\rho / 4$, as stated
in the subsequent text.) 
Since we neglect mass splittings within an isomultiplet, the sum
over charge states acts as a simple multiplier of the 
partial width into the specific charge channel 
used as our example; this multiplier is quoted as ${\cal F}$ in
Table~A1. This ${\cal F}$ also incorporates the $1/ 2!$ statistical factor
present if B and C are identical. 
The actual light meson masses used here are
$m_\pi=138$~MeV,
$m_K=496$~MeV,
$m_\eta=547$~MeV,
$m_\rho=770$~MeV,
$m_\omega=782$~MeV, 
$m_{K^*}=894$~MeV,
$m_{\eta'}=958$~MeV, 
$m_\phi=1019$~MeV,
$m_{f_2}=1275$~MeV, 
$m_{f_1}=1282$~MeV, 
$m_{f_0}=1370$~MeV, 
$m_{h_1}=1170$~MeV, 
$m_{a_2}=1318$~MeV,
$m_{a_1}=1230$~MeV, 
$m_{a_0}=1450$~MeV 
and
$m_{b_1}=1230$~MeV. 
For the less familiar higher-mass states we used the 
resonance label to display the assumed mass. For example, the 
$K^*(1414)$ entries in the decay tables imply that
we assumed a $K^*$ mass of 1414~MeV in our decay calculations.

\begin{figure}
$$\epsfxsize=4truein\epsffile{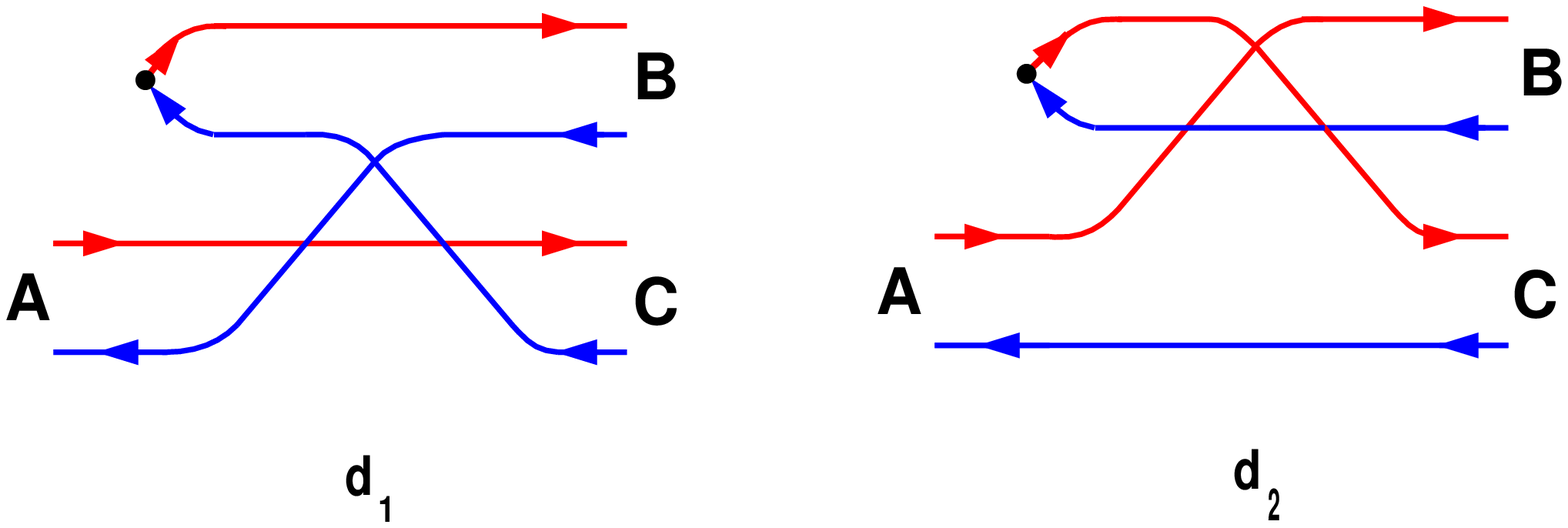}$$
\center{Figure~A1.
The two meson decay diagrams in the \3P0 model.}
\end{figure}

\newpage

\renewcommand{\theequation}{B\arabic{equation}}
\setcounter{equation}{0}  
\section*{Appendix B: Selection rules for decays to
$\eta K^{(*)}$ and $\eta' K^{(*)}$}  

\subsection*{B1. Introduction}

The relative branching fractions of decays of an excited kaon to the
pairs of modes $(\eta K, \eta' K)$ and $(\eta K^*, \eta' K^*)$ are 
very interesting, in that they involve constructive or destructive 
interference between the $|n\bar n \rangle$ and $|s\bar s \rangle$
internal 
components of the $\eta $ or $\eta' $ meson. 
Lipkin \cite{Lip81, Lip97} has previously discussed this effect in the context of
heavy-quark (D and B) nonleptonic weak decays.
In this appendix we consider the application to excited kaon
strong decays, and 
derive the associated selection rules. 
The results are counterintuitive, in that the higher-mass
$\eta'$ decay mode is often favored over the $\eta$ mode.

\subsection*{B2. $\eta K$ and $\eta' K$ final states}

To illustrate these selection rules, we first
consider the decay of a generic excited kaon $K'^{*+}$,
with flavor state $-|u\bar s\rangle$, to
$\eta K^+$. 
(The $K'^{*+}$ must have $S_{u\bar s}=1$ because the 
spin matrix element $S_\A=0\to (S_\B, S_\C) =(0,0) $ 
vanishes in the \3P0 model.)
We first attach the flavor state vectors
\begin{eqnarray}
&&|\A\rangle 
= |K'^{*+}\rangle 
= -|u\bar s\rangle \\
&&|\B\rangle 
= |\eta\rangle 
\hskip 0.5cm
= \cos(\phi)\, {1\over \sqrt{2}}\, 
\Big(|u\bar u\rangle + |d\bar d\rangle \Big) 
-\sin(\phi)\, |s\bar s\rangle \\
&&|\C\rangle 
= |K^+\rangle 
= -|u\bar s\rangle 
\end{eqnarray}
to the two Feynman diagrams in Fig.A1. 
Evidently diagram $d_1$ only couples to the $|s\bar s\rangle$
component of the $\eta$, requires $s\bar s$ pair production,
and gives a flavor factor (flavor matrix element) of 
$-\sin(\phi)$. 
Following Lipkin \cite{Lip81}, we also assume
an $s\bar s$ 
pair-production suppression factor of $\xi$.
Diagram $d_2$ instead requires $u\bar u$ pair production and 
only couples to the $|u\bar u\rangle$ component of the $\eta$,
giving a flavor factor of
$+\cos(\phi)/\sqrt{2}$.
Combining these factors, and carrying out this exercise for the
$\eta'$ as well, we find the generalized flavor factors
given in Table B1.

\begin{center}
\begin{tabular}{||l|c|c||}  \hline
\multicolumn{3}{||c||}{Table B1. 
Generalized $\eta$ and $\eta'$ flavor factors.}\\ 
\hline
$\ $ Channel $\ $  
& 
$\ I_{flavor}(d_1)\ $
& 
$\ I_{flavor}(d_2)\ $  
\\ 
\hline
\hline
$K'^{*+}\to \eta K^+$
& 
$-\sin(\phi)\, \xi $    
& 
$+\cos(\phi)/\sqrt{2}$ 
\\  \hline
$K'^{*-}\to \eta' K^-$
& 
$+\cos(\phi)\, \xi $    
& 
$+\sin(\phi)/\sqrt{2}$ 
\\  \hline
\end{tabular}
\end{center}
Since $\sin(\phi)\approx \cos(\phi)$ in practice, it is clear
from the table that decays to $\eta K$ would experience destructive
interference between diagrams $d_1$ and $d_2$ if the flavor factors were 
the only relevant variables. Conversely, decays to $\eta' K$ experience
constructive flavor interference. If the amplitudes associated with
diagrams $d_1$ and $d_2$ were equal, neglecting phase space differences
the branching fraction ratio would be
\begin{equation}
{B_{\eta K} \over B_{\eta' K}}
=
\Bigg(
{
1-\sqrt{2}\, \xi \tan(\phi)
\over
\tan(\phi) + \sqrt{2}\, \xi 
}
\Bigg)^2
\ .
\end{equation}
For maximally mixed states $(\tan(\phi)=1)$ without $s\bar s$ pair-production 
suppression $(\xi=1)$ this ratio is
\begin{equation}
{B_{\eta K} \over B_{\eta' K}}
\ = \
\Bigg(
{
1-\sqrt{2}
\over
1 + \sqrt{2} 
}
\Bigg)^2 \ \approx \ 0.029 \ ,
\end{equation}
which shows that this interference can have a dramatic effect.

Of course the amplitudes associated with diagrams $d_1$ and $d_2$ are
{\it not} equal in general. They instead have diagram-dependent, 
coupled spin 
factors and spatial overlap integrals, which in the \3P0 model are 
CG-weighted sums of terms of the form
\eject
\begin{equation}
I_{\rm spin+space}(d_1)
=
\langle s\bar s| \vec \sigma | 0 \rangle \Big|_{d_1} \cdot
\int \, d^3k \;
\phi_A(2\vec k   - 2\vec B\, )\;
\phi^*_B(2\vec k - \vec B\, )\;
\phi^*_C(2\vec k - \vec B\, )
\ \vec k 
\end{equation}
and
\begin{equation}
I_{\rm spin+space}(d_2)
=
\langle s\bar s| \vec \sigma | 0 \rangle \Big|_{d_2} \cdot
\int \, d^3k \;
\phi_A(2\vec k   + 2\vec B\, )\;
\phi^*_B(2\vec k + \vec B\, )\;
\phi^*_C(2\vec k + \vec B\, )
\ \vec k  \ .
\end{equation}
Here $\langle s\bar s| \vec \sigma | 0 \rangle$ is the spin-vector
matrix element of the $q\bar q$ pair produced in the spin state
implied by each diagram.
We can take the $K'^{*+}\to\eta K^+$ and 
$K'^{*+}\to\eta' K^+$ overlaps $I_{\rm spin+space}$
to be single terms (B6) and (B7) for each diagram
without loss of generality
because the final mesons factor as
$\phi_{_{LL_z=00}}(\vec p\, )\, 
\chi_{_{SS_z=00}}$, and the initial meson $K'^{*+}$
is the sum of $\phi_{_{LL_z}}(\vec p\, )\, \chi_{_{1S_z}}$ 
factored components that can be treated
individually as the initial $K'^{*+}$ . Each $K'^{*+}$ component 
gives a single 
$\langle s\bar s|\, \vec \sigma \, | 0 \rangle
= \langle \chi_{_{00}} \chi_{_{00}} |\, \vec \sigma \, | \chi_{_{1S_z}} 
\rangle $
matrix element, which is the same for both diagrams because the
final $\eta K$ or $\eta' K$ spin state $|\chi_{_{00}} \chi_{_{00}} \rangle $
is symmetrical.   

The $d_1$ and $d_2$ 
spatial overlap integrals evidently satisfy
\begin{equation}
\vec I_{space}^{\ (d_1)}(\vec B\, ) 
=
\vec I_{space}^{\ (d_2)}(-\vec B\, ) 
\end{equation}
for {\it any} set of meson spatial wavefunctions. Since these 
integrals are
related by parity, and the final states we are considering
have definite parity $(-1)^{\L_{BC}}$, we may remove a common factor
and find for the $K'^{*+}\to\eta K^+$ decay 
amplitude 
\begin{equation}
{\cal A}_{K'^{*+}\to\eta K^+} \propto 
\langle s\bar s| \vec \sigma | 0 \rangle 
\cdot \vec I_{space}^{\ (d_1)}(\vec B\, )
\
\bigg\{ 
- \xi \, \sin(\phi) 
+(-1)^{\L_{BC}}
{1\over \sqrt{2}}\, \cos(\phi)
\bigg\}
\end{equation}
and similarly for $K'^{*+}\to\eta' K^+$
\begin{equation}
{\cal A}_{K'^{*+}\to\eta' K^+} \propto
\langle s\bar s| \vec \sigma | 0 \rangle
\cdot \vec I_{space}^{\ (d_1)}(\vec B\, )
\
\bigg\{ 
+ \xi \, \cos(\phi) 
+(-1)^{\L_{BC}}
{1\over \sqrt{2}}\, \sin(\phi)
\bigg\}\ .
\end{equation}
Neglecting phase space differences, the $\eta K / \eta' K$
branching fraction ratio is again the amplitude ratio 
squared, 
\begin{equation}
{
B_{\eta K}
\over   
B_{\eta' K}
}
=
\Bigg(
{
1 
-(-1)^{\L_{BC}}
\sqrt{2}\, \xi \tan(\phi) 
\over
\tan(\phi)
+(-1)^{\L_{BC}}
\sqrt{2}\, \xi  
}
\Bigg)^2 \ , 
\end{equation}
\vskip 0.5cm
\noindent
which generalizes the S-wave result (B4). This agrees with Lipkin's
Eq.(10b) in Ref.\cite{Lip81}, in which the maximally-mixed case
$\tan(\phi)=1$ was assumed.
Evidently the general rule 
is that 
odd-L final states favor $\eta K$, 
and
even-L favors $\eta' K$. 

Observation of the enhanced $\eta' K$ modes
requires the study of even-J kaonia with
masses well above
$M_{\eta'} +  M_K = 1.45$~GeV. 
Only four of the states we have considered 
in this paper satisfy these requirements,
the 2$^3$P$_2$,  2$^3$P$_0$, 1$^3$F$_4$ and 1$^3$F$_2$. Of these
only the 1$^3$F$_4$ has a widely accepted experimental candidate, the 
$K^*_4(2045)$. Unfortunately the $K^*_4(2045)\to\eta' K$ 
branching fraction is predicted to be quite small, due to the
G-wave centrifical barrier. Identification of these 
as yet unknown 2P 
and 1$^3$F$_2$ states could prove difficult because they are all expected to
be rather broad, with $\Gamma_{tot}\approx 300$-$400$~MeV. 
In part because of these large widths
the theoretical branching fractions of these states to the enhanced 
$\eta' K$ modes are unfortunately not especially
large; all are
3-5$\% $. 

It will be easier to test the selection rule on
$\eta K$ states alone, since this mode has a much lower threshold 
of $\approx 1.04$~GeV. The branching fraction ratio relative to $\pi K$,
which provides a convenient reference, is
\begin{equation}
{
B_{\eta K}
\over   
B_{\pi K}
}
=
{1\over 3} 
\Big(
\cos(\phi)  
-(-1)^{\L_{BC}}
\sqrt{2}\, \xi \sin(\phi) 
\Big)^2 \ , 
\end{equation}
if we again neglect phase space differences. For the 
maximally-mixed case with no $s\bar s$ suppression this becomes
\begin{equation}
{
B_{\eta K}
\over   
B_{\pi K}
}
=\
{ 1 \over 6}\;
\Big(\;  3 -(-1)^{\L_{BC}} 2\sqrt{2} \; \Big)
\ , 
\end{equation}
\vskip 0.5cm
\noindent
which shows that $\eta K$ will be comparable to $\pi K$ in
strength in odd-L$_{\eta K}$ modes
(decays of odd-J kaons),
but a factor of $\approx 35$ smaller 
than $\pi K$ 
in decays of even-J kaons. Of course this simple estimate should
be corrected for phase space, which usually leads to additional
suppression of the $\eta K$
mode relative to $\pi K$.

A very weak $\eta K$ mode has already been 
reported in the decay 
of the even-J $K_2^*(1429)$, as expected (Table 4).
This suppression of $\eta K$ in even-L final states
could also be
tested in $K^*_4(2045)$ decays. In contrast we should see
large $\eta K$ branching fractions in odd-L final states,
arising for example from decays of the odd-J spin-triplet
states $K^*(1414)$ (assuming this actually {\it is} the $2^3$S$_1$ kaon),
the $K_3^*(1776)$ and the $K^*(1717)$. 
The $K_3^*$ is especially attractive for this study
because it is relatively narrow, and there is already evidence
from LASS \cite{Ast88d} that the $K_3^*\to\eta K$ mode 
is enhanced approximately as expected;
the PDG width and $\pi K$ branching fraction, combined with the 
LASS $B_{\eta K} / B_{\pi K}$ ratio \cite{Ast88d} correspond to
$\Gamma_{K^*_3\to \eta K} = 15\pm 6$~MeV, consistent 
with
our theoretical prediction of 19~MeV (see Table 6). 

\subsection*{B3. $\eta K^*$ and $\eta' K^*$ final states}

\subsubsection*{\it B3.1 Derivation}

The selection rules for decays to $\eta$ and $\eta'$ and an
$S_{q\bar q}=1$ ``$K^*$" kaon are rather more complicated, as they depend
on the $S_{q\bar q}$ of the initial kaon as well.
The difference from the previous case is due to the modified spin
matrix elements. The decays $K^*\to \eta K$ and $\eta' K$
involved the spin matrix elements
$\langle s\bar s|\, \vec \sigma \, | 0 \rangle
= \langle \chi_{_{00}} \chi_{_{00}} |\, \vec \sigma \, | \chi_{_{1S_z}} 
\rangle $, 
which were identical for diagrams $d_1$ and $d_2$,
\begin{equation}
I_{spin}^{(d_1)}\bigg|_{K^*\to \eta^{(\, ')} K}
=
I_{spin}^{(d_1)}\bigg|_{K^*\to \eta^{(\, ')} K} \ .
\end{equation}  
For a transition of the type
$K\to \eta^{(\, ')} K^*$ we have a spin matrix element 
$\langle s\bar s|\, \vec \sigma \, | 0 \rangle
= \langle \chi_{_{00}}  \chi_{_{1S_z}}|\, \vec \sigma \, | \chi_{_{00}} 
\rangle $, which is again identical for each diagram,
\begin{equation}
I_{spin}^{(d_1)}\bigg|_{K\to \eta^{(\, ')} K^*}
=
I_{spin}^{(d_2)}\bigg|_{K\to \eta^{(\, ')} K^*} \ .
\end{equation}  
Since the spatial matrix elements are identical, in this case we
have a result analogous to the previous $\eta K$ and $\eta' K$ result
(again neglecting phase space differences for illustration),
\begin{equation}
{
B_{K\to \eta K^*}
\over   
B_{K\to \eta' K^*}
}
=
\Bigg(
{
1 
-(-1)^{\L_{BC}}
\sqrt{2}\, \xi \tan(\phi) 
\over
\tan(\phi)
+(-1)^{\L_{BC}}
\sqrt{2}\, \xi  
}
\Bigg)^2 \ , 
\end{equation}
\vskip 0.5cm
\noindent
so that 
decays of a spin-singlet ``$K$" 
favor $\eta K^*$ in odd-L$_{BC}$ channels
and $\eta' K^*$ in even-L$_{BC}$. 

For transitions from spin-triplet ``$K^*$" states to $\eta K^*$ 
and $\eta' K^*$ the rule is inverted; in this case the spin matrix
elements are
$\langle s\bar s|\, \vec \sigma \, | 0 \rangle
= \langle \chi_{_{00}}  \chi_{_{1S'_z}}|\, \vec \sigma \, | \chi_{_{1S_s}} 
\rangle $, which are opposite in sign between diagrams,
\begin{equation}
I_{spin}^{(d_1)}\bigg|_{K^*\to \eta^{(\, ')} K^*}
=
- I_{spin}^{(d_2)}\bigg|_{K^*\to \eta^{(\, ')} K^*} \ .
\end{equation}  
This change of relative sign between diagrams generalizes the
reduced branching fraction ratio to
\begin{equation}
{
B_{K^*\to \eta K^*}
\over   
B_{K^*\to \eta' K^*}
}
=
\Bigg(
{
1 
+(-1)^{\L_{BC}}
\sqrt{2}\, \xi \tan(\phi) 
\over
\tan(\phi)
-(-1)^{\L_{BC}}
\sqrt{2}\, \xi  
}
\Bigg)^2 \ , 
\end{equation}
\vskip 0.5cm
\noindent
so
decays of a spin-triplet ``$K^*$" instead
favor 
$\eta K^*$ in even-L$_{BC}$
and 
$\eta' K^*$ in odd-L$_{BC}$. 

Our results for these $\eta$ and $\eta'$ modes in all cases 
are summarized in Table B2. Again $K$ and $K^*$ refer
to any spin-singlet and spin-triplet state, respectively.

\begin{center}
\begin{tabular}{||c|c|c||}  \hline
\multicolumn{3}{||c||}
{Table B2. 
Summary of dominant $K^{(*)} \to \eta K^{(*)}, \eta' K^{(*)}$ 
transitions.}\\ 
\hline
Transition type A$\to \eta\!\! $~C, $\eta'\!\! $~C  
& 
even-L$_{BC}$ dominant
& 
odd-L$_{BC}$ dominant
\\ 
\hline
\hline
$K^*\to K^{\phantom{*}}$
& 
$\eta' K^{\phantom{*}}$ 
& 
$\eta\phantom{'}\! K^{\phantom{*}}$ 
\\  \hline
$K^*\to  K^*$
& 
$\eta\phantom{'}\! K^*\, $ 
& 
$\eta'\! K^*$ 
\\  
\hline
$K^{\phantom{*}} \to  K^*$
& 
$\eta' K^*$ 
& 
$\eta\phantom{'}\! K^*$ 
\\  
\hline
\end{tabular}
\end{center}

\subsubsection*{\it B3.2 Application to B decays}

Lipkin's rules for B decays, $B_{\eta'\! K} >> B_{\eta K}$ but
$B_{\eta K^*} >> B_{\eta'\! K^*}$ \cite{Lip97,Lip02} 
follow from the first and third of our
strong selection rules in Table~B2. 
In this decay mechanism
an initial J$^{\rm P}=0^-$ $n\bar b$ B meson is assumed to evolve weakly
into a virtual $n\bar s$ system, which is a superposition
of J$^{\rm P} = 0^-$ and $0^+$ states. This 
J=0 intermediary then decays strongly into the
observed $\eta^{(\, ')} K^{(*)}$ final states. 
The $\eta K$ and $\eta' K$
must be in S-wave since they arise from an initial J=0 meson, 
and therefore have 
J$^{\rm P} = 0^+$, as does their excited $n\bar s$ precursor.
This J$^{\rm P} = 0^+$ $s\bar n$ precursor
must be a spin-triplet.
The first selection rule, for spin-triplet to spin-singlet transitions,
$``K^*" \to ``K"$,
says that for 
even-L$_{BC}$ this transition is dominated by $\eta' K$. 
In contrast the
$\eta K^*$ and $\eta' K^*$ final states must be in P-wave, since they
also must have total
J=0. This J=0 final state can only have J$^{\rm P} = 0^-$. The
$\eta K^*$ and $\eta' K^*$  
$s\bar n$ precursor thus has J$^{\rm P} = 0^-$, making it a
spin-singlet. The third selection rule, for
$``K" \to ``K^*"$ transitions, states that in odd-L$_{BC}$ 
this final state is dominated by $\eta K^*$. 
These two cases in combination
agree with Lipkin \cite{Lip97}. Our second strong decay 
selection rule in Table~B2, for
$``K^*" \to ``K^*"$, is not accessed in B decays to 
$\eta^{(\, ')} K^{(*)}$ modes.

It is possible to test the second ($``K^*" \to ``K^*"$) 
selection rule in other B meson decays. Assuming an initial
J$^{\rm P} = 0^+$ spin-triplet source and a final 
state of an $\eta$ or $\eta'$ combined with a
spin-triplet kaon, it follows that
the final kaon must have 
J = L$_{BC}$ and unnatural spin parity.
The spin-triplet excited kaons accessible in
$``K^*" \to ``K^*"$ B decays
therefore have
J$^{\rm P} = 1^+, 2^-,
3^+ \dots$. Unfortunately these are just those kaon systems that experience 
singlet-triplet mixing, so we cannot observe a pure $``K^*" \to ``K^*"$
transition in isolation. Experimentally we must instead study decays 
to singlet-triplet mixed states, in which
both $``K^*" \to ``K"$ and $``K^*" \to ``K^*"$ transition amplitudes
are present. 

We may illustrate this using transitions to
$\eta^{(\, ')} K_1(1273)$ as an example.
The 
decays $B\to \eta^{(\, ')} K_1$ create a P-wave (odd-L$_{BC}$), 
J$^{\rm P} = 0^+$
final states, and hence are driven by an initial excited
J$^{\rm P} = 0^+$ $^3$P$_0$  $``K^*"$ source.
From Table~B2 the 
$``K^*" \to ``K^*"$ amplitude, which couples the $B$ to the 
$\sin(\theta) |{}^3$P$_1\rangle$ 
$``K^*"$ component of the $K_1(1273)$,
strongly favors the $\eta'$.
However the
$\cos(\theta) |{}^1$P$_1\rangle$ 
$``K"$ component of the $K_1(1273)$ is also excited,
through a
$``K^*" \to ``K"$ transition, which in odd-L$_{BC}$ 
dominantly favors the $\eta$.
Thus the relative branching fraction
$B_{B\to \eta' K_1(1273) / \eta K_1(1273)}$ 
follows from a competition between
these two processes,
and is given by
\begin{equation}
B_{B\to \eta' K_1(1273) / \eta K_1(1273) }
= 
\bigg|
{
\sin(\theta)\; 
{\cal A}({ {}^3{\rm P}_0 \to {}^1{\rm S}_0 + {}^3{\rm P}_1 })
\over
\cos(\theta)\;
{\cal A}({ {}^3{\rm P}_0 \to {}^1{\rm S}_0 + {}^1{\rm P}_1 })
}
\bigg|^2\  
\end{equation}
assuming complete dominance by the
larger amplitudes in Table~B2.
Unfortunately these two amplitudes are different functions
of momenta, so 
the 
$\tan^2(\theta)$ state ratio is 
multiplied by a squared ratio of unknown transition amplitudes. 
A simplification is possible, since 
the corresponding result for the $K_1(1402)$ has
$\sin(\theta)$ and $\cos(\theta)$ exchanged; if we take the
ratio of relative branching fractions, we find a result
that only involves the mixing angle $\theta$ (neglecting
phase space differences between modes),
\begin{equation}
{
B_{B\to \eta' K_1(1273) /  \eta K_1(1273)}
\over   
B_{B\to \eta' K_1(1402) /  \eta K_1(1402)}
}
= 
\tan^4(\theta) \ .
\end{equation}
For higher-mass states such as the 
$K_2$ kaons this ``mixed" selection rule may thus allow an estimate
of the corresponding 
singlet-triplet mixing angle, using the analogous 
$B\to \eta K_2$ and $\eta' K_2$ decays.

\subsubsection*{\it B3.3 Application to strong excited kaon decays}

To test these $\eta(\, ') K^*$ strong selection rules directly in 
excited kaon decays
we would ideally prefer a parent resonance with sufficient mass to
populate both $\eta K^*$ and $\eta' K^*$. This requires a mass above
$M_{\eta'} +  M_{K^*} = 1.90$~GeV; of the excited kaons we have considered
only the 1F states satisfy this constraint. Since these 1F states
all have P = (+) they will populate even-L$_{BC}$ $\eta(\, ') K^*$
final states. The $1^3$F$_4$ and $1^3$F$_2$ initial states are 
pure spin-triplet, which from Table~B2 decay preferentially to
$\eta K^*$ in this even-L$_{BC}$ case. 
This is evident at the amplitude level in Table~K14, with
$\eta K^*$ preferred over $\eta' K^*$ by an order of magnitude
in the $1^3$F$_2$ $K_2^*(2050)$ case. Unfortunately there are
important 
centrifical barriers, especially in the decays of the
$1^3$F$_4$ $K_2^*(2045)$, which restrict the $\eta K^*$ branching fractions
of these 1F states to a few percent. A measurement of the ratio
$B_{\eta K^*} / B_{\pi K^*}$ might be feasible for the $1^3$F$_2$ $K_2^*(2050)$
state, once this resonance is identified.

Just as we found previously in 
considering $\eta K$ and $\eta' K$ decays, it will be easier to study
the strength of the lower mass mode $\eta K^*$,
since the lower threshold 
of $\approx 1.44$~GeV makes more resonance couplings 
accessible. The ratio $B_{\eta K^*} / B_{\pi K^*}$
can provide a normalized measure of the strength of the $\eta K^*$
mode. Of the states we have considered, the single spin-triplet
excited kaon below $\eta' K^*$ threshold that is expected to have a large
$\eta K^*$ branching fraction 
is the as yet unidentified $2^3$P$_2$ $K^*_2(1850)$;
this has a theoretical branching fraction of
$B_{\eta K^*} = 7\%$, and a branching fraction ratio of
$B_{\eta K^*} / B_{\pi K^*} = 0.55$. The remaining interesting 
excited kaon states
below $\eta' K^*$ threshold are the known 
$1^3$D$_3$ $K_3^*(1776)$ 
and
$1^3$D$_1$ $K^*(1717)$; these 
have suppressed $\eta K^*$ modes since they are
odd-L$_{BC}$,
and their $\eta K^*$ branching fractions are predicted to be 
smaller 
than the corresponding $\pi K^*$ branching fractions
by factors of about 160 (for $K_3^*(1776)$) 
and 50 (for $K^*(1717)$) respectively. 

Decay amplitudes of the mixed singlet-triplet mesons to these
modes are quite sensitive to the singlet-triplet mixing angles,
and may be useful in determining these parameters 
more accurately; even in the relatively 
well-studied 1P
case ($K_1(1273)$ and $K_1(1402)$) the mixing angle is not well determined 
by the existing decay measurements (see Fig.4). Since the mixed-spin kaons
all have unnatural ${\rm J}^{\rm P}$, the lightest final state of relevance
here is $\eta K^*$,
which can be used to estimate the singlet-triplet mixing for example
in the expected $2^1$P$_1$-$2^3$P$_1$ $K_1(1800)$ 
and $1^1$F$_3$-$1^3$F$_3$ $K_3(2050)$
pairs. The spin-triplet component in each 
of these resonances strongly
favors $\eta K^*$, as shown in Tables~K10 and K15.

\newpage

\newcommand{\m}{\multicolumn}
\newcommand{\ba}{\begin{array}}
\newcommand{\ea}{\end{array}}

\vskip 2cm

\begin{tabular}{||c||c|l||}
\hline
\multicolumn{3}{||c||}{Table S1. 1S $s\bar s$ } \\
\hline 
& 
\m{2}{c||}{$\phi(1019)$}  \\
\hline 
Mode
& $\Gamma_i($MeV$)$  & \m{1}{c||}{Amps.}  \\
\hline \hline
$K K$ 
& 
2.5
& 
\small $\ba{l} 
^1P_1=-0.081
\ea $ 
\\
\hline
\hline 
& 
\m{2}{c||}{$\Gamma_{thy} = 2.5$ MeV}  \\
\hline 
& 
\m{2}{c||}{$\Gamma_{expt} = 4.26\pm 0.05$ MeV}  \\
\hline 
\end{tabular}

\vskip 1cm

\begin{tabular}{||c||c|l||c|l||}
\hline
\multicolumn{5}{||c||}{Table S2. 2S $s\bar s$ } \\
\hline 
& 
\m{2}{c||}{$\phi(1680)$} 
& 
\m{2}{c||}{$\eta_s(1415\to 1500)$}  \\
\hline 
Mode
& $\Gamma_i($MeV$)$  & \m{1}{c||}{Amps.}  
& $\Gamma_i($MeV$)$  & \m{1}{c||}{Amps.}  \\
\hline \hline
$K K$ 
& 
89 
& 
\small $\ba{l} 
^1P_1=-0.16
\ea $ 
& 
-
& 
-
\\
\hline
$K K^*$ 
& 
245
& 
\small $\ba{l} 
^3P_1=-0.23
\ea $ 
& 
$11 \to 100$
& 
\small $\ba{l} 
^3P_0=+0.10\to +0.20
\ea $ 
\\
\hline
\hline
$\eta \phi $ 
& 
44
& 
\small $\ba{l} 
^3P_1=+0.25 
\ea $ 
&  
-
& 
-
\\
\hline
\hline 
& 
\m{2}{c||}{$\Gamma_{thy} = 378$ MeV}
& 
\m{2}{c||}{$\Gamma_{thy} = 11\to 100$ MeV}  \\
\hline 
& 
\m{2}{c||}{$\Gamma_{expt} = 150\pm 50$ MeV}
& 
\m{2}{c||}{$\Gamma_{expt}(\eta(1440)) \approx 50$-$80$ MeV}  \\
\hline 
\end{tabular}

\vskip 1cm

\begin{tabular}{||c||c|l||c|l||}
\hline
\multicolumn{5}{||c||}{Table S3. 3S $s\bar s$ } \\
\hline 
& 
\m{2}{c||}{$\phi(2050)$} 
& 
\m{2}{c||}{$\eta_s(1950)$}  \\
\hline 
Mode
& $\Gamma_i($MeV$)$  & \m{1}{c||}{Amps.}  
& $\Gamma_i($MeV$)$  & \m{1}{c||}{Amps.}  \\
\hline \hline

$K K$ 
& 
0 
& 
\small $\ba{l} 
^1P_1=+0.0029 
\ea $ 
& 
-
& 
-
\\
\hline

$K K^* $ 
& 
20 
& 
\small $\ba{l} 
^3P_1=-0.047 
\ea $ 
& 
53 
& 
\small $\ba{l} 
^3P_0=+0.081 
\ea $ 
\\
\hline

$K^* K^*$ 
& 
102
& 
\small $\ba{l} 
^1P_1=-0.039  \\
^5P_1=+0.17  
\ea $ 
& 
67
& 
\small $\ba{l} 
^3P_0=-0.17
\ea $ 
\\
\hline

$K K_1(1273) $ 
& 
58 
& 
\small $\ba{l} 
^3S_1\equiv 0 \\
^3D_1= -0.10  
\ea $ 
& 
-
& 
-
\\
\hline

$K K_1(1402) $ 
& 
26 
& 
\small $\ba{l} 
^3S_1= +0.083  \\
^3D_1\equiv 0
\ea $ 
& 
-
& 
-
\\
\hline

$K K^*_0(1412) $ 
& 
-
& 
-
& 
30
& 
\small $\ba{l} 
^1S_0=-0.13 
\ea $ 
\\
\hline

$KK^*_2(1429) $ 
& 
9
& 
\small $\ba{l} 
^5D_1=+0.053
\ea $ 
& 
0
& 
\small $\ba{l} 
^5D_0=-0.016 
\ea $ 
\\
\hline

$K K^*(1414) $ 
& 
93
& 
\small $\ba{l} 
^3P_1=-0.16
\ea $ 
& 
25
& 
\small $\ba{l} 
^3P_0=+0.12
\ea $ 
\\
\hline

$K K(1460) $ 
& 
29
& 
\small $\ba{l} 
^1P_1=-0.10
\ea $ 
& 
-
& 
-
\\
\hline
\hline

$\eta \phi $ 
& 
21
& 
\small $\ba{l} 
^3P_1=+0.10  
\ea $ 
&  
-
& 
-
\\
\hline

$\eta' \phi $ 
& 
11
& 
\small $\ba{l} 
^3P_1=-0.11 
\ea $ 
&  
-
& 
-
\\
\hline

$\eta h_1(1386) $ 
& 
8
& 
\small $\ba{l} 
^3S_1=-0.078  \\ 
^3D_1=-0.060  
\ea $ 
&  
-
& 
-
\\
\hline
\hline 
& 
\m{2}{c||}{$\Gamma_{thy} = 378 $ MeV}
& 
\m{2}{c||}{$\Gamma_{thy} = 175 $ MeV}  \\
\hline 
\end{tabular}

\begin{tabular}{||c||c|l||c|l||}
\hline
\multicolumn{5}{||c||}{Table S4. 1$^3$P$_2$ and 1$^3$P$_0$ $s\bar s$ } \\
\hline 
     & \m{2}{c||}{$f'_2(1525)$} &  
       \m{2}{c||}{$f_0(1500)$}  \\
\hline 
Mode
& $\Gamma_i($MeV$)$  & \m{1}{c||}{Amps.}  
& $\Gamma_i($MeV$)$  & \m{1}{c||}{Amps.}  \\
\hline \hline
$K K$ 
& 
61 
& 
\small $\ba{l} 
^1D_2=+0.15
\ea $ 
& 
214 
& 
\small $\ba{l} 
^1S_0=+0.28
\ea $ 
\\
\hline
$KK^* $ 
& 
9 
& 
\small $\ba{l} 
^3D_2=+0.056
\ea $ 
& 
- 
& 
-
\\
\hline
\hline
$\eta \eta $ 
& 
10 
& 
\small $\ba{l} 
^1D_2=-0.13
\ea $ 
&  
66
& 
\small $\ba{l} 
^1S_0=-0.33 
\ea $ 
\\
\hline
$\eta \eta' $ 
& 
0 
& 
\small $\ba{l} 
^1D_2=+0.0073 
\ea $ 
& 
- 
& 
-
\\
\hline
\hline 
& 
\m{2}{c||}{$\Gamma_{thy} = 80$ MeV} 
& 
\m{2}{c||}{$\Gamma_{thy} = 279 $ MeV}  \\
\hline 
& 
\m{2}{c||}{$\Gamma_{expt} = 76\pm 10$ MeV} 
& 
\m{2}{c||}{see text}  \\
\hline 
\end{tabular}

\small $\ba{l} 
\ea $ 

\vskip 1cm

\begin{tabular}{||c||c|l||}
\hline
\multicolumn{3}{||c||}{Table S5. 1$^3$P$_1$ $s\bar s$ } \\
\hline 
& 
\m{2}{c||}{$f_1(1420\to 1530)$}  \\
\hline 
Mode
& $\Gamma_i($MeV$)$  & \m{1}{c||}{Amps.}  \\
\hline \hline
$KK^* $ 
& 
$ 254\to 459 $
& 
\small $\ba{l} 
^3S_1=-0.47\to -0.40 \\ 
^3D_1=+0.0092\to +0.043
\ea $ 
\\
\hline
\hline 
& 
\m{2}{c||}{$\Gamma_{thy} = 254\to 459 $ MeV}  \\
\hline 
& 
\m{2}{c||}{see text}  \\
\hline 
\end{tabular}

\vskip 1cm

\begin{tabular}{||c||c|l||}
\hline
\multicolumn{3}{||c||}{Table S6. 1$^1$P$_1$ $s\bar s$ } \\
\hline 
& 
\m{2}{c||}{$h_1(1390\to 1440)$}  \\
\hline 
Mode
& $\Gamma_i($MeV$)$  & \m{1}{c||}{Amps.}  \\
\hline 
\hline
$KK^* $ 
& 
$0\to 160 $
& 
\small $\ba{l} 
^3S_1=-0.34\to -0.32  \\
^3D_1=0 \to -0.022
\ea $ 
\\
\hline
\hline 
& 
\m{2}{c||}{$\Gamma_{thy} = 0\to 160$ MeV}  \\
\hline 
& 
\m{2}{c||}{$\Gamma_{expt} = 91\pm 30$ MeV}  \\
\hline 
\end{tabular}

\vskip 1cm

\begin{tabular}{||c||c|c||}
\hline
\multicolumn{3}{||c||}{Table S8. 2$^1$P$_1$ $s\bar s$ } \\
\hline 
& 
       \m{2}{c||}{$h_1(1850)$}  \\
\hline 
Mode
& $\Gamma_i($MeV$)$  & \m{1}{c||}{Amps.}  \\
\hline \hline
$KK^* $ 
& 
95
&   
\small $\ba{l} 
^3S_1=-0.024 \\ 
^3D_1=-0.12
\ea $ 
\\
\hline
$K^* K^*$ 
& 
64 
& 
\small $\ba{l} 
^3S_1=+0.21 \\
^3D_1=+0.037 
\ea $ 
\\
\hline
$K K_1(1273) $ 
& 
1  
&  
\small $\ba{l} 
^3P_1=+0.024
\ea$ 
\\
\hline
\hline
$\eta \phi $ 
& 
33 
& 
\small $\ba{l} 
^3S_1=+0.098  \\
^3D_1=+0.12
\ea$ 
\\
\hline
\hline
& 
\m{2}{c||}{$\Gamma_{thy} = 193 $ MeV}  \\
\hline 
\end{tabular}

\vskip 1cm

\begin{tabular}{||c||c|l||c|l||c|l||}
\hline
\multicolumn{7}{||c||}{Table S9. 1$^3$D$_\J$ $s\bar s$ } \\
\hline 
& 
\m{2}{c||}{$\phi_3(1854)$} 
& 
\m{2}{c||}{$\phi_2(1850)$} 
& 
\m{2}{c||}{$\phi(1850)$}  \\
\hline 
Mode
& $\Gamma_i($MeV$)$  & \m{1}{c||}{Amps.}  
& $\Gamma_i($MeV$)$  & \m{1}{c||}{Amps.}  
& $\Gamma_i($MeV$)$  & \m{1}{c||}{Amps.}  \\
\hline \hline
$K K$ 
& 
45 
& 
\small $\ba{l} 
^1F_3=-0.10
\ea $ 
& - 
& - 
& 
65 
& 
\small $\ba{l} 
^1P_1=-0.12
\ea $ 
\\
\hline
$KK^* $ 
& 
24
& 
\small $\ba{l} 
^3F_3=-0.059
\ea $ 
& 
151 
&   
\small $\ba{l} 
^3P_2=+0.14 \\
^3F_2=-0.049
\ea $ 
& 
75
& 
\small $\ba{l} 
^3P_1=+0.11
\ea $ 
\\
\hline
$K^* K^*$
&
32
& 
\small $\ba{l}
^5P_3=+0.15 \\
^1F_3=-0.0024 \\
^5F_3=+0.0052 \\
^5H_3\equiv 0
\ea $
&
7
&
\small $\ba{l}
^5P_2=+0.073 \\
^5F_2=+0.0072
\ea $
&
5
& 
\small $\ba{l}
^1P_1=-0.054 \\
^5P_1=+0.024 \\
^5F_1=+0.0088
\ea $
\\
\hline
$K K_1(1273) $
&
0
&
\small $\ba{l}
^3D_3=-0.012 \\
^3G_3=-0.00061
\ea$
&
2 
& 
\small $\ba{l}
^3D_2=+0.028
\ea$
&
478
& 
\small $\ba{l}
^3S_1=-0.45 \\
^3D_1=-0.019
\ea$
\\
\hline
\hline
$\eta \phi $
&
3
&
\small $\ba{l}
^3F_3=+0.046
\ea $
&
53
&
\small $\ba{l}
^3P_2=-0.20 \\
^3F_2=+0.038
\ea $
& 
29
&
\small $\ba{l}
^3P_1=-0.15
\ea $
\\
\hline
\hline
& 
\m{2}{c||}{$\Gamma_{thy} = 104$ MeV} 
& 
\m{2}{c||}{$\Gamma_{thy} = 214$ MeV}
& 
\m{2}{c||}{$\Gamma_{thy} = 652$ MeV}  \\
\hline 
& 
\m{2}{c||}{$\Gamma_{expt} = 87 {+28 \atop -23} $ MeV} 
& 
\m{2}{c||}{}
& 
\m{2}{c||}{}  \\
\hline 
\end{tabular}

\vskip 1cm

\begin{tabular}{||c||c|l||}
\hline
\multicolumn{3}{||c||}{Table S10. 1$^1$D$_2$ $s\bar s$ } \\
\hline 
& 
\m{2}{c||}{$\eta_2(1850)$}  \\
\hline 
Mode
& $\Gamma_i($MeV$)$  & \m{1}{c||}{Amps.}  \\
\hline \hline
$KK^* $ 
& 
114
& 
\small $\ba{l} 
^3P_2=+0.12 \\
^3F_2=+0.060
\ea $ 
\\
\hline
$K^* K^*$ 
& 
15 
&  
\small $\ba{l} 
^3P_2=-0.10 \\
^3F_2=-0.0062 
\ea $ 
\\
\hline
$K K_1(1273) $ 
& 
0
&  
\small $\ba{l} 
^3D_2=-0.011 
\ea$ 
\\
\hline
\hline 
& 
\m{2}{c||}{$\Gamma_{thy} = 129$ MeV}  \\
\hline 
\end{tabular}
\vskip 1cm

\begin{tabular}{||c||c|l||}
\hline
\multicolumn{3}{||c||}{Table S11. 1$^1$D$_2$  ({\it alt.})} \\
\hline 
& 
\m{2}{c||}{$\eta_2(1842)\ =\ s\, |n\bar n\rangle\ + c\, |s\bar s\rangle $}  \\
\hline 
Mode
& $\Gamma_i($MeV$)$  & \m{1}{c||}{Amps.}  \\
\hline \hline
$KK^* $ 
& 
$
111.7 c^2 
+ 
158.0 sc 
+ 
55.9 s^2 
$
& 
\small $\ba{l} 
^3P_2= -0.116 c - 0.0820 s \\
^3F_2=  -0.0580 c - 0.0410 s
\ea $ 
\\
\hline
$K^* K^*$ 
& 
$
12.1 c^2 
- 
17.1 sc 
+ 
6.1 s^2 
$
&  
\small $\ba{l} 
^3P_2=-0.10 \\
^3F_2= +0.00506 c - 0.00358 s 
\ea $ 
\\
\hline
$K K_1(1273) $ 
& 
$
0.2 c^2 
- 
0.3 sc 
+ 
0.1 s^2 
$
&  
\small $\ba{l} 
^3D_2 = +0.00977 c - 0.00691 s 
\ea$ 
\\
\hline

$\rho \rho $ 
& 
$
129.9 s^2 
$
&  
\small $\ba{l} 
^3P_2=+0.231 s \\
^3F_2= - 0.0765 s 
\ea $ 
\\
\hline

$\omega \omega $ 
& 
$
40.3 s^2 
$
&  
\small $\ba{l} 
^3P_2=-0.229 s \\
^3F_2= - 0.0689 s 
\ea $ 
\\
\hline

$\pi a_1 $ 
& 
$
33.8 s^2 
$
&  
\small $\ba{l} 
^3D_2= + 0.0986  s 
\ea $ 
\\
\hline

$\eta f_1 $ 
& 
$
0.0  
$
&  
\small $\ba{l} 
^3D_2= - 0.00321 s 
\ea $ 
\\
\hline

$\pi a_0(1450) $ 
& 
$
0.6  s^2 
$
&  
\small $\ba{l} 
^1D_2= + 0.0182 s 
\ea $ 
\\
\hline

$\pi a_2 $ 
& 
$
260.7 s^2 
$
&  
\small $\ba{l} 
^5S_2=+0.285 s \\
^5D_2= + 0.109 s \\
^5G_2= + 0.00681 s 
\ea $ 
\\
\hline

$\eta f_2 $ 
& 
$
30.0 s^2 
$
&  
\small $\ba{l} 
^5S_2= - 0.314 s \\
^5D_2= - 0.00721 s \\
^5G_2= - 0.0000358 s 
\ea $ 
\\
\hline
\hline 
& 
\m{2}{c||}{$\Gamma_{thy} = 141 c^2 + 124 sc + 557 s^2$ MeV} \\ 
\hline 
& 
\m{2}{c||}{$\Gamma_{expt} = 225\pm 14 $ MeV}  \\
\hline 
\end{tabular}

\vskip 1cm

\begin{tabular}{||c||c|l||}
\hline
\multicolumn{3}{||c||}{Table S12. 1$^1$D$_2$  partner ({\it alt.})} \\
\hline 
& 
\m{2}{c||}{$\eta_2(1617)\ =\ c\, |n\bar n\rangle\ - s\, |s\bar s\rangle $}  \\
\hline 
Mode
& $\Gamma_i($MeV$)$  & \m{1}{c||}{Amps.}  \\
\hline \hline
$KK^* $ 
& 
$
22.6 c^2 
-
63.8 sc 
+ 
45.1 s^2 
$
& 
\small $\ba{l} 
^3P_2= -0.0749 c + 0.106 s \\
^3F_2=  -0.0141 c + 0.0200 s
\ea $ 
\\
\hline
$\rho \rho $ 
& 
$
21.4  c^2 
$
&  
\small $\ba{l} 
^3P_2 = +0.150 c \\
^3F_2= + 0.00983 c 
\ea $ 
\\
\hline
$\omega \omega $ 
& 
$
6.2 c^2 
$
&  
\small $\ba{l} 
^3P_2= - 0.129 c \\
^3F_2= - 0.00573 c 
\ea $ 
\\
\hline

$\pi a_1 $ 
& 
$
3.1 c^2 
$
&  
\small $\ba{l} 
^3D_2= + 0.0436  c 
\ea $ 
\\
\hline

$\pi a_0(1450) $ 
& 
$
0.0  
$
&  
\small $\ba{l} 
^1D_2= + 0.00158  c 
\ea $ 
\\
\hline

$\pi a_2 $ 
& 
$
169.4 c^2 
$
&  
\small $\ba{l} 
^5S_2= + 0.402 c \\
^5D_2= + 0.0369 c \\
^5G_2= + 0.000692 c 
\ea $ 
\\
\hline
\hline 
& 
\m{2}{c||}{$\Gamma_{thy} =  221 c^2 - 64 sc + 45 s^2$ MeV} \\ 
\hline 
& 
\m{2}{c||}{$\Gamma_{expt} = 181\pm 11 $ MeV}  \\
\hline 
\end{tabular}

\vskip 1cm

\begin{tabular}{||c||c|l||}
\hline
\multicolumn{3}{||c||}{Table S14. 1$^1$F$_3$ $s\bar s$ } \\
\hline 
& 
       \m{2}{c||}{$h_3(2200)$}  \\
\hline 
Mode
& $\Gamma_i($MeV$)$  & \m{1}{c||}{Amps.}  \\
\hline \hline

$KK^* $ 
& 
71
&   
\small $\ba{l} 
^3D_3=-0.061 \\ 
^3G_3=-0.050 
\ea $ 
\\
\hline

$K^* K^*$ 
& 29 
& 
\small $\ba{l} 
^3D_3=+0.076 \\
^3G_3=+0.028 
\ea $ 
\\
\hline

$K K_1(1273) $ 
& 
4  
&  
\small $\ba{l} 
^3F_3=+0.022
\ea$ 
\\
\hline

$KK_1(1402) $ 
& 
2  
&  
\small $\ba{l} 
^3F_3=+0.019 
\ea$ 
\\
\hline

$KK^*_0(1412) $ 
& 
1 
& 
\small $\ba{l} 
^1F_3=+0.014
\ea$ 
\\
\hline

$KK^*_2(1429) $ 
& 
94
& 
\small $\ba{l} 
^5P_3=+0.13 \\ 
^5F_3=+0.025 \\ 
^5H_3=+0.0015
\ea$ 
\\
\hline

$KK^*(1414) $ 
& 
1 
& 
\small $\ba{l} 
^3D_3=-0.010\\ 
^3G_3=-0.0083 
\ea $ 
\\
\hline

$K^* K_1(1273) $ 
& 
17   
&  
\small $\ba{l} 
^5P_3=-0.083 \\
^1F_3\equiv 0 \\
^3F_3=+0.00053 \\
^5F_3=-0.0020 \\
^5H_3=-0.000018
\ea$ 
\\
\hline
\hline 

$\eta \phi $ 
& 
25 
& 
\small $\ba{l} 
^3D_3=+0.085  \\
^3G_3=+0.050
\ea$ 
\\
\hline

$\eta'\phi  $ 
& 
5 
& 
\small $\ba{l} 
^3D_3=-0.054 \\
^3G_3=-0.010
\ea$ 
\\
\hline

$\eta h_1(1386) $ 
& 
0 
& 
\small $\ba{l} 
^3F_3\equiv 0
\ea$ 
\\
\hline
\hline
& 
\m{2}{c||}{$\Gamma_{thy} = 249 $ MeV}  \\
\hline 
\end{tabular}

\newpage


\begin{tabular}{||c||c|l||}
\hline
\multicolumn{3}{||c||}{Table K1. 1S $kaons$ } \\
\hline 
& 
\m{2}{c||}{$K^*(894)$}  \\
\hline 
Mode
& $\Gamma_i($MeV$)$  & \m{1}{c||}{Amps.}  \\
\hline \hline
$\pi K$ 
& 
21
& 
\small $\ba{l} 
^1P_1=+0.14
\ea $ 
\\
\hline
\hline 
& 
\m{2}{c||}{$\Gamma_{thy} = 21$ MeV}  \\
\hline 
& 
\m{2}{c||}{$\Gamma_{expt} = 51(1)$ MeV}  \\
\hline 
\end{tabular}

\vskip 1cm

\begin{tabular}{||c||c|l||c|l||}
\hline
\multicolumn{5}{||c||}{Table K2. 2S $kaons$ } \\
\hline 
& 
\m{2}{c||}{$K^*(1414)$} 
& 
\m{2}{c||}{$K(1460)$}  \\
\hline 
Mode
& $\Gamma_i($MeV$)$  & \m{1}{c||}{Amps.}  
& $\Gamma_i($MeV$)$  & \m{1}{c||}{Amps.}  \\
\hline \hline
$\pi K $ 
& 
55 
& 
\small $\ba{l} 
^1P_1=+0.12
\ea $ 
& 
-
& 
-
\\
\hline
$\eta K$ 
& 
42
& 
\small $\ba{l} 
^1P_1=+0.20
\ea $ 
& 
-
& 
-
\\
\hline
$\rho K$ 
& 
34
& 
\small $\ba{l} 
^3P_1=+0.13 
\ea $ 
&  
73
& 
\small $\ba{l} 
^3P_0=-0.18 
\ea $ 
\\
\hline 
$\omega K $ 
& 
10
& 
\small $\ba{l} 
^3P_1=+0.13 
\ea $ 
&  
23
& 
\small $\ba{l} 
^3P_0=-0.17 
\ea $ 
\\
\hline
$\pi K^* $ 
& 
55
& 
\small $\ba{l} 
^3P_1=-0.15 
\ea $ 
&  
101
& 
\small $\ba{l} 
^3P_0=-0.20 
\ea $ 
\\
\hline
$\eta K^* $ 
& 
-
& 
-
&  
3
& 
\small $\ba{l} 
^3P_0=-0.11  
\ea $ 
\\
\hline
$\pi K_1(1273) $ 
& 
0.0
& 
\small $\ba{l} 
^3S_1\equiv 0  \\ 
^3D_1=+0.00039 
\ea $ 
&  
-
& 
-
\\
\hline
\hline 
& 
\m{2}{c||}{$\Gamma_{thy} = 196$ MeV}
& 
\m{2}{c||}{$\Gamma_{thy} = 200$ MeV}  \\
\hline 
& 
\m{2}{c||}{$\Gamma_{expt} = 232\pm 21$ MeV}
& 
\m{2}{c||}{$\Gamma_{expt} = 250-260$ MeV}  \\
\hline 
\end{tabular}

\vskip 1cm

\begin{tabular}{||c||c|l||}
\hline
\multicolumn{3}{||c||}{Table K3. 2S $kaons$ $(alt.)$} \\
\hline 
& 
\m{2}{c||}{$K^*(1580)$} \\
\hline 
Mode
& $\Gamma_i($MeV$)$  & \m{1}{c||}{Amps.}  \\
\hline \hline

$\pi K $ 
& 
61  
& 
\small $\ba{l} 
^1P_1=+0.11  
\ea $ 
\\
\hline

$\eta K$ 
& 
60  
& 
\small $\ba{l} 
^1P_1= + 0.20 
\ea $ 
\\
\hline

$\eta' K$ 
& 
0.5 
& 
\small $\ba{l} 
^1P_1= - 0.026 
\ea $ 
\\
\hline

$\rho K$ 
& 
90
& 
\small $\ba{l} 
^3P_1= +0.16 
\ea $ 
\\
\hline 

$\omega K $ 
& 
29
& 
\small $\ba{l} 
^3P_1=+0.16 
\ea $ 
\\
\hline

$\phi K $ 
& 
8.6
& 
\small $\ba{l} 
^3P_1=-0.13 
\ea $ 
\\
\hline

$\pi K^* $ 
& 
99
& 
\small $\ba{l} 
^3P_1=-0.17 
\ea $ 
\\
\hline

$\eta K^* $ 
& 
1.1
& 
\small $\ba{l} 
^3P_1=+0.039
\ea $ 
\\
\hline

$\pi K_1(1273) $ 
& 
1.0
& 
\small $\ba{l} 
^3S_1\equiv 0 \\ 
^3D_1=+0.030 
\ea $ 
\\
\hline

$\pi K_1(1402) $ 
& 
0.9
& 
\small $\ba{l} 
^3S_1=-0.054 \\ 
^3D_1\equiv 0  
\ea $ 
\\
\hline

$\pi K_2^*(1429) $ 
& 
0.0
& 
\small $\ba{l} 
^5D_1= - 0.0018  
\ea $ 
\\
\hline

\hline 
\hline 
& 
\m{2}{c||}{$\Gamma_{thy} = 352$ MeV} \\
\hline 
\end{tabular}

\vskip 1cm

\newpage

\begin{tabular}{||c||c|l||c|l||}
\hline
\multicolumn{5}{||c||}{Table K4. 3S $kaons$ } \\
\hline 
& 
\m{2}{c||}{$K^*(1950)$} 
& 
\m{2}{c||}{$K(1830)$}  \\
\hline 
Mode
& $\Gamma_i($MeV$)$  & \m{1}{c||}{Amps.}  
& $\Gamma_i($MeV$)$  & \m{1}{c||}{Amps.}  \\
\hline \hline

$\pi K$ 
& 
0
& 
\small $\ba{l} 
  ^1P_1=-0.0036
\ea $ 
&
-
&
-
\\
\hline

$\eta K$ 
& 
1
&
\small $\ba{l} 
  ^1P_1=+0.017
\ea $ 
&
-
& 
-
\\
\hline

$\eta' K$ 
& 
0
&
\small $\ba{l} 
  ^1P_1=-0.011
\ea $ 
&
-
& 
-
\\
\hline

$\rho K$ 
&
8
&
\small $\ba{l} 
  ^3P_1=+0.034
\ea $ 
& 
21
& 
\small $\ba{l} 
  ^3P_0=-0.061
\ea $ 
\\
\hline

$\omega K$ 
&
3
&
\small $\ba{l} 
  ^3P_1=+0.035
\ea $ 
& 
7
& 
\small $\ba{l} 
  ^3P_0=+0.062
\ea $ 
\\
\hline

\end{tabular}
\vskip 1cm

\begin{tabular}{||c||c|l||c|l||}
\hline
\multicolumn{5}{||c||}{Table K4. 3S $kaons$ $(cont.)$} \\
\hline 
& 
\m{2}{c||}{$K^*(1950)$} 
& 
\m{2}{c||}{$K(1830)$}  \\
\hline 
Mode
& $\Gamma_i($MeV$)$  & \m{1}{c||}{Amps.}  
& $\Gamma_i($MeV$)$  & \m{1}{c||}{Amps.}  \\
\hline \hline

$\pi K^*(1414) $ 
&
46
&
\small $\ba{l} 
  ^3P_1=-0.12
\ea $ 
& 
40
& 
\small $\ba{l} 
  ^3P_0=-0.14
\ea $ 
\\
\hline

$\pi K(1460) $ 
&
20
&
\small $\ba{l} 
  ^1P_1=+0.087
\ea $ 
& 
-
& 
-
\\
\hline

$\pi K^*(1717) $ 
&
0
&
\small $\ba{l} 
  ^3P_1=+0.0012
\ea $ 
& 
-
& 
-
\\
\hline
$\pi K_2(1773) $ 
&
0
&
\small $\ba{l} 
  ^5P_1 = +0.0013 c_D - 0.0016 s_D  \\
  ^5F_1 = +0.00035 c_D + 0.00029 s_D 
\ea $ 
& 
-
& 
-
\\
\hline

$\pi K^*_3(1776) $ 
&
0
&
\small $\ba{l} 
  ^7F_1=-0.00036
\ea $ 
& 
-
& 
-
\\
\hline
\hline
& 
\m{2}{c||}{$\Gamma_{thy} = 320$ MeV}
& 
\m{2}{c||}{$\Gamma_{thy} = 201$ MeV}  \\
\hline 
& 
\m{2}{c||}{}
& 
\m{2}{c||}{$\Gamma_{expt} \sim 250$ MeV}  \\
\hline 
\end{tabular}
\vskip 1cm

\begin{tabular}{||c||c|l||c|l||}
\hline
\multicolumn{5}{||c||}{Table K5. 1$^3$P$_\J$ $kaons$ } \\
\hline 
& 
\m{2}{c||}{$K_2^*(1429)$} 
& 
\m{2}{c||}{$K_0^*(1412)$}  \\
\hline 
Mode
& $\Gamma_i($MeV$)$  & \m{1}{c||}{Amps.}  
& $\Gamma_i($MeV$)$  & \m{1}{c||}{Amps.}  \\
\hline \hline
$\pi K $ 
& 
56 
& 
\small $\ba{l} 
  ^1D_2=+0.12
\ea $ 
& 
113
& 
\small $\ba{l} 
  ^1S_0=+0.17
\ea $ 
\\
\hline
$\eta K $ 
& 
1 
& 
\small $\ba{l} 
  ^1D_2=-0.023
\ea $ 
& 
6
& 
\small $\ba{l} 
  ^1S_0=-0.075
\ea $ 
\\
\hline
$\rho K $ 
& 
4
& 
\small $\ba{l} 
  ^3D_2=+0.045 
\ea $ 
&  
-
& 
-
\\
\hline
$\omega K $ 
& 
1
& 
\small $\ba{l} 
  ^3D_2=+0.042 
\ea $ 
&  
-
& 
-
\\
\hline
$\pi K^* $ 
& 
13
& 
\small $\ba{l} 
  ^3P_1=-0.073 
\ea $ 
&  
-
& 
-
\\
\hline
$\pi K_1(1273) $ 
& 
0
& 
\small $\ba{l} 
  ^3P_2=+0.0093 \\ 
  ^3F_2=+0.000067 
\ea $ 
&  
0
& 
\small $\ba{l} 
  ^3P_0=+0.012 
\ea $ \\
\hline
\hline 
& 
\m{2}{c||}{$\Gamma_{thy} = 75 $ MeV}
& 
\m{2}{c||}{$\Gamma_{thy} = 119$ MeV}  \\
\hline 
& 
\m{2}{c||}{$\Gamma_{expt} = 99\pm 3$ MeV}
& 
\m{2}{c||}{$\Gamma_{expt} = 294\pm 23$ MeV}  \\
\hline 
\end{tabular}

\vskip 1cm

\begin{tabular}{||c||c|l||}
\hline
\multicolumn{3}{||c||}{Table K6. 1$^{2S+1}$P$_1$ $kaons$, general mixing } \\
\hline 
& 
\m{2}{c||}{$K_1(1273)  = c | 1^1{\rm P}_1\rangle + s | 1^3{\rm P}_1\rangle $} 
\\
\hline 
Mode
& $\Gamma_i($MeV$)$  & \m{1}{c||}{Amps.}  
\\
\hline \hline
$\rho K$ 
& 
$21.8c^2+ 61.6sc + 43.6s^2 $ 
& 
\small $\ba{l} 
  ^3S_1=-0.242c - 0.342s \\ 
  ^3D_1=-0.00202c + 0.00143s 
\ea $ 
\\
\hline

$\pi K^* $ 
& 
$59.6 c^2 - 158.7 sc + 115.7 s^2 $
& 
\small $\ba{l} 
  ^3S_1 -0.204 c + 0.288 s \\ 
  ^3D_1=-0.0411 c - 0.290 s  
\ea $ 
\\
\hline
\hline
& 
\m{2}{c||}{$\Gamma_{thy} = 81 c^2 - 97 sc  + 159 s^2 $ MeV}
\\
\hline 
& 
\m{2}{c||}{$\Gamma_{expt} = 90\pm 20$ MeV}
\\
\hline 
\end{tabular}
\vskip 1cm

\begin{tabular}{||c||c|l||}
\hline
\multicolumn{3}{||c||}{Table K7. 1$^{2S+1}$P$_1$ $kaons$, general mixing } \\
\hline 
& 
\m{2}{c||}{$K_1(1402)  = -s | 1^1{\rm P}_1\rangle + c | 1^3{\rm P}_1\rangle $} 
\\
\hline 
Mode
& $\Gamma_i($MeV$)$  & \m{1}{c||}{Amps.}  \\
\hline \hline
$\rho K$ 
& 
$160.0 c^2 - 219.9 sc + 82.3 s^2 $
& 
\small $\ba{l} 
  ^3S_1=-0.290 c + 0.205 s  \\ 
  ^3D_1=+0.0282 c + 0.0399 s 
\ea $ 
\\
\hline

$\omega K$ 
& 
$52.3  c^2 - 72.3  sc + 26.8 s^2 $ 
& 
\small $\ba{l} 
  ^3S_1=-0.294 c + 0.208 s  \\ 
  ^3D_1=+0.0259 c + 0.0366 s
\ea $ 
\\
\hline

$\pi K^* $ 
& 
$141.1 c^2 + 176.2 sc + 78.8 s^2 $
& 
\small $\ba{l} 
  ^3S_1=+0.247c + 0.175 s \\ 
  ^3D_1 = -0.0498 c + 0.0704 s  
\ea $ 
\\
\hline
\hline
& 
\m{2}{c||}{$\Gamma_{thy} = 353 c^2 - 116 sc + 188 s^2 $ MeV}  \\
\hline 
& 
\m{2}{c||}{$\Gamma_{expt} = 174\pm 13$ MeV}  \\
\hline 
\end{tabular}

\vskip 1cm

\begin{tabular}{||c||c|l||c|l||}
\hline
\multicolumn{5}{||c||}{Table K8. 1$^{2S+1}$P$_1$ $kaons$, HQET magic mixing} \\
\hline 
& 
\m{2}{c||}{$K_1(1273)$} 
& 
\m{2}{c||}{$K_1(1402)$}  \\
\hline 
Mode
& $\Gamma_i($MeV$)$  & \m{1}{c||}{Amps.}  
& $\Gamma_i($MeV$)$  & \m{1}{c||}{Amps.}  \\
\hline \hline
$\rho K$ 
& 
58 
& 
\small $\ba{l} 
  ^3S_1=-0.39 \\ 
  ^3D_1=-0.00083
\ea $ 
& 
30
& 
\small $\ba{l} 
  ^3S_1=-0.12 \\ 
  ^3D_1=+0.046
\ea $ 
\\
\hline

$\omega K$ 
& 
- 
& 
-
& 
10
& 
\small $\ba{l} 
  ^3S_1=-0.12 \\ 
  ^3D_1=+0.042
\ea $ 
\\
\hline

$\pi K^* $ 
& 
3
& 
\small $\ba{l} 
  ^3S_1\equiv 0 \\ 
  ^3D_1=-0.050
\ea $ 
& 
203
& 
\small $\ba{l} 
  ^3S_1=+0.30 \\ 
  ^3D_1\equiv 0 
\ea $ 
\\
\hline
\hline
& 
\m{2}{c||}{$\Gamma_{thy} = 62$ MeV}
& 
\m{2}{c||}{$\Gamma_{thy} = 244$ MeV}  \\
\hline 
& 
\m{2}{c||}{$\Gamma_{expt} = 90\pm 20$ MeV}
& 
\m{2}{c||}{$\Gamma_{expt} = 174\pm 13$ MeV}  \\
\hline 
\end{tabular}

\vskip 1cm

\begin{tabular}{||c||c|l||c|l||}
\hline
\multicolumn{5}{||c||}{Table K9. 2$^3$P$_\J$ $kaons$ } \\
\hline 
& 
\m{2}{c||}{$K_2^*(1850)$} 
& 
\m{2}{c||}{$K_0^*(1850)$}  \\
\hline 
Mode
& $\Gamma_i($MeV$)$  & \m{1}{c||}{Amps.}  
& $\Gamma_i($MeV$)$  & \m{1}{c||}{Amps.}  \\
\hline \hline

$\pi K $ 
& 
44 
& 
\small $\ba{l} 
  ^1D_2=+0.077
\ea $ 
& 
29
& 
\small $\ba{l} 
  ^1S_0=-0.062
\ea $ 
\\
\hline

$\eta K $ 
& 
1 
& 
\small $\ba{l} 
  ^1D_2=-0.023
\ea $ 
& 
0
& 
\small $\ba{l} 
  ^1S_0=+0.011
\ea $ 
\\
\hline

$\eta' K $ 
& 
15 
& 
\small $\ba{l} 
  ^1D_2=+0.10
\ea $ 
& 
11
& 
\small $\ba{l} 
  ^1S_0=+0.083
\ea $ 
\\
\hline

$\rho K $ 
& 
44 
& 
\small $\ba{l} 
  ^3D_2=+0.087
\ea $ 
& 
-
& 
-
\\
\hline

$\omega K $ 
& 
14 
& 
\small $\ba{l} 
  ^3D_2=+0.086
\ea $ 
& 
-
& 
-
\\
\hline 
\end{tabular}
\vskip 1cm

\begin{tabular}{||c||c|l||c|l||}
\hline
\multicolumn{5}{||c||}{Table K11. 1$^3$D$_\J$ $kaons$ } \\
\hline 
& 
\m{2}{c||}{$K_3^*(1776)$} 
& 
\m{2}{c||}{$K^*(1717)$}  \\
\hline 
Mode
& $\Gamma_i($MeV$)$  & \m{1}{c||}{Amps.}  
& $\Gamma_i($MeV$)$  & \m{1}{c||}{Amps.}  \\
\hline \hline

$\pi K $ 
& 
40 
& 
\small $\ba{l} 
  ^1F_3=+0.077
\ea $ 
& 
45
& 
\small $\ba{l} 
  ^1P_1=+0.085
\ea $ 
\\
\hline

$\eta K $ 
& 
19 
& 
\small $\ba{l} 
  ^1F_3=+0.098
\ea $ 
& 
53
& 
\small $\ba{l} 
  ^1P_1=+0.17 
\ea $ 
\\
\hline

$\eta' K $ 
& 
0 
& 
\small $\ba{l} 
  ^1F_3=-0.0062
\ea $ 
& 
1
& 
\small $\ba{l} 
  ^1P_1=-0.030 
\ea $ 
\\
\hline

$\rho K $ 
& 
10 
& 
\small $\ba{l} 
  ^3F_3=+0.045
\ea $ 
& 
26
& 
\small $\ba{l} 
  ^3P_1=-0.075
\ea $ 
\\
\hline

$\omega K $ 
& 
3 
& 
\small $\ba{l} 
  ^3F_3=+0.043
\ea $ 
& 
8
& 
\small $\ba{l} 
  ^3P_1=-0.075
\ea $ 
\\
\hline 
\end{tabular}
\vskip 1cm

\begin{tabular}{||c||c|l||c|l||}
\hline
\multicolumn{5}{||c||}{Table K14. 1$^3$F$_\J$ $kaons$ } \\
\hline 
& 
\m{2}{c||}{$K_4^*(2045)$} 
& 
\m{2}{c||}{$K_2^*(2050)$}  \\
\hline 
Mode
& $\Gamma_i($MeV$)$  & \m{1}{c||}{Amps.}  
& $\Gamma_i($MeV$)$  & \m{1}{c||}{Amps.}  \\
\hline \hline

$ \pi K$
& 
21
& 
\small $\ba{l} 
  ^1G_4=+0.048
\ea $ 
& 
20
& 
\small $\ba{l} 
  ^1D_2=+0.046
\ea $ 
\\
\hline

$\eta K $
& 
0
& 
\small $\ba{l} 
  ^1G_4=-0.011
\ea $ 
& 
1
& 
\small $\ba{l} 
  ^1D_2=-0.015
\ea $ 
\\
\hline

$\eta' K $
& 
2
& 
\small $\ba{l} 
  ^1G_4=+0.029
\ea $ 
& 
15
& 
\small $\ba{l} 
  ^1D_2=+0.083
\ea $ 
\\
\hline

$\rho K $
& 
7
& 
\small $\ba{l} 
  ^3G_4=+0.030
\ea $ 
& 
13
& 
\small $\ba{l} 
  ^3D_2=-0.042
\ea $ 
\\
\hline

$ \omega K $
& 
2
& 
\small $\ba{l} 
  ^3G_4=+0.029
\ea $ 
& 
4
& 
\small $\ba{l} 
  ^3D_2=-0.041
\ea $ 
\\
\hline
\end{tabular}
\vskip 1cm

\begin{tabular}{||c||c|l||c|l||}
\hline
\multicolumn{5}{||c||}{Table K14. 1$^3$F$_\J$ $kaons$  $(cont.)$} \\
\hline 
& 
\m{2}{c||}{$K_4^*(2045)$} 
& 
\m{2}{c||}{$K_2^*(2050)$}  \\
\hline 
Mode
& $\Gamma_i($MeV$)$  & \m{1}{c||}{Amps.}  
& $\Gamma_i($MeV$)$  & \m{1}{c||}{Amps.}  \\
\hline \hline

$ \pi(1300) K$
& 
0
& 
\small $\ba{l} 
  ^1G_4=+0.0036
\ea $ 
& 
0
& 
\small $\ba{l} 
  ^1D_2=+0.0083
\ea $ 
\\
\hline

$\eta(1293) K$
& 
0
& 
\small $\ba{l} 
  ^1G_4=+0.0038
\ea $ 
& 
0
& 
\small $\ba{l} 
  ^1D_2=+0.0083
\ea $ 
\\
\hline

$\eta_s(1415) K$
& 
0
& 
\small $\ba{l} 
  ^1G_4=-0.0015
\ea $ 
& 
0
& 
\small $\ba{l} 
  ^1D_2=-0.0085
\ea $ 
\\
\hline

$ \rho(1465) K$
& 
0
& 
\small $\ba{l} 
  ^3G_4=+0.00046
\ea $ 
& 
0
& 
\small $\ba{l} 
  ^3D_2=-0.0035
\ea $ 
\\
\hline

$\omega(1419) K$
& 
0
& 
\small $\ba{l} 
  ^3G_4=+0.0011
\ea $ 
& 
0
& 
\small $\ba{l} 
  ^3D_2=-0.0048
\ea $ 
\\
\hline 
\end{tabular}
\vskip 1cm

\begin{tabular}{||c||c|l||}
\hline
\multicolumn{3}{||c||}{Table K15. 1$^{2S+1}$F$_3$ $kaons$, general mixing  $(cont.)$} \\
\hline 
& 
\m{2}{c||}{$K_3^a(2050)$}  \\
\hline 
Mode
& $\Gamma_i($MeV$)$  & \m{1}{c||}{Amps.}  \\
\hline \hline

$\rho(1465) K$ 
& 
$   0.1 sc $
& 
\small $\ba{l} 
  ^3D_3 =  -0.00358  c - 0.00414 s  \\
  ^3G_3 =  -0.000520 c + 0.000451 s 
\ea $ 
\\
\hline

$\omega(1419) K$ 
& 
$0.0 $ 
& 
\small $\ba{l} 
  ^3D_3 =  -0.00497  c - 0.00573  s  \\
  ^3G_3 =  -0.00120  c + 0.00104  s 
\ea $ 
\\
\hline

$\pi K_1(1273) $ 
& 
$    1.3  c^2 + 5.7  sc + 6.1 s^2   $
& 
\small $\ba{l} 
  ^3F_3 = - 0.0161 c  - 0.0346 s
\ea $ 
\\
\hline 
\end{tabular}

\vfill\eject


\end{document}